\documentclass[traditabstract]{aa} 
\usepackage{epsfig,amsmath}
\usepackage[dvips]{color}
\usepackage{verbatim}
\usepackage{graphicx}
\usepackage{txfonts}
\usepackage{natbib}
\bibpunct{(}{)}{;}{a}{}{,}
\usepackage{hyperref}
\usepackage[hyphenbreaks]{breakurl}
\usepackage{longtable}
\usepackage{lscape}
\usepackage[toc,page]{appendix}
\usepackage{url}

\newcommand\fion[2]{$[$#1{\sc #2}$]$}
\definecolor{Black}{named}{Black}
\definecolor{Red}{named}{Red}
\definecolor{Green}{named}{Green}
\definecolor{Blue}{named}{Blue}

\begin{document}
\title{Emission line selected galaxies at $z=0.6-2$ in GOODS South: Stellar masses, SFRs, and large scale structure}

\titlerunning{Emission line selected galaxies at $z=0.6-2$ in GOODS-S}
\authorrunning{Kochiashvili et.al.}

\author{I. Kochiashvili\inst{1,2},
P. M\o ller \inst{3},
B. Milvang-Jensen \inst{1},
L. Christensen\inst{1},
J.P.U. Fynbo \inst{1}, 
W. Freudling \inst{3},
B. Cl\'ement \inst{4,5},
J.-G. Cuby \inst{6},
J. Zabl \inst{1},
S. Zibetti \inst{7}
}

\institute{Dark Cosmology Centre, Niels Bohr Institute, University of Copenhagen,
Juliane Maries Vej 30, DK-2100 Copenhagen \O, Denmark.
\and Abastumani Astrophysical Observatory, Ilia State University, Kakutsa Cholokashvili Ave 3/5, Tbilisi 0162, Georgia
\and European Southern Observatory, Karl Schwarzschild Strasse 2, D-85748 Garching bei M\"unchen, Germany
\and
Steward Observatory, University of Arizona, 933 North Cherry Avenue, Tucson, AZ, 85721, USA
\and
CRAL, Observatoire de Lyon, Universite ́ Lyon 1, 9 Avenue Ch. Andre ́, 69561 Saint Genis Laval Cedex, France
\and
Aix Marseille Universit\'e, CNRS, LAM (Laboratoire d'Astrophysique de Marseille) UMR 7326, 13388, Marseille, France
\and
INAF-Osservatorio Astrofisico di Arcetri, Largo Enrico Fermi 5, I-50125 Firenze, Italy
}  

\offprints{ia@dark-cosmology.dk}

\date{Received  / Accepted }

\abstract{We have obtained deep NIR narrow and broad (J and Y) band
imaging data of the GOODS-South field. The narrow band filter is
centered at 1060 nm
corresponding to redshifts $z = 0.62, 1.15, 1.85$ for the strong emission
lines H$\alpha$, \fion{O}{iii}/H$\beta$ and \fion{O}{ii}, respectively.
From those data we extract a well defined sample ($M(AB)=24.8$ in the
narrow band) of 
objects with large emission line equivalent widths in the narrow band. 
Via SED fits to published broad band data we identify which of the three lines 
we have detected and assign redshifts accordingly. This results in a well defined, 
strong emission line selected sample of galaxies down to lower masses than
can easily be obtained with only continuum flux limited selection techniques.
We compare the (SED fitting-derived) main sequence of star-formation (MS) 
of our sample to previous works and find that it has a steeper slope than that of
samples of more massive galaxies. We conclude that the MS steepens
at lower (below $M_{\star} = 10^{9.4} M_{\odot}$) galaxy masses.
We also show that the SFR at any redshift is higher in our
sample. We attribute this to the targeted selection of galaxies
with large emission line equivalent widths, and conclude that our
sample presumably forms the upper boundary of the MS.

We briefly investigate and outline how samples with accurate redshifts
down to those low stellar masses open a new window to study the formation
of large scale structure in the early universe. In particular we report
on the detection of a young galaxy cluster at $z=1.85$ which
features a central massive galaxy which is the candidate of an early stage
cD galaxy, and we identify a likely filament mapped out by
 \fion{O}{III} and $H\beta$ emitting galaxies at $z=1.15$.
}


\maketitle{}

\section{Introduction}

The study of galaxies at both intermediate and high redshifts has
gained tremendous momentum from the concerted efforts to gather deep
imaging of large fields, and from the ensuing high quality photometry
covering large spectral ranges. Analyses exploiting those data to
derive prime observables such as star-formation rates (SFRs) and
stellar masses $M_{\star}$ have revealed that galaxies follow scaling
relations that evolve with redshifts \citep{Brinchmann04, Noeske07,Daddi07}. 
The most comprehensive investigations are based on multi-band photometry, 
and the ability to obtain redshift information via fitting of theoretical model data 
is a critical component \citep{Daddi07, Karim11,Bayliss'11, Koyama13}. 
The photometric redshift accuracy also places a fundamental limitation 
on the results from the unavoidable uncertainty in the assignment of redshifts 
to each galaxy, an uncertainty which propagates to all the derived physical parameters
of the galaxies.

There are different methods of addressing the galaxy formation and
evolution quest. Galaxy samples are selected differently and therefore
probe different aspects of galaxy evolution. Intensively starforming galaxies
have been studied for nearly two decades thanks to the Lyman-break
selection technique \citep{Steidel03,Shapley11}. Flux limited high-redshift 
samples selected at primarily red wavelengths include Luminous Infrared 
Galaxies (LIRGs), Ultra Luminous Infrared Galaxies (ULIRGs), and massive 
($M_{\star}\sim10^{10.7} M_{\odot}$) red ellipticals \citep{Jacobs11}. Sub-mm 
selected samples target high-redshift galaxies with unprecedented 
star-formation rates \citep{Michalowski10,Hodge13}. Long-duration 
gamma-ray bursts (GRBs) select fainter and bluer star-forming galaxies 
\citep{LeFloc'h03, Christensen04}. Also here selection effects play a
role, as GRB hosts have been suggested to have low stellar masses
\citep[e.g.,][]{Castro10}, while dusty GRBs occur preferentially in more 
massive host galaxies \citep{Kruhler11}. Absorption-line selected samples 
allow us to study the gas content of galaxies and can be used to probe the
mass-metallicity relation \citep{Ledoux06,Moller13, Christensen14}. 
In a nutshell, these methods all address different populations of galaxies 
and have different advantages and disadvantages for particular science goals.

In order to investigate the $M_{\star}$ vs SFR relation for galaxies found in
isolation and in clusters, none of these methods will simultaneously
probe the low-mass end of the star-forming main sequence and cover 
intermediate-to-high redshifts. An alternative method that can
help us in achieving this goal is the narrow-band imaging technique
\citep[e.g.,][]{Pritchet1987}. Emission-line selected samples are
smaller, but the advantage is that they allow us to probe fainter
objects than broad-band selected samples do and still have a much more
accurate photometric redshift determination \citep{Ly12,Sobral14}. Narrow-band selected
objects have excess flux in the narrow-band filter compared to a
broad-band filter that covers adjacent wavelengths. Primarily, this
technique has been used to detect high redshift Lyman-$\alpha$ (Ly$\alpha$) 
emission lines because $Ly\alpha$ is a good tracer of galaxies at the 
beginning of the reionization era \citep[][]{PP67, Malhotra04,Nilsson2007}.

The scope of this paper is to fill in the knowledge gap concerning the
low-mass end of the main-sequence of star-forming galaxies in a broad
redshift range.  We analyse emission-line sources selected from deep
1060 nm narrow-band ($NB1060$ hereafter) and $Y$- and $J$-band observations
of the GOODS-South field from \citet{Clement2011}. The GOODS-South
field is ideal for our objective as the field has been observed in a
wide range of wavelengths and with a good photometric accuracy
\citep{GOODS'03} allowing for very detailed photometric
scrutiny of sources in the field. When searching for emission-line
galaxies at redshifts $z\sim7.7$, we also detect galaxies with emission
lines other than $Ly\alpha$ falling within the narrow-band filter. In this
way, we can probe the universe in four independent redshift slices:
besides the high-redshift $Ly\alpha$ line, we detect galaxies at $z = 0.6$ 
from strong $H\alpha$ emission lines, at $z = 1.12/1.18$ from \fion{O}{iii}/$H\beta$ 
emission lines, and $z = 1.85$ where galaxies with strong \fion{O}{ii} emission lines lie. 
We perform multi-band photometry SED fitting and derive masses and SFRs of 40 
emission-line galaxies at three different redshift slices. We analyse the redshift evolution 
of the $M_{\star}$-SFR relation spanning more than four decades in stellar mass from a 
unique data set.

The paper is organized as follows: in Sect.~2 we describe candidate selection process and datasets used for this
project. Sect.~3 characterizes spectroscopic and photometric properties of the
selected galaxies and compare with redshifts from the MUSYC
survey. Sect.~4 and ~5 present the results and a discussion of
these. 

Throughout this paper, we assume a flat cosmology with
$\Omega_{\Lambda}=0.70$, $\Omega_m=0.30$ and a Hubble constant of
$H_0=70$~km~s$^{-1}$~Mpc$^{-1}$.

\section{Selection of emission-line galaxies}

\begin{figure}
\begin{flushleft}
\epsfig{file=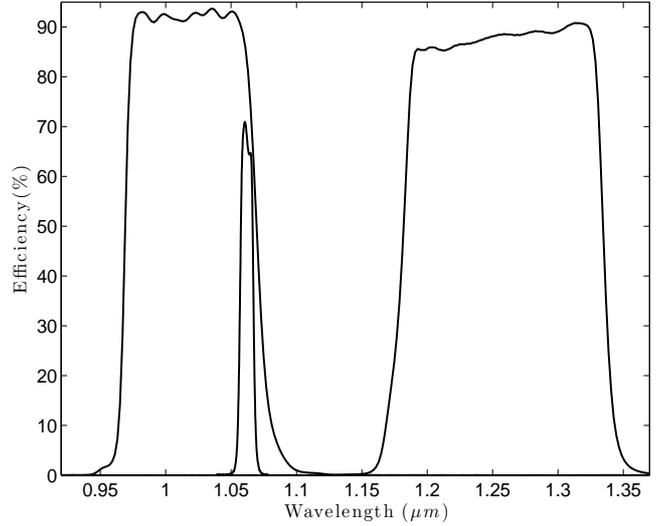, height=7cm,width=8.5cm}
\caption{The transmission curves for the $NB1060$, $Y$, and $J$-band
filters.
The narrow filter transmission is located in the red wing of the $Y$-band
filter and is entirely outside the $J$-band transmission range.
}
\label{filters}
\end{flushleft}
\end{figure}

\begin{figure}
\begin{center}
\epsfig{file=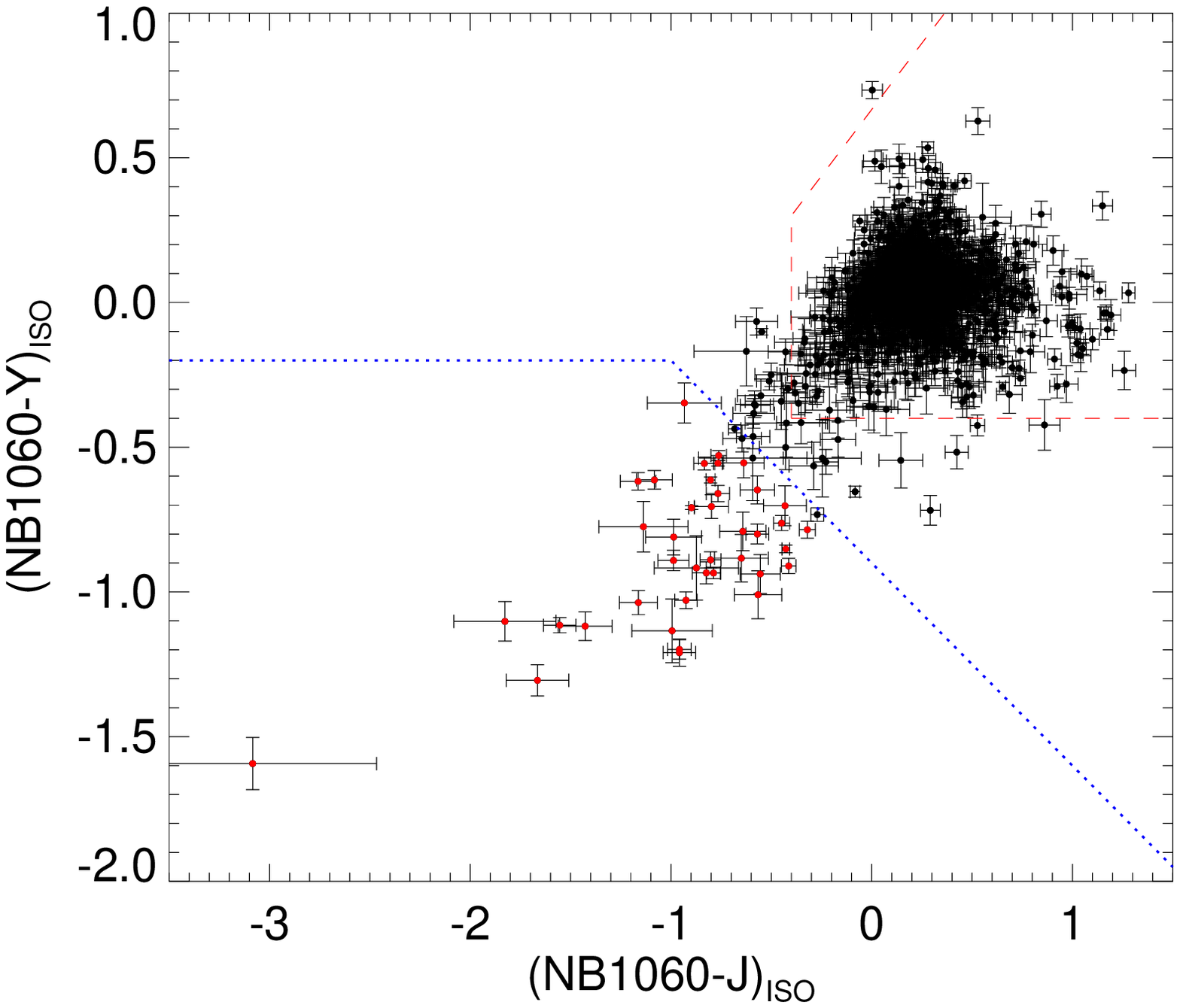,width=8.5cm}\\
\epsfig{file=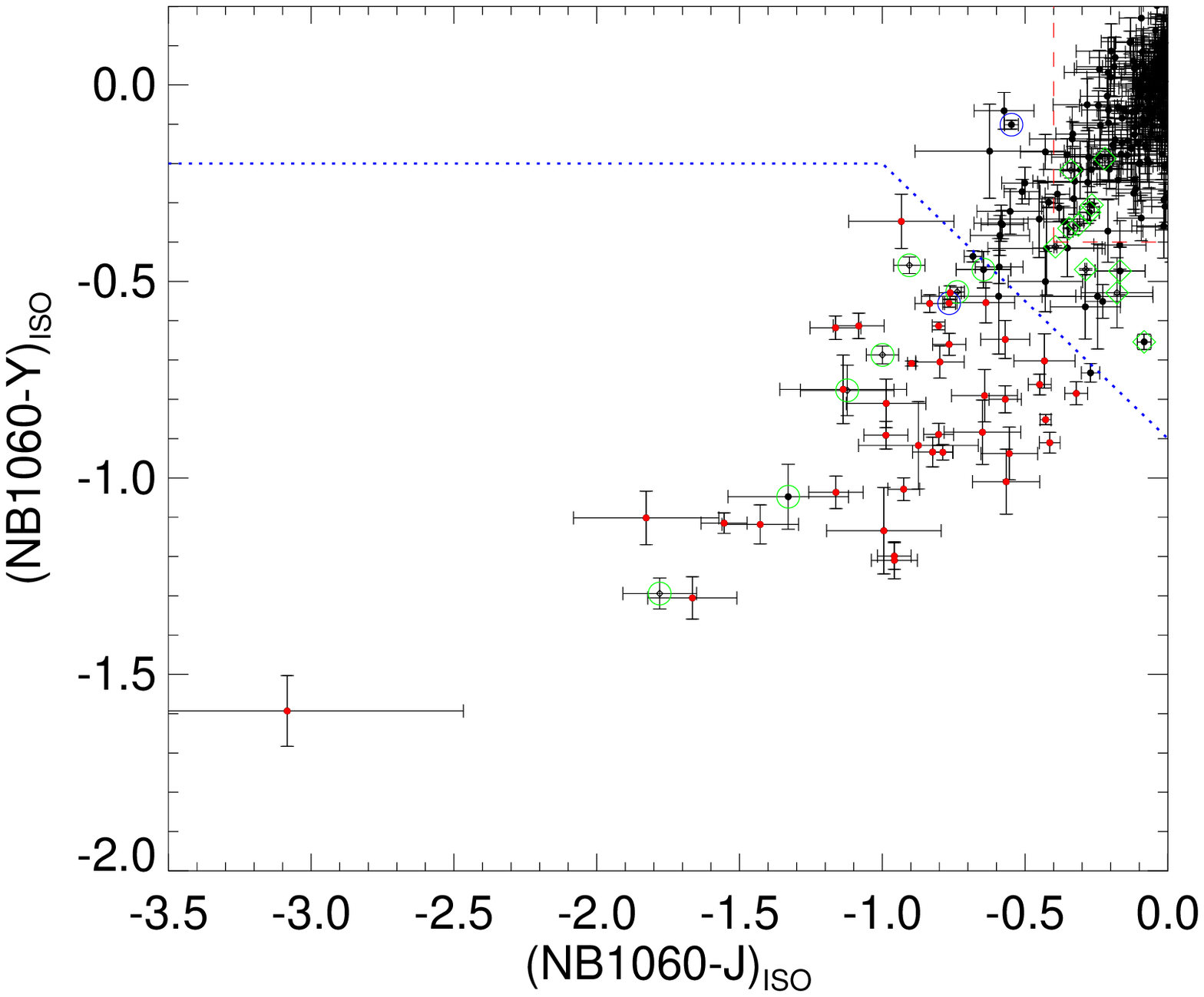,width=8.5cm}
\caption{
The colour-colour diagram for objects detected in the NB1060 image and
brighter than NB1060(AUTO) = 24.8. The top panel represents the colour
distribution of continuum and emission-line galaxies. The expected region
occupied by continuum emitters is enclosed by a red dashed line, whereas the
region we use to select candidate line-emitters lies below the blue dotted
line. Red dots represent objects from the basic sample, i.e.\ objects that
meet the selection criteria.  The lower panel additionally shows in green
circles and green diamonds objects that have emission-lines but do not enter
our basic sample due to either being masked or being outside conservatively
defined selection area (therefore above the blue dotted line). \label{colplot}
}
\end{center}
\end{figure}

\subsection{Imaging observations}
The GOODS South field was observed with VLT/HAWK-I in the 1060 nm narrow-band
and broad J- and Y-band filters (see filter transmission curves in
Fig.\ref{filters})
as part of a Large ESO Programme (Prog-Id: 181.A-0485, PI: Cuby) and
a HAWK-I science verification programme (Prog-Id: 60.A-9284(B), PI: Fontana).
For details on the observations and data reduction we refer
to \citet{Castellano10} and \citet{Clement2011}. 
The field is in the northern half of the GOODS-S field (centred at
RA,Dec = $03^{h}32^{m}29^{s}, $-$27^{d}44^{m}42^{s}$, J2000).

\subsection{Candidate selection}
\label{LEGOs}
For object detection and photometry, we use the software package SExtractor
(Bertin \& Arnouts 1996). For the actual selection of candidate emission
line galaxies we rely only on the Hawk-I $NB1060$, $Y$ and $J$-band images.
As a detection image we use the narrow-band image, and photometry is
subsequently done in all three images with aperture sizes defined in the
$NB1060$ image.  Before object detection the detection image is convolved
with a Gaussian filter function having a FWHM equal to that of point
sources. We use a detection threshold of 1.5 times the background
sky-noise in the unfiltered detection image and a
minimum area of 15 connected pixels above the detection threshold in the
filtered image. Isophotal
apertures are defined on the detection image and those same isophotal
apertures are used in the different bands ($NB1060$, $Y$, $J$).
We reject objects close to the chip gap and the edge of the image where
the noise is higher. The regions of the field masked out in this way
are shaded grey in Fig.~\ref{SKYMAP}. In total, we detect 2700
objects at a signal-to-noise ratio greater than 5 in the
narrow-band.  We measure the flux of all objects in the isophotal
aperture which is suitable for precise colour measurement
as the effective seeing of the images are very similar.
To get a measure of the total magnitudes we use the so-called AUTO
aperture in SExtractor. The AUTO aperture is an elliptical aperture defined
by the isophotal shape of the object. For objects blended with neighbours a
scaled isophotal flux is used to estimate the total flux.
Our final catalog is complete   (10$\sigma$ detection) down to $M(AB)=24.8$
in the narrow-band.

In order to select objects with excess flux in the narrow-band we employ
the method introduced by \citet{MoellerP93} and refined by
\citet{Fynbo03}. This method uses two broad band filters which
bracket the narrow-band. Plotting the two
narrow-minus-broad colours against each other causes objects with an
emission line within the narrow pass-band to drop diagonally down to the
left (Fig.~\ref{colplot} upper panel). We compute the distribution of the
cloud of continuum emitters using theoretical spectral energy distributions
from Bruzual \& Charlot (2003), and enclose the region where the model
galaxies fall with a red dashed line in
Fig.~\ref{colplot} (for details see \citet{Fynbo03}). All objects in our
catalog are plotted in Fig.~\ref{colplot}, upper panel, and it is seen
that most objects do indeed fall inside the red dashed line. The dotted
blue line marks the selection window we have adopted below and to the left
side of the main locus of continuum objects.  For $NB1060 - J < -1$
we select objects with $NB1060 - Y < -0.2$. For $NB1060 - J > -1$ we use
$NB1060 - Y < -0.7\times(NB1060 - J) -0.9$. The 40 objects found inside
this area, and at least 1$\sigma$ from the border, make up our ''basic
sample'', they are listed in Table \ref{Table1} and are highlighted
red in Fig.~\ref{colplot}. The basic sample is complete in the
sense that we have included all objects within the unmasked area of the
observed field down to $NB1060=24.8$, and it is therefore
suitable for statistical studies within the unmasked area which spans
38.7 square arcminutes on the sky.

We searched the NED/IPAC\footnote{The NASA/IPAC Extragalactic Database (NED) is operated by
the Jet Propulsion Laboratory, California Institute of Technology, under
contract with the National Aeronautics and Space Administration.}
and SIMBAD \citep{SIMBAD07} databases, and found spectroscopic, secure
redshifts for a subset of the basic sample, as listed in Table~\ref{Table1}.

As a check of the selection, the images were inspected in ds9 in RGB mode,
with blue=Y, green=NB1060, red=J\@. 
Objects that looked green (i.e.\ showed
some degree of narrow-band excess) and which looked like galaxies and not
artifacts or noise were marked.
The mask used in defining the basic sample was not used, i.e.\ also objects
located in higher noise regions of the image were included.
After removing the basic sample of 40 galaxies and the ELG00 galaxy,
this visually-identified narrow-band excess sample comprised
58 objects. There were 3 not necessarily mutually exclusive reasons
why these galaxies were not part of the basic sample:
(1)~their colours were outside the selection region,
i.e.\ the observed EW was too low,
(2)~they were in a masked part of the image, or
(3)~they were fainter than NB1060(AUTO) = 24.8.
SIMBAD was searched, and 18 of the 58 objects had a spectroscopic, secure
redshift. For all 18 galaxies (named x01 to x18), the redshift matched an
emission line, see Table~\ref{Table1X}.
These 18 galaxies, as well as ELG00 (see below), do not fulfill our
selection criteria and thus cannot be used in our basic sample, but
together with the basic sample they form an ``extended sample''.

In addition we obtained spectra and determined redshifts of two objects
as described in Sect.~\ref{specsec}. The two objects are highlighted by
blue circles in Fig.~\ref{colplot}, where one is seen to be in our basic
sample (ELG55) while the other is directly to the left of the large cloud
of galaxies. This is an intriguingly strange position since it shows that
it has an emission line in the ($NB1060 - J)$ colour, but no line in the
($NB1060 - Y)$ colour. It is not in the basic sample so we have named it
ELG00 and it is listed in the first line of Table \ref{Table1X}.

In Fig.\ref{clipouts} we show $NB1060$, $Y$ and {\it HST} $F606$W-band
thumbnails (the latter is the deepest optical band we have) for all 40
galaxies in the basic sample, and also including ELG00 of the extended
sample. As seen, all are indeed detected
in the $F606$W-band and hence are not consistent with being Ly$\alpha$
emitters at $z=7.7$. The candidates have very mixed morphologies ranging
from bright spirals over irregular galaxies with multiple cores, to very
faint compact systems.

\section{Characterization of the candidate emission-line galaxies}
\label{Sec.3}

\subsection{Spectroscopic observations}
\label{specsec}

\begin{figure}
\begin{center}
\epsfig{file=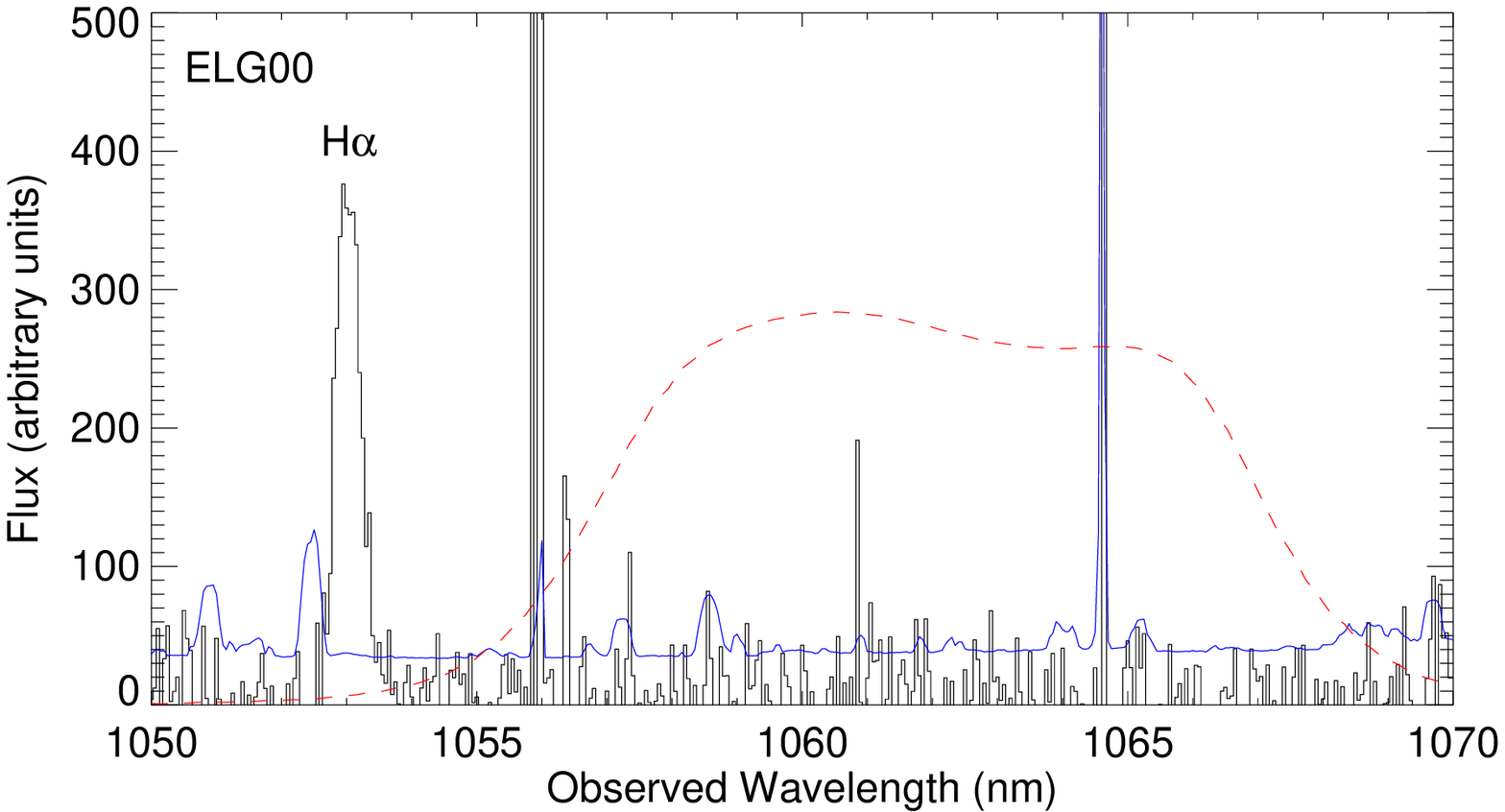,width=8.5cm}
\epsfig{file=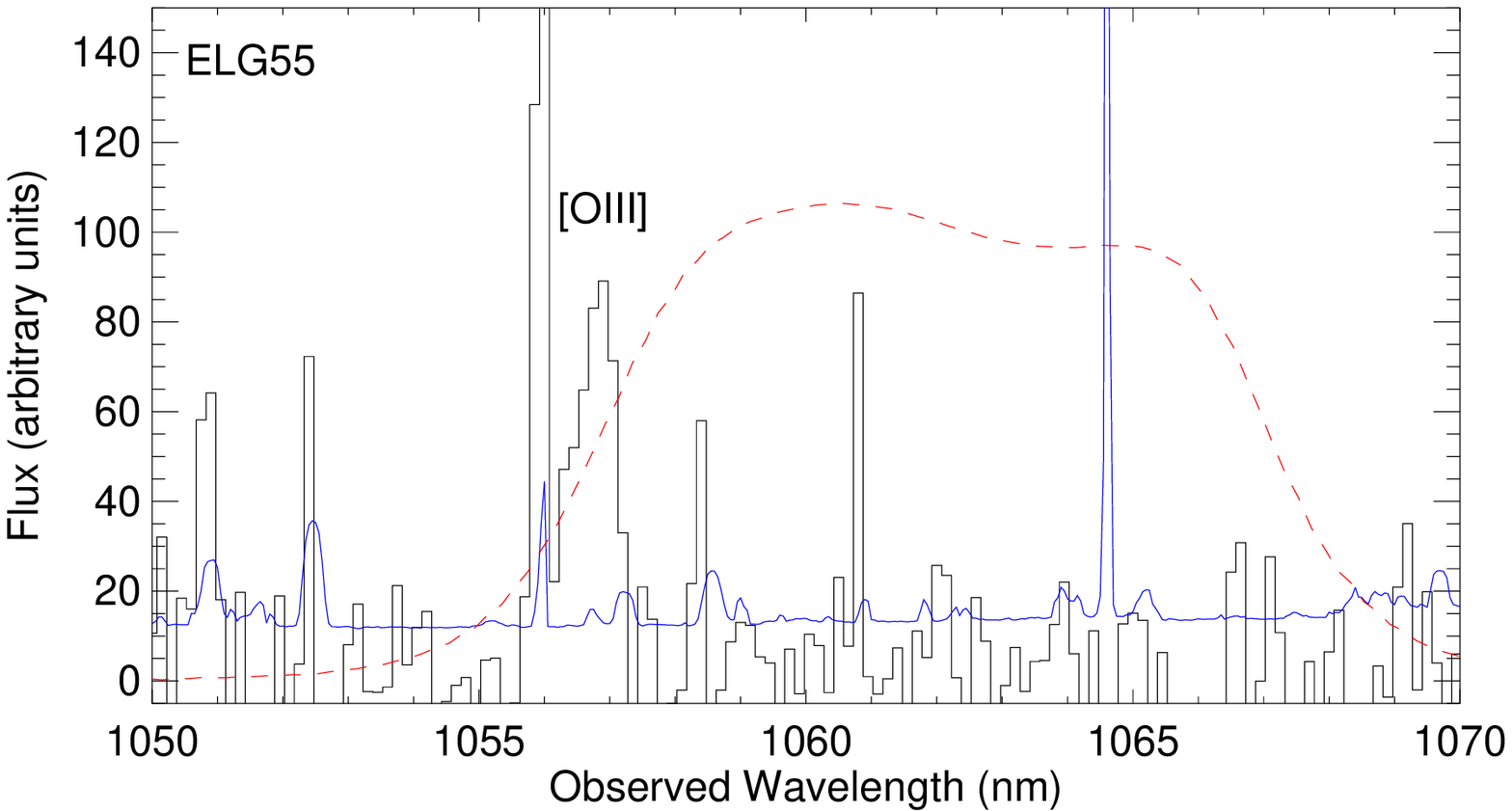,width=8.5cm}
\caption{
\label{xshooter}
X-shooter spectra of ELG00 (top) and ELG55 (bottom). The red dashed
line shows the NB1060 filter transmission curve, the blue solid line is
the error spectrum. H$\alpha$ of ELG00 is seen to be out of the narrow
band pass transmission causing its peculiar colours.
}
\end{center}
\end{figure}

On March 15 and 16 2013 we secured redshift measurements for two objects in
our catalog.
The spectra were obtained with the X-shooter spectrograph
\citep{2011A&A...536A.105V} installed at
the Cassegrain focus of the Very Large Telescope (VLT), Unit 2 -- Kueyen,
operated by the European Southern Observatory (ESO) on Cerro Paranal in
Chile (prog.\ ID 090.A-0147).
The spectra were reduced with the ESO X-shooter pipeline 2.0
\citep{2011AN....332..227G}. In Fig.~\ref{xshooter}
we show the X-shooter spectra around the region of the $NB1060$ filter.

One of the object (ELG55, lower panel of Fig.~\ref{xshooter}) is
belonging to the basic sample, and we see that the line is confirmed to
\fion{O}{iii}$\lambda$5007 based on the detection of
\fion{O}{iii}$\lambda$4959 and \fion{O}{ii}$\lambda$3727 and the derived
redshift is 1.1107.

\begin{landscape}
\begin{table}
\caption{The 40 objects of our statistically complete ``basic sample''.
In the first column we present our ID numbers for the candidate Emission-line galaxies.
 Next we list RA \& Dec, NB magnitudes, colors and redshift from our work accordingly. 
 In columns 7 and 8 we present redshifts reported in the MUSYC catalogue. Namely, $z$[peak] 
 corresponds to the best assigned redshift by the survey and [zmin] and [zmax] represent 
 $1\sigma$ minimum and maximum redshift values. Column 9 lists the emission lines 
 observed in the narrow band filter; here \fion{O}{iii} means \fion{O}{iii}/H$\beta$. 
 For 5 objects we could not uniquely assign a redshift; for four of them we have preferred 
 value, which is listed first, while for ELG30 we do not have a preferred redshift identification and
 we consider all the three listed values possible.
 Column 10 lists emission-line fluxes and column 11 and 12 corresposnd to the 
 observed frame equivalent width and references to the spectroscopic redshift literature, respectively.}
\label{Table1}
\begin{tabular}{l c c c c c c c c c c c}
\hline\hline

ID &  RA $\&$ DEC & $NB$ & $NB-Y$& $NB-J$  & Redshift & $z$[Peak] & $z$[min/max] & line ID & Em. Line flux & Eq.Width & Ref \\ [0.5ex] 
\hline
$ELG\# $&  (2000.0)            & mag (AUTO)  & mag (ISO)    &  mag (ISO)   & This work  &MUSYC & MUSYC  & & $[10^{-17} erg/s/cm^2]$ & $\AA$ \\  [0.5ex] 
\hline
03  & 03:32:40.32 -27:47:22.71 & $22.11\pm0.01$ & $-0.61\pm0.01$ &$-0.80\pm0.02$ &  $0.619$ &0.61 & 0.60/0.63& H${\alpha}$ & $17.54\pm0.14$ & $188.3\pm 2.0$ & (1)  \\
04 & 03:32:44.30 -27:46:59.99 & $23.60\pm0.03$  & $-0.62\pm0.03$ &$-1.16\pm0.09 $ & $1.144$ & 1.12 & 1.10/1.14& \fion{O}{iii} &$9.03\pm0.13$ & $288.1\pm 9.1$ & (2) \\
05  & 03:32:36.30 -27:47:32.63 & $23.43\pm0.02$ & $-1.20\pm0.03$ &$-0.96\pm0.06$ & $1.86$ & 2.29 & 2.05/2.48 & \fion{O}{ii} & $28.55\pm0.12$ & $533.8\pm11.3$ & (3) \\
06  & 03:32:37.20 -27:47:25.56 & $23.88\pm0.06$ & $-0.70\pm0.07$ &$-0.43\pm0.11$ & 1.85 &--&-- &\fion{O}{ii} & $13.28\pm0.22$ & $405.5\pm25.7$ \\
09  & 03:32:41.34 -27:46:46.23 & $24.27\pm0.05$ & $-0.55\pm0.05$ &$-0.64\pm0.10 $ & $0.62$ & 0.60 & 0.56/0.64 & H$\alpha$ &  $1.98\pm0.08$ & $155.3\pm 8.2$ \\
10 & 03:32:37.97 -27:46:51.86 & $21.03\pm0.01$ & $-0.71\pm0.01$ & $-0.90\pm0.01$ &  $0.62$ & 0.63 & 0.62/0.63 & H$\alpha$ & $43.86\pm0.41$ & $235.8\pm 2.5$ & (4)  \\
11 & 03:32:42.76 -27:46:33.19 & $24.45\pm0.05$ & $-0.81\pm0.06$ &$-0.99\pm0.14$ & $0.62$ &  0.64 & 0.62/0.66 & H$\alpha$ & $1.65\pm0.08$ & $201.2\pm10.6$ \\
12  & 03:32:37.36 -27:46:45.52 & $21.91\pm0.01$ & $-0.85\pm0.01$ & $-0.43\pm0.02$ & $1.843$ & 2.26 & 2.22/2.31 & \fion{O}{ii} & $155.23\pm0.19$ & $257.3\pm 2.7$ & (5)\\
14  & 03:32:36.83 -27:46:51.52 & $24.51\pm0.06$ & $-0.88\pm0.08$ & $-0.65\pm0.13$ & $1.85$ & 1.80 & 1.72/1.89 & \fion{O}{ii} & $2.85\pm0.08$ & $153.6\pm 9.7$ \\
15 & 03:32:37.08 -27:46:47.03 & $23.10\pm0.02$ & $-1.03\pm0.03$ & $-0.93\pm0.06$ &  $1.85$ &  1.93 & 1.84/2.04 & \fion{O}{ii} & $24.09\pm0.13$ & $287.1\pm 6.1$ \\
16  & 03:32:35.81 -27:46:43.62 & $23.26\pm0.03$ & $-0.79\pm0.03$ &$-0.32\pm0.04$ & $1.85$ &2.13 & 1.88/2.32 & \fion{O}{ii} & $21.17\pm0.13$ & $163.8\pm 5.2$ & (6) \\
20  & 03:32:36.69 -27:46:20.98 & $23.74\pm0.02$ & $-0.90\pm0.04$ & $-0.99\pm0.08$ & $1.85$ & 1.90 & 1.80/2.00 & \fion{O}{ii} & $9.42\pm0.07$ & $234.3\pm 5.0$ & (3) \\
21 & 03:32:37.45 -27:46:15.34 & $24.55\pm0.07$ & $-0.79\pm0.07$ & $-0.64\pm0.12$ & $1.85$ & 1.94 & 1.83/2.04 & \fion{O}{ii} & $4.52\pm0.09$ & $141.6\pm10.5$ \\
22 & 03:32:36.55 -27:46:12.28 & $22.60\pm0.01$ & $-0.93\pm0.02$ & $-0.79\pm0.04$ & $1.85$ & 1.94 & 1.90/1.98  & \fion{O}{ii} & $53.91\pm0.11$ & $351.3\pm 3.7$ & (3)\\
23 & 03:32:39.52 -27:45:59.75 & $24.59\pm0.08$ & $-1.13\pm0.11$ & $-0.99\pm0.20$ & $1.85$ &  1.86 & 1.43/2.31 & \fion{O}{ii} & $6.32\pm0.14$ & $314.9\pm26.7$ \\
25 & 03:32:39.33 -27:45:55.14 & $23.38\pm0.04$ & $-1.21\pm0.05$ & $-0.96\pm0.08$ & $1.85$ & 1.90 & 1.75/2.04 & \fion{O}{ii} & $22.05\pm0.24$ & $474.7\pm20.1$ \\
26 & 03:32:45.73 -27:45:24.97 & $23.72\pm0.04$ & $-0.71\pm0.04$ & $-0.80\pm0.08$ & $1.15$ & 1.09 & 1.05/1.12& \fion{O}{iii} & $5.29\pm0.12$ & $  178.7\pm 7.6$ \\
28  & 03:32:27.82 -27:46:35.07 & $24.02\pm0.03$ & $-1.31\pm0.05$ & $-1.67\pm0.16$ & $1.15$ & -- & -- & \fion{O}{iii} & $6.50\pm0.11$ & $699.8\pm22.2$ \\
30 & 03:32:30.03 -27:46:04.24 & $24.35\pm0.05$ & $-1.59\pm0.09$ & $-3.08\pm0.61$ & $1.15/1.85/0.62$ & -- & -- & \fion{O}{iii}/\fion{O}{ii}/H$\alpha$ & $2.96\pm0.16$ & $1851.6\pm97.9$ \\
34 & 03:32:26.60 -27:46:05.02 & $24.76\pm0.08$ & $-0.77\pm0.09$ & $-1.14\pm0.22$  & $1.15$ & 1.28 & 1.06/1.55 & \fion{O}{iii} & $1.38\pm0.12$ & $294.7\pm24.9$ \\
35 & 03:32:21.53 -27:46:18.71 & $23.31\pm0.03$ & $-0.80\pm0.03$ & $-0.57\pm0.06$ & $1.85$ & 1.74 & 1.50/1.96 & \fion{O}{ii} & $29.74\pm0.19$ & $420.8\pm13.4$ \\  
36 & 03:32:26.68 -27:45:54.79 & $23.98\pm0.04$ & $-1.12\pm0.05$ & $-1.43\pm0.13$ & $1.15$ & 1.18 & 1.09/1.30 & \fion{O}{iii} & $4.60\pm0.15$ & $566.7\pm24.0$ \\
37 & 03:32:21.69 -27:46:16.57 & $24.70\pm0.07$ & $-1.01\pm0.08$ & $ -0.57\pm0.12$ & $1.85$ & 2.45 & 2.32/2.59 & \fion{O}{ii} & $4.16\pm0.10$ & $274.7\pm20.3$ \\  
41 & 03:32:21.26 -27:46:02.55 & $23.68\pm0.03$ & $-0.93\pm0.04$ & $-0.82\pm0.07$ & $1.85$ & 1.53 & 1.12/1.74 & \fion{O}{ii} & $8.11\pm0.11$ & $272.6\pm 8.7$ \\  
43 & 03:32:19.60 -27:46:08.31 & $23.63\pm0.03$ & $-0.61\pm0.03$ & $-1.08\pm0.09$ & $1.15$& 1.07 & 1.05/1.09 & \fion{O}{iii} & $5.43\pm0.16$ & $683.2\pm21.7$ \\
45& 03:32:42.51 -27:44:15.55 & $23.38\pm0.03$ & $-0.66\pm0.03$ & $-0.77\pm0.06$ & $0.62$ & 0.61 & 0.60/0.63 & H${\alpha}$ & $4.73\pm0.12$ & $169.6\pm 5.4$  \\  
51 & 03:32:16.50 -27:44:45.04 & $22.49\pm0.02$ & $-0.53\pm0.02$ & $ -0.76\pm0.04$ & $1.15/0.62$ & 1.11 & 1.10/1.12  & \fion{O}{iii}/H$\alpha$ & $19.61\pm0.31$ & $809.5\pm17.1$  \\
52 & 03:32:41.59 -27:42:50.68 & $24.62\pm0.08$ & $-0.35\pm 0.08$ & $ -0.93\pm0.18$ & $1.15$ & --& -- & \fion{O}{iii} & $1.44\pm0.12$ & $238.3\pm20.2$ \\
53 & 03:32:13.15 -27:45:01.19 & $23.18\pm0.02$ & $-0.91\pm0.03$  & $-0.41\pm0.04$ & $1.85$ & 2.06 & 1.95/2.17 & \fion{O}{ii} & $27.63\pm0.12$ & $249.4\pm 5.3$ \\
54 & 03:32:12.98 -27:44:59.81 & $23.16\pm0.02$ & $-0.89\pm0.03$ & $-0.80\pm0.05$ & $1.85$ & -- & -- & \fion{O}{ii} & $35.58\pm0.15$ & $476.8\pm10.1$ \\
55 & 03:32:16.31 -27:44:41.93 & $22.03\pm0.01$ & $-0.56\pm0.01$ & $-0.77\pm0.02$ & $1.1107 $ & 1.12 & 1.11/1.12 & \fion{O}{iii} & $43.90\pm0.14$ & $159.8\pm 1.7$ & (*)\\
58 & 03:32:41.68 -27:42:04.45 & $23.50\pm0.03$ & $-1.04\pm0.04$ & $-1.16\pm0.10$ & $1.15/1.85$  &  1.25 & 1.21/1.30 & \fion{O}{iii}/\fion{O}{ii} & $4.81\pm0.15$ & $384.9\pm12.2$ \\
62 & 03:32:34.22 -27:42:31.37 & $24.37\pm0.04$ & $-1.10\pm0.07$ & $-1.83\pm0.25$ & $0.62$ & 0.61 & 0.60/0.63 & H${\alpha}$ & $1.90\pm0.07$ & $202.4\pm 8.6$ \\
65 & 03:32:39.20 -27:41:44.69 & $23.05\pm0.02$ & $-1.12\pm0.03$ & $-1.55\pm0.08$ & $1.15$ & 1.14 & 1.12/1.15 & \fion{O}{iii} & $9.54\pm0.17$ & $546.4\pm11.6$ \\
66 & 03:32:38.22 -27:41:45.51 & $24.31\pm0.07$ & $-0.92\pm0.11$ & $-0.87\pm0.21$ & $1.85/1.15$ & -- & -- & \fion{O}{ii}/\fion{O}{iii} & $5.67\pm0.19$ & $545.0\pm40.4$ \\
68 & 03:32:23.88 -27:42:11.56 & $24.07\pm0.05$ & $-0.65\pm0.05$ & $-0.57\pm0.09$ & $1.85$ & 1.60 & 1.50/1.70 & \fion{O}{ii} & $3.94\pm0.12$ & $219.5\pm11.6$ \\
70 & 03:32:33.03 -27:40:48.06 & $23.14\pm0.02$ & $-0.56\pm0.02$ & $-0.83\pm0.06$ & $1.15$ & 1.13 & 1.11/1.14 & \fion{O}{iii} & $7.26\pm0.10$ & $174.9\pm 3.7$ \\
75 & 03:32:30.36 -27:41:46.66 & $24.37\pm0.06$ & $-0.94\pm0.07$ & $-0.56\pm0.10$ & $1.85/1.15$ &  -- & -- & \fion{O}{ii}/\fion{O}{iii}\ & $  5.99\pm0.13$ & $  321.0\pm20.4$ \\
76 & 03:32:22.75 -27:42:11.59 & $23.17\pm0.02$ & $-0.76\pm0.03$ & $-0.45\pm0.04$ & $1.85$& 1.89 & 1.76/2.00 & \fion{O}{ii} & $28.31\pm0.12$ & $258.4\pm 5.5$  \\
78 & 03:32:33.89 -27:42:37:92 &$20.13\pm0.01$ & $-0.73\pm0.01$ & $-0.74\pm0.01$ & $0.624$ & 0.64 & 0.64/0.65 & H$\alpha$ & $183.81\pm0.90$ & $210.9\pm 2.2$ & (7) \\
\hline 
\hline 
\end{tabular}

(1) \citet{Ravikumar07}; (2) \citet{Xu07}; (3) \citet{WFC3GRISM11};
(4) \citet{ELG10Rodrigues08}; (5) \citet{Mignoli05};
(6) \citet{Guo12}; (*) This work
(7) \citet{Balestra10}; \\

\end{table} 
\end{landscape}

\newpage

\begin{landscape}
\begin{table}
\centering
\caption{Continuation of Table \ref{Table1}. Here we present candidates from the extended sample. First column lists ID numbers, second lists coordinates of the objects. Redshift and line IDs are listed in third and fourth columns respectively. Column 5 lists narrow-band magnitudes and magnitude errors. Columns 6 and 7 present colors and color errors for $Y$ and $J$ filters respectively.  And final three columns are emission-line fluxes, observed frame equivalent widths and references to the literature where we obtain spectroscopic redshift from.}
\label{Table1X}
\begin{tabular}{l c c c c c c c c c}
\hline\hline

ID &  RA $\&$ DEC & Redshift & line ID & $NB$ & $NB-Y$& $NB-J$  &  Em.line flux & Eq.Width  & Ref. \\ [0.5ex] 
\hline
$ $&  (2000.0)     &   spectroscpic  &  & mag (AUTO)  & mag (ISO)    &  mag (ISO)    & $[10^{-17} erg/s/cm^2]$ & $\AA$ \\  [0.5ex] 
\hline
ELG00 & 03:32:18.57 -27:42:29.50 & $0.6045$ & H$\alpha$& $22.46\pm0.01$ &$-0.10\pm0.01$ & $-0.55\pm0.02$ & $5.07\pm0.05$ & $69.6\pm0.7$ & (*) \\
x01   & 03:32:13.24 -27:42:40.03 & $0.6072$  & H$\alpha$ & $18.88\pm0.01$ & $-0.36\pm0.01$ & $-0.35\pm0.01$ & $227.56\pm0.21$ & $152.8\pm0.1$ & (1)\\
x02   & 03:32:23.40 -27:43:16.58 & $0.615$   & H$\alpha$& $19.72\pm0.01$ & $-0.47\pm0.01$ & $-0.29\pm0.01$ & $98.15\pm0.16$ & $135.4\pm0.2$ & (2) \\
x03   & 03:32:41.83 -27:40:42.31 & $0.6162$  & H$\alpha$ & $23.35\pm0.02$ & $-1.29\pm0.04$ & $-1.78\pm0.13$ & $6.50\pm0.16$ & $721.0\pm17.8$ & (1)\\
x04   & 03:32:38.59 -27:46:31.36 & $0.625$   & H$\alpha$& $20.87\pm0.01$ & $-0.31\pm0.01$ & $-0.27\pm0.01$ & $25.03\pm0.13$ &  $83.4\pm0.4$ & (3)\\
x05   & 03:32:31.50 -27:41:58.04 & $0.620$   & H$\alpha$& $23.32\pm0.02$ & $-0.32\pm0.02$ & $-0.27\pm0.03$ & $2.48\pm0.05$ &  $77.0\pm1.6$ & (1)\\
x06   & 03:32:45.65 -27:44:05.80 & $0.6206$  & H$\alpha$ & $20.15\pm0.01$ & $-0.41\pm0.01$ & $-0.39\pm0.01$ & $63.40\pm0.17$ & $127.7\pm0.3$ & (1) \\
x07   & 03:32:28.01 -27:43:57.44 & $0.6207$  & H$\alpha$ & $21.93\pm0.01$ & $-0.53\pm0.01$ & $-0.78\pm0.03$ & $12.70\pm0.16$ & $133.4\pm1.7$ & (4) \\
x08   & 03:32:40.79 -27:46:15.70 & $0.6218$  & H$\alpha$ & $19.57\pm0.01$ & $-0.36\pm0.01$ & $-0.31\pm0.01$ & $85.02\pm0.20$ &  $86.6\pm0.2$ & (1) \\
x09   & 03:32:46.75 -27:46:24.02 & $0.6250$  & H$\alpha$	& $24.86\pm0.06$ & $-0.78\pm0.06$ & $-1.12\pm0.16$ & $0.98\pm0.07$ & $169.8\pm11.5$ & (5) \\
x10   & 03:32:22.25 -27:49:01.47 & $1.109$   & \fion{O}{iii} & $22.98\pm0.02$   & $-0.19\pm0.02$ & $-0.22\pm0.03$ & $2.09\pm0.04$ &  $41.3\pm0.8$ & (6)\\
x11   & 03:32:27.66 -27:45:05.77 & $1.110$   & \fion{O}{iii} & $23.02\pm0.02$   & $-0.22\pm0.02$ & $-0.34\pm0.04$ & $3.55\pm0.09$ &  $86.8\pm2.2$ & (6)\\
x12   & 03:32:26.77 -27:45:30.63 & $1.122$   & \fion{O}{iii} & $22.97\pm0.02$   & $-0.69\pm0.02$ & $-1.00\pm0.06$ & $7.11\pm0.17$ & $290.4\pm6.8$ & (6)\\
x13   & 03:32:18.81 -27:49:08.59 & $1.128$   & \fion{O}{iii} & $23.19\pm0.02$   & $-0.46\pm0.02$ & $-0.91\pm0.05$ & $4.66\pm0.10$ & $177.2\pm3.7$ & (7) \\
x14   & 03:32:49.83 -27:46:58.30 & $1.174$   & H$\beta$ & $24.70\pm0.06$   & $-1.05\pm0.08$ & $-1.33\pm0.21$ & $1.44\pm0.09$ & $293.3\pm17.8$ & (1) \\
x15   & 03:32:17.11 -27:42:20.95 & $1.749$   & \fion{Ne}{iii} & $24.52\pm0.05$   & $-0.47\pm0.05$& $-0.65\pm0.10$ & $0.51\pm0.03$ &  $41.2\pm2.2$ & (8) \\
x16   & 03:32:38.80 -27:47:14.82 & $1.836$   & \fion{O}{ii} & $22.67\pm0.02$   & $-0.65\pm 0.02$& $-0.08\pm0.02$ & $8.17\pm0.16$ & $209.6\pm4.0$ & (9) \\
x17   & 03:32:18.43 -27:42:51.95 & $1.846$   & \fion{O}{ii} & $25.08\pm0.09$   & $-0.53\pm0.09$ & $-0.18\pm0.13$ & $0.56\pm0.05$ &  $93.5\pm9.2$ & (8) \\
x18   & 03:32:36.69 -27:46:48.48 & $1.86 $   & \fion{O}{ii} & $24.60\pm0.07$ & $-0.47\pm0.06$ & $-0.17\pm0.09$ & $0.91\pm0.07$ & $101.8\pm7.9$ & (10) \\
\hline
\end{tabular}
\begin{flushleft}
(*) This work; (1) \citet{Balestra10}; (2) \citet{Vanzella05}; (3) \citet{Szoloky04}; 
 (4) \citet{LeFevre04}; (5)\citet{Xia11}; (6) \citet{Vanzella06};
  (7) \citet{Villforth12}; (8) \citet{Straughn11}; (9) \citet{Guo12}; (10) \citet{WFC3GRISM11}; 
\end{flushleft}
\end{table} 
\end{landscape}

The other object (ELG00, upper panel of Fig.~\ref{xshooter}) is not in
the basic sample but was observed because of its strange position in the
colour-colour plot as described in section 2.2 above. Here we see a strong
H$\alpha$ line (based on the detection of a wide range of other lines in
the visual spectral region) and the derived redshift is 0.6045.
The strong H$\alpha$ line is located in the very wing of the
filter curve as given by the
ESO web page \footnote{\url{http://www.eso.org/sci/facilities/paranal/instruments/hawki/inst/filters/hawki_NB1060.dat}}
We do not detect the \fion{N}{ii}$\lambda$6583 line in the spectrum.

\subsection{Photometric redshifts}
\label{LePhare}

The very conservative selection criteria employed for our basic sample
definition ensures
that a strong emission line is present in the narrow-band filter.
Therefore the task of redshift determination of our narrow band selected
sample is reduced to determining which of the three most likely redshift
groups each object belongs to, H$\alpha$, \fion{O}{iii}/H$\beta$,
or \fion{O}{ii}. In a few cases we already
have spectroscopic confirmations, for the remainder we
rely on photometric redshift analysis. For this we take
advantage of the variety of photometric data available for the GOODS field.
We explored a wide range of available data sets, and in the end we
concluded that the most robust results are obtained using primarily the
available photometry from the CANDELS survey \citep{Guo13CANDELS_GOODS} (G13
hereafter).  CANDELS is a survey including nearly 35000 sources combining data
from among others {\it HST}-WFC3 and {\it HST}-ACS, VLT-VIMOS, VLT-HawkI, 
VLT-ISAAC and Spitzer/IRAC, spanning wavelengths from the UV to the near-infrared. The CANDELS
catalogue contains magnitudes and magnitude errors for in total 17 different
bands. To construct the catalogue a careful and complete source detection
algorithm as well as flux derivation methods including aperture
corrections were employed. However, no photometric or spectroscopic
redshift information is provided in the catalogue.

For a subset of the objects Y-band (F105W) photometry was not available in G13 (the
last 14 in Table~\ref{Table1}). For these targets we added our own Y-band
photometry (from HAWK-I) to the data sets before the SED fitting and redshift determination.  For
these objects we performed aperture photometry in circular apertures. The
aperture size was matched to the apparent extension of the object on the sky.
For each used aperture size we determine aperture corrections measured on
isolated, unsaturated point-sources.

For the Spectral Energy Distribution (SED) fits we use the LePhare
code \citep{Arnouts99,Ilbert06}. Those fits provide also a first
photometric redshift probability distribution which we use to guide us
towards the final ``redshift slice'' assignments for each object.

To construct the model SED we use the Bruzual and Charlot (BC03) spectral
library \citep{BC03}.
The library uses stellar evolutionary tracks for different metallicities
and Helium abundances from the Padova 1994 stellar synthesis models.
It generates spectra in the wavelength range from 3200 to 9500
$\AA$ at higher resolution and across wider wavelength range, 91 $\AA$ to $160\mu m$ with lower resolution, assuming Chabrier initial mass function (IMF) \citep{Chabrier03} and the Calzetti extinction
law \citep{Calzetti00}. The ages for the model galaxies range from$10^{5}$ to
$2\times10^{10}$ yr. The code is based on the exponentially declining
star formation history (SFH). We also include contribution from the
emission lines in the models. For this, LePhare uses a simple recipe
based on the \cite{Kennicutt98} SFR and UV luminosity relation. The code includes the strongest emission lines, like  $Ly\alpha$, $H\alpha$, $H\beta$, \fion{O}{iii} doublet - $\lambda\lambda4959, 5007 \AA$ and \fion{O}{ii}, varying the ratio of the above-mentioned lines with \fion{O}{ii}. For further details on LePhare code characteristics, see \cite{Ilbert06} and the LePhare manual.

\begin{figure}

\begin{center}
\epsfig{file=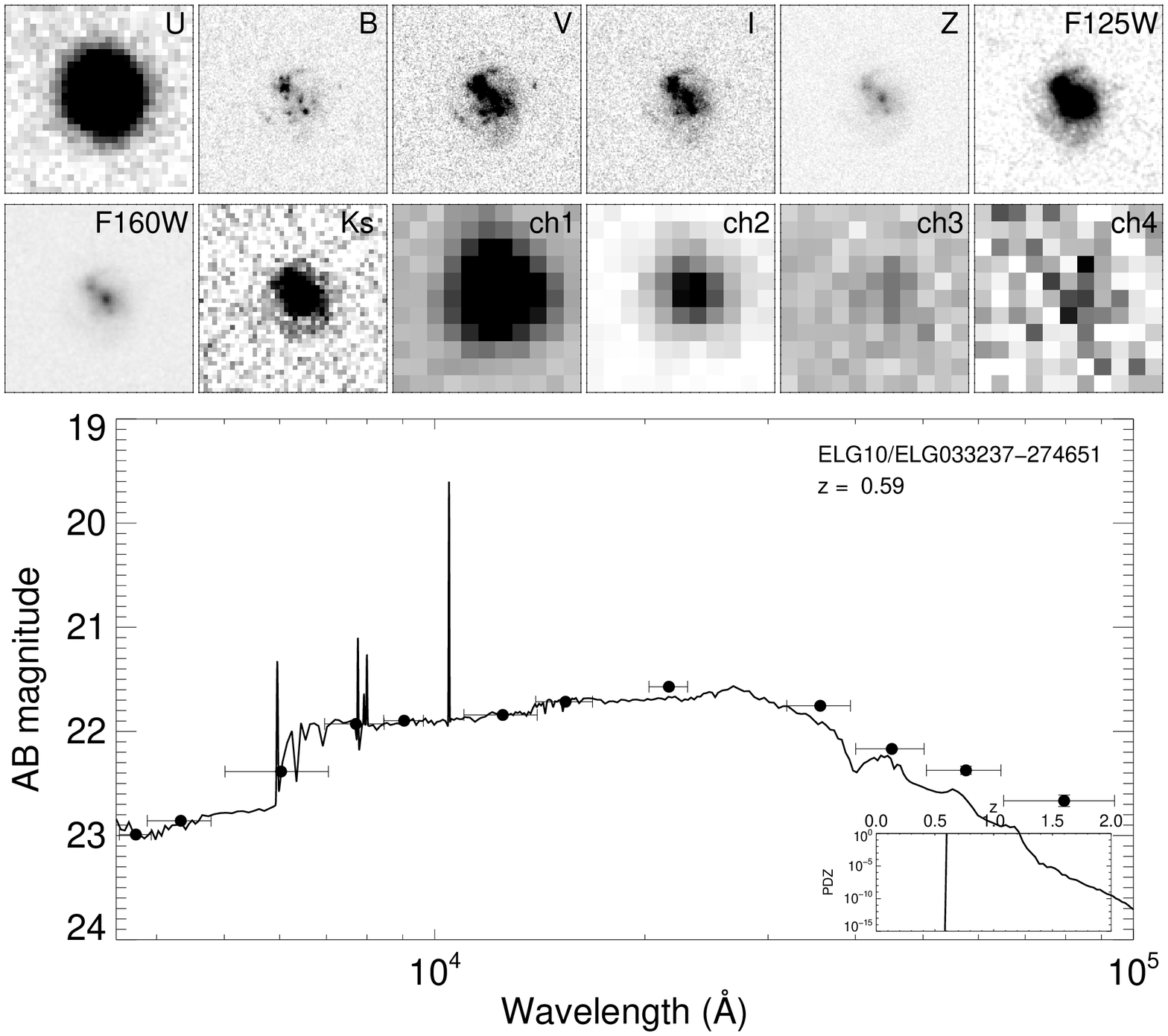,width=0.9\columnwidth}
\epsfig{file=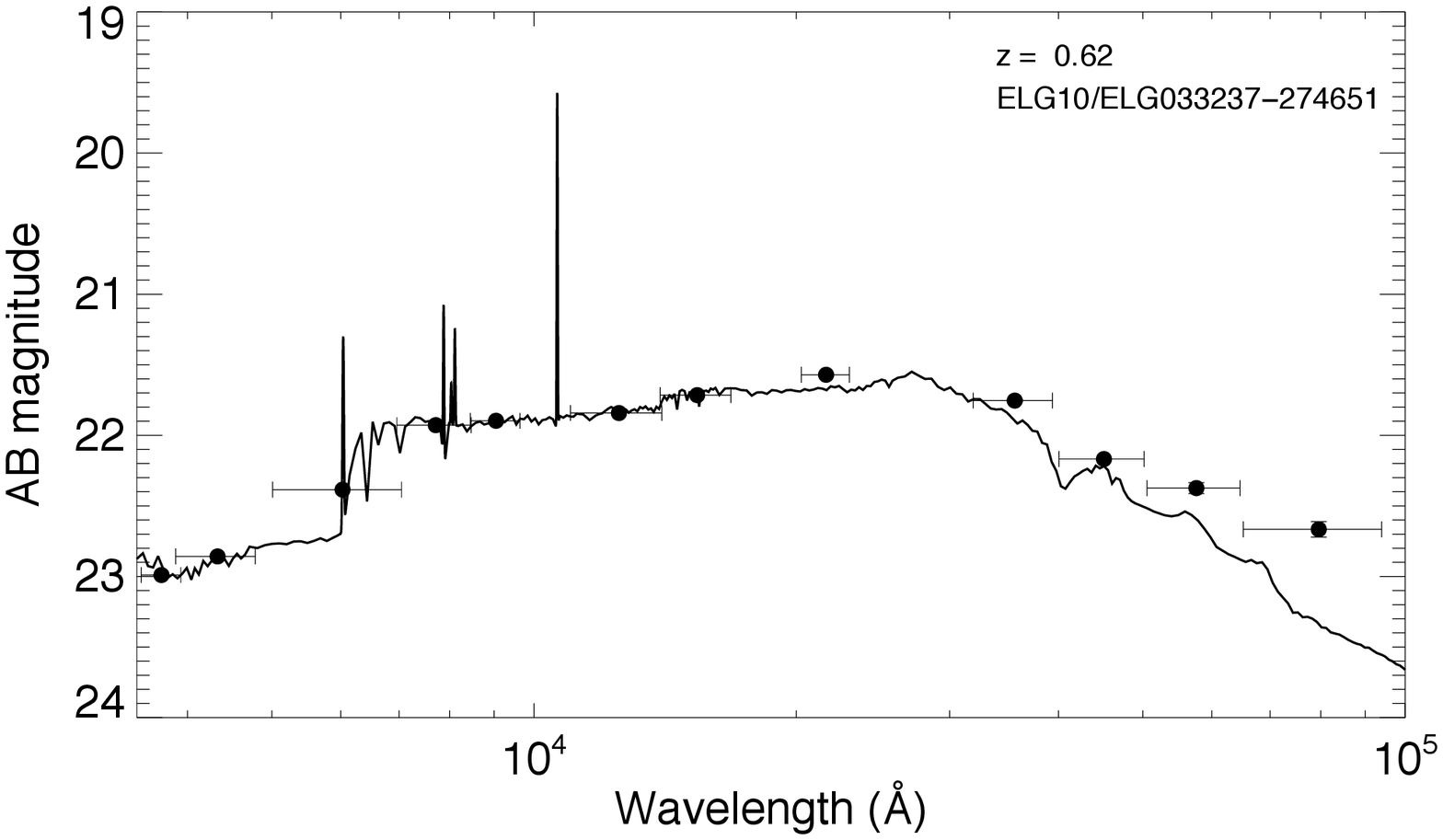,height=4cm,width=0.88\columnwidth}
\epsfig{file=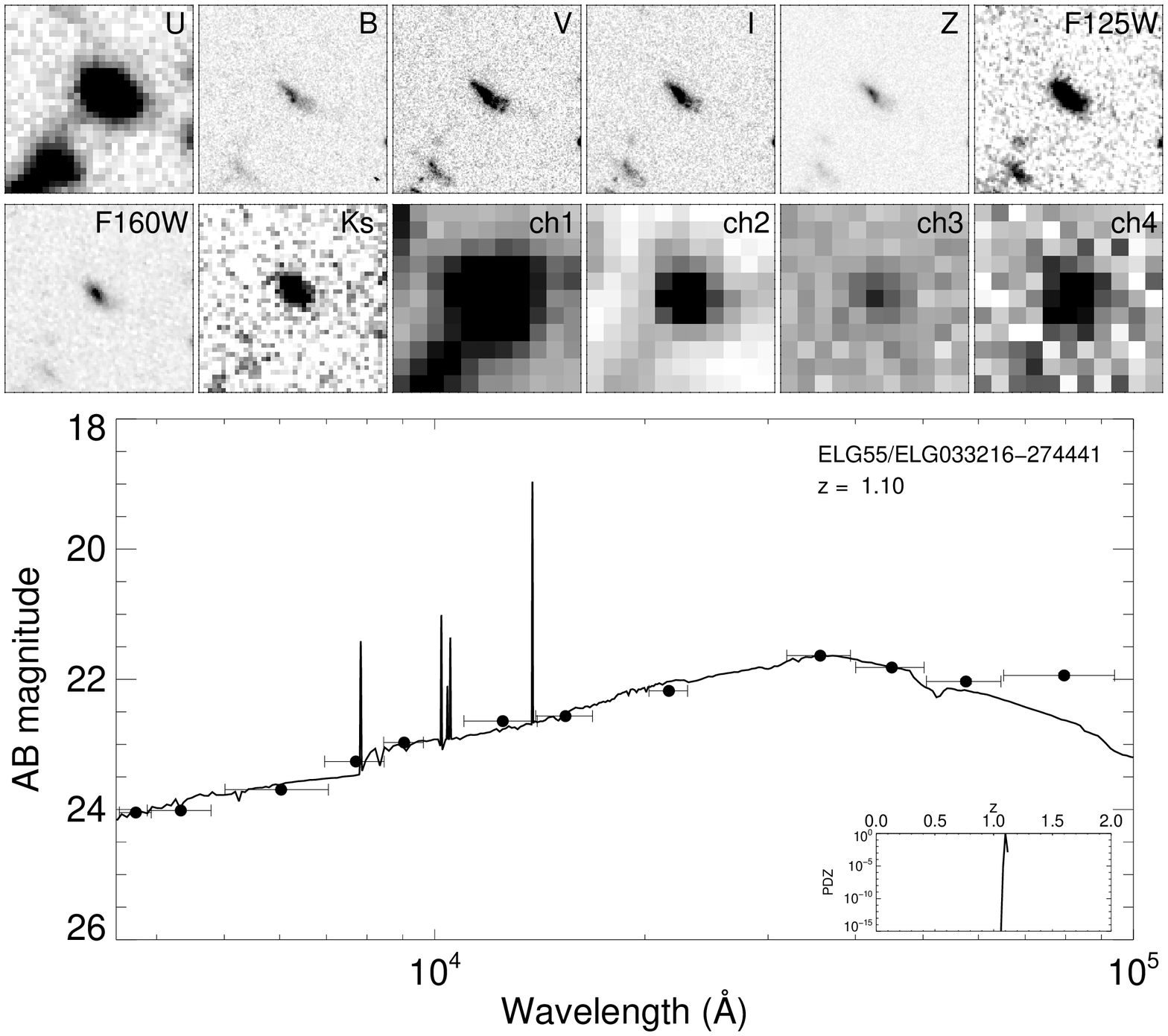,width=0.9\columnwidth}
\epsfig{file=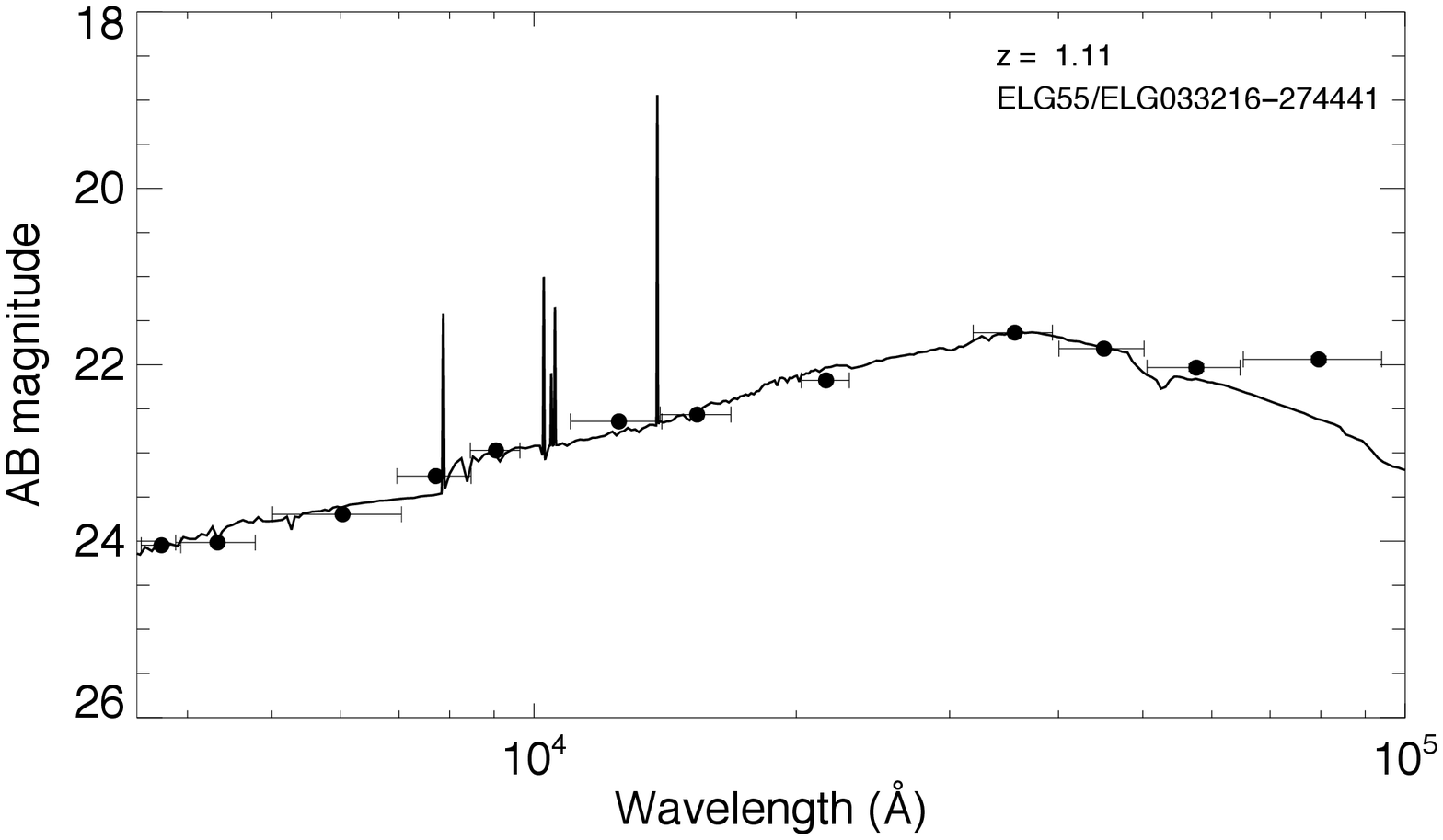,height=4cm,width=0.88\columnwidth}
\caption{\label{SEDs} Illustration of our redshift assignment
procedure, we show an example for each z slice. We first fit the SED
leaving z as a free parameter (upper fit for each slice), based on the
z-probability density from that fit we then assign a slice and fit for
that z value (lower fit for each slice). We also provide thumbnail images
covering 6$\times$6 arcsec$^2$ around each object in broad band filter
images from $U$ through Spitzer channel 4. Errors on the photometry are
included in the figure, but are in almost all cases too small to be
visible.}
\end{center}
\end{figure}

\begin{figure}
\begin{center}
\addtocounter{figure}{-1}
\epsfig{file=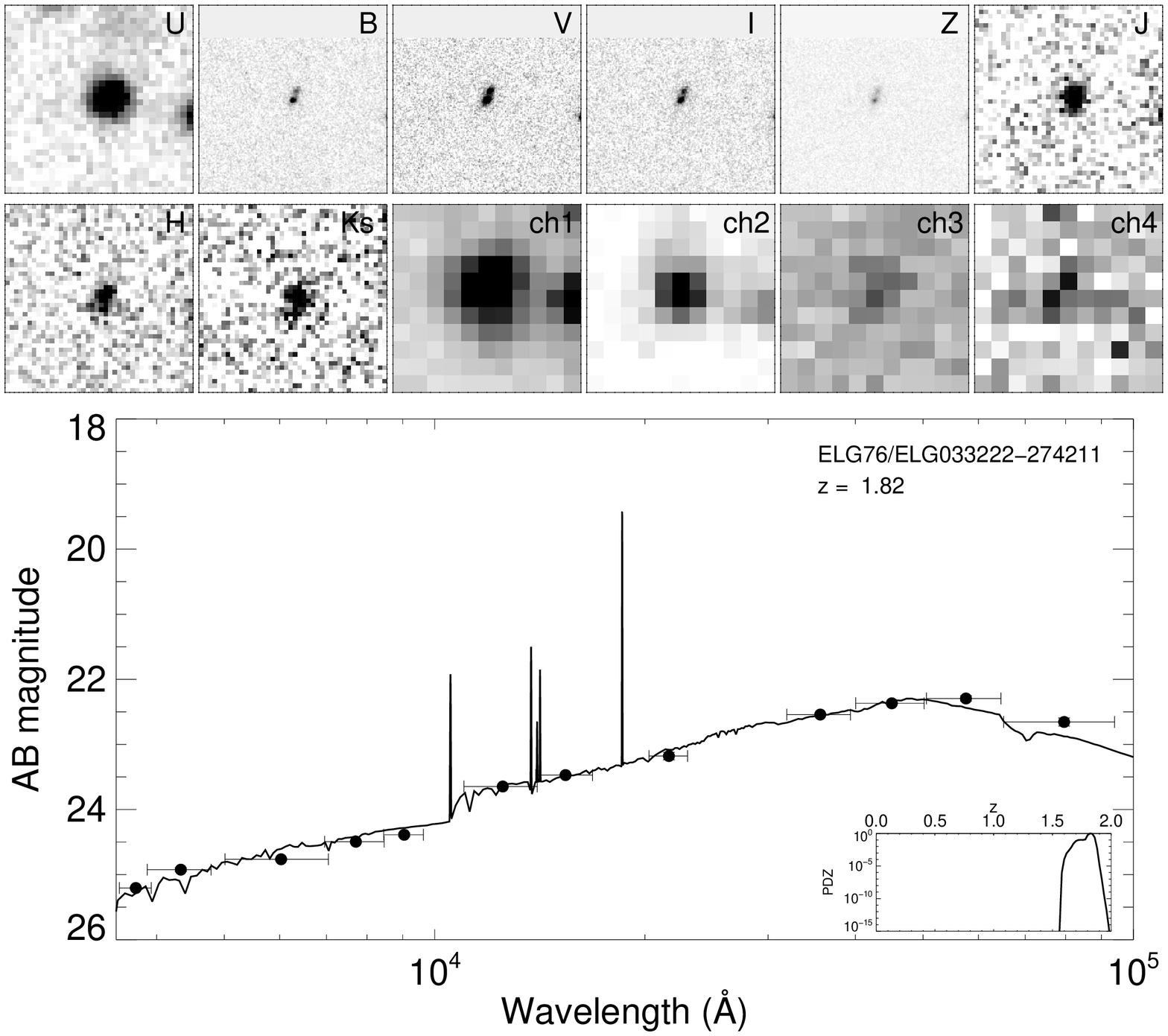,width=0.9\columnwidth}
\epsfig{file=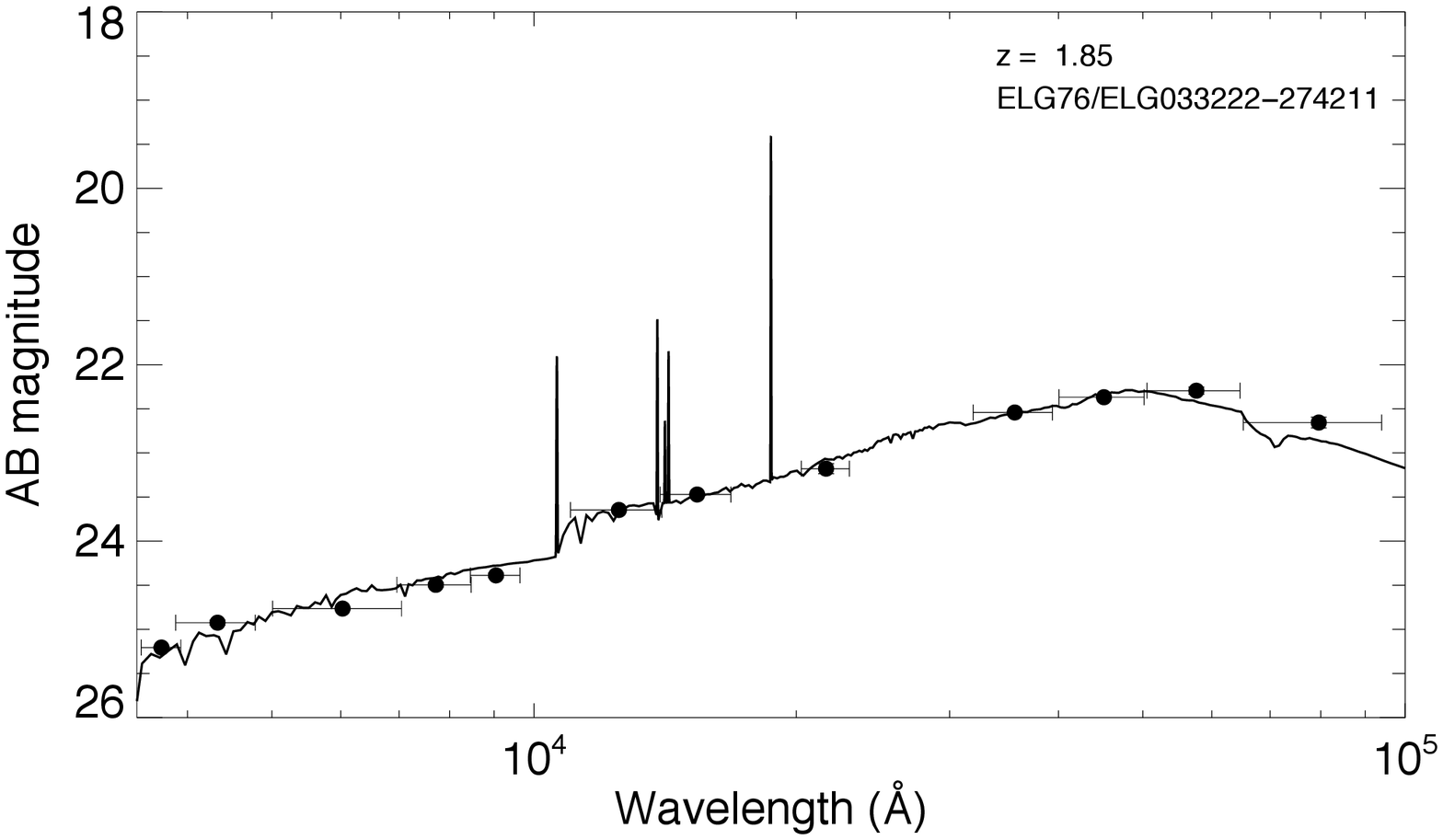,height=4cm,width=0.88\columnwidth}
\caption{continued}
\end{center}
\end{figure}

For each object we go through the following steps. We fit to the
full set of photometric data twice, once using all data points, and
once where we exclude the narrow band and the Y-band since they are
both dominated by the emission line which may skew the
fit. We then decide, after visual inspection of each individual fit, if
there is a unique solution, or if two or even all three redshift solutions
are possible. This is done independently by 4 of us and redshifts are
only assigned if we all 4 agree. For most (35) objects there clearly is a
unique solution, but for the remainder 5 objects no unique redshift
assignment is possible this way. In 4 cases there is a best solution
(dubbed ``primary redshift'' and listed first in Table~\ref{Table1})
but also a possible secondary solution. In one case (ELG30) all three
solutions are possible and none of them are preferred. ELG30 is the
object which is in the lowest left corner of Fig.~\ref{colplot}, i.e.
it has larger emission line equivalent width than any other object in
our sample. Presumably the strong emission lines are confusing the
SED fit. All redshifts assigned in this way are provided in 
Table~\ref{Table1}.
As a final step we then repeat the SED fit but this time locking the
redshift to the spectroscopic redshift (when available) or to the
assigned redshift based on the identification of the emission line. The
purpose of this last fit is to obtain the best fitted values for
stellar mass and star formation rate.

In Fig.~\ref{SEDs} we show examples of fits to three of the objects with
unique solutions, one belonging to each redshift slice. We show both the
first fit where the redshift was left as a free fitting parameter,
and the final fit with assigned redshift.

\subsubsection{The V-I vs Z-J redshift diagnostic plot}

\begin{figure}
\begin{center}
\epsfig{file=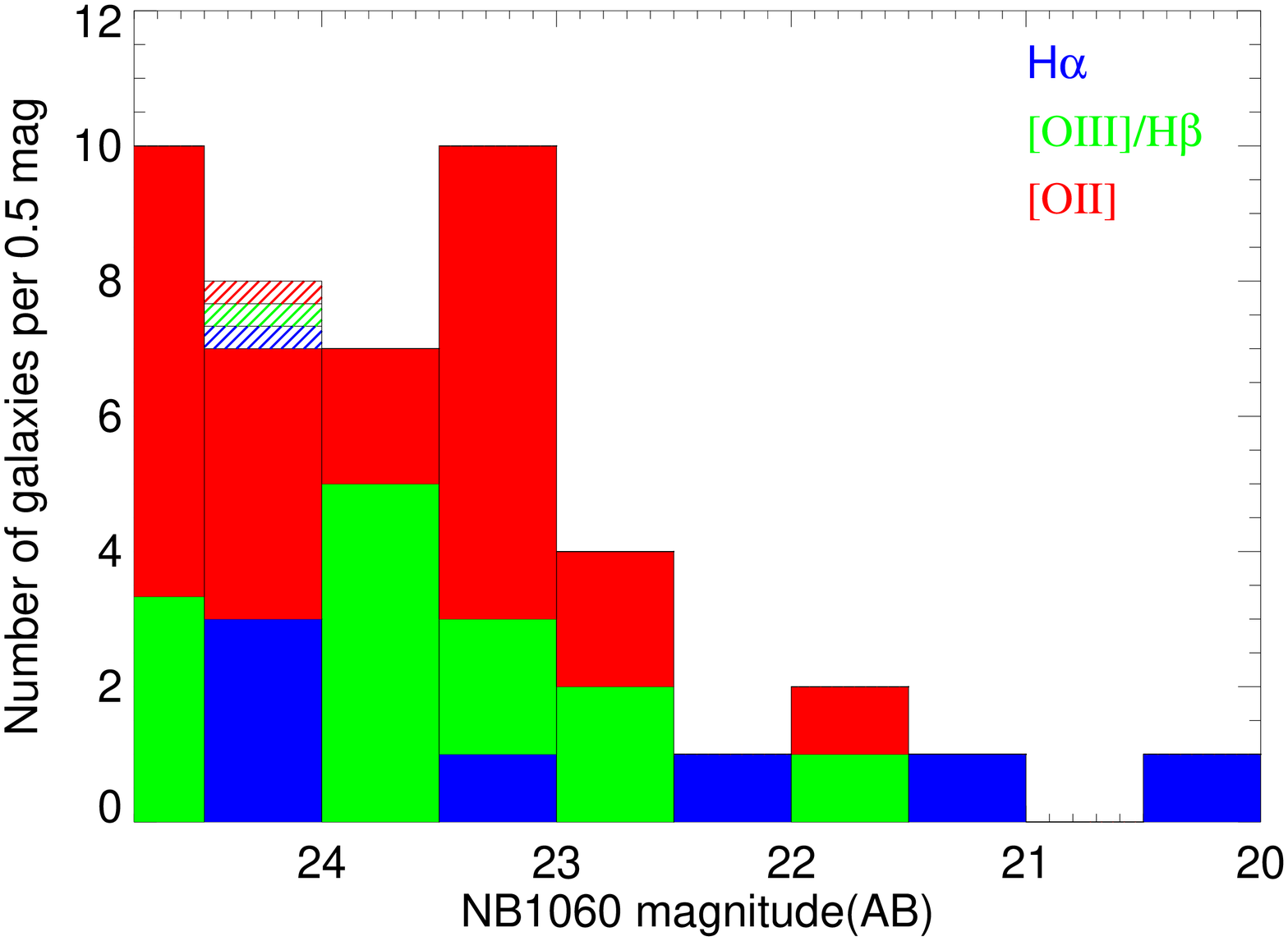,width=9.0cm}
\caption{
\label{histogram} Emission line flux distribution of objects in our
three redshift slices. It is seen that the median narrow band
magnitude is roughly 23.5 for all slices. ELG~30 is marked as the
hashed object with undecided redshift. The last bin size
($M_{AB}> 24.5$) is 0.3 instead of 0.5 and has been scaled accordingly.
}
\end{center}
\end{figure}

\begin{figure}
\begin{flushleft}
\epsfig{file=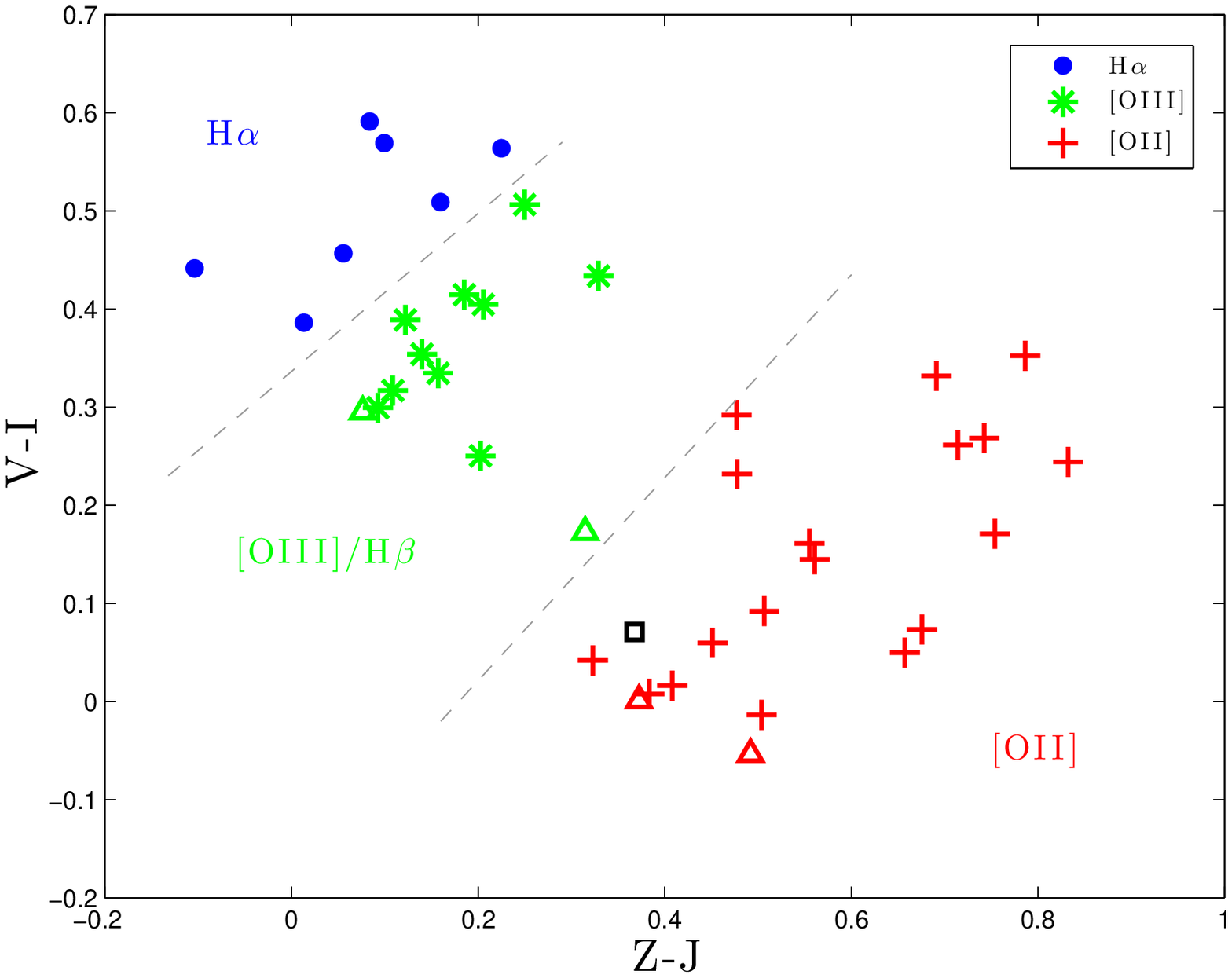, height=7.5cm,width=9cm}
\caption{$Z-J$ versus $V-I$ colour distribution for our basic sample,
with
$V$ = F606W,
$I$ = F814W,
$z$ = F850LP and
$J$ = F125W, as taken from G13.
Solid dots are secure redshifts, open triangles are primary redshift
solutions, the black square marks ELG30 for which there is no preferred
redshift. As \citet{Bayliss'11} we see a clear separation of redshifts
into separate colour groupings making this diagram useful as a redshift
diagnostic for emission line selected samples.}
\label{ZJ-VI}
\end{flushleft}
\end{figure}

In Fig.~\ref{ZJ-VI} we plot the
V-I colour versus the Z-J colour for all the unique object redshifts
and the four primary but non-unique redshift solutions
(open triangles). The objects are colour coded according to redshift
slice (H$\alpha$ blue, \fion{O}{iii}/H$\beta$ green, and \fion{O}{ii}
red). It is seen that the points separate out quite clearly in this
diagram in agreement with the work by \citet{Bayliss'11}. Galaxies
move from the lower right towards the upper left in this diagram as
they move to lower redshifts, and it is a coincidence that the
internal scatter of the distribution at any given redshift forms
a perfect match to the separation in redshift forced by the wavelengths
of the three transitions.  It is therefore possible to use this figure
as a diagnostic plot to assist slice identification in cases where no
unique solution can be found. Our primary redshifts are seen to agree
well with this plot which is further support that those assignments
are correct.  We have also plotted the last object without redshift
assignment (ELG30) as a black square, and we see that it is mostly
embedded in the region occupied by \fion{O}{ii} emitters, close also
to \fion{O}{iii} emitters, but far away from H$\alpha$ emitters.

\begin{figure}
\centering
\epsfig{file=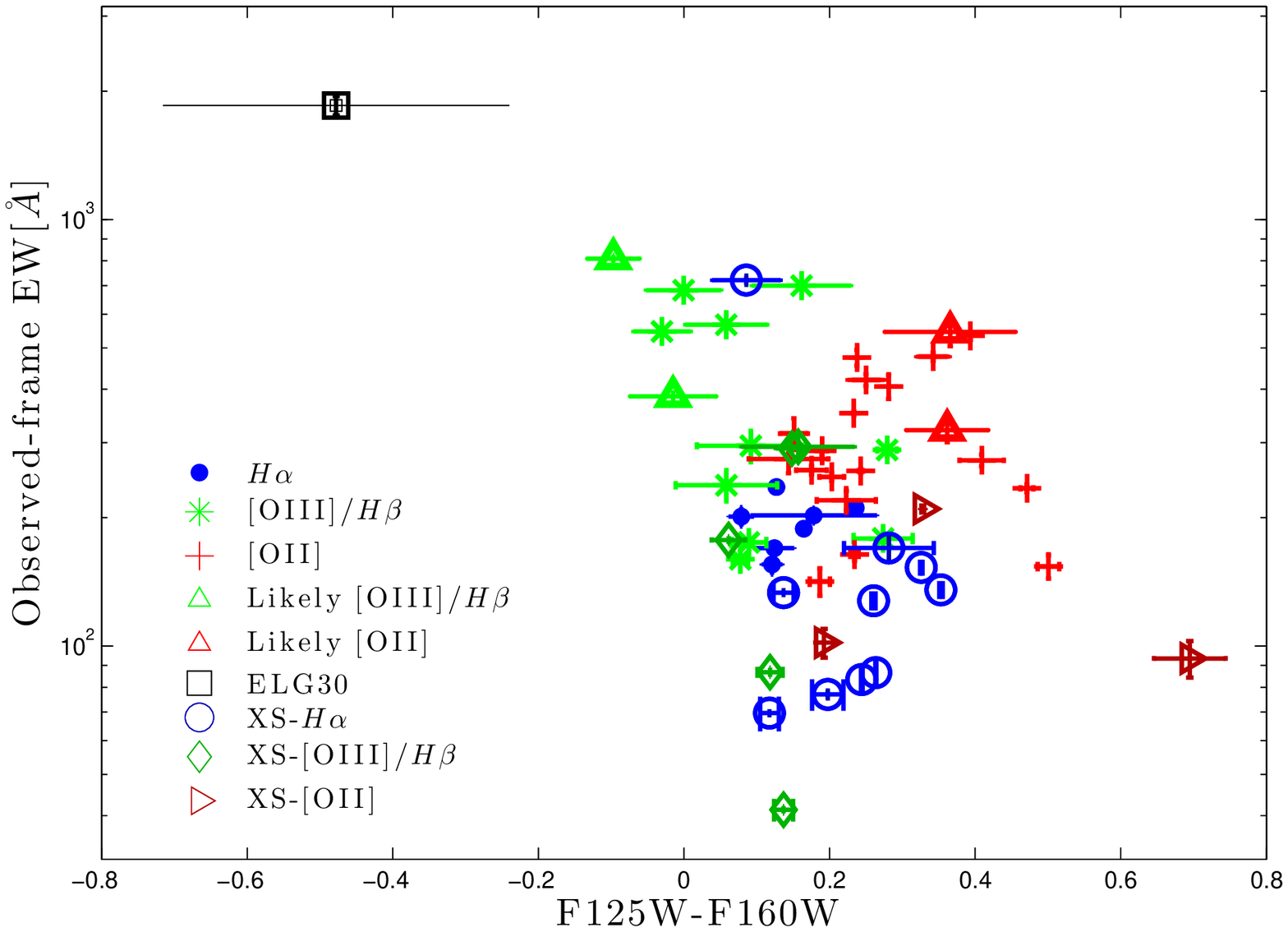,width=9cm}
\caption{\label{EWs}Observed-frame EW of the line in the NB1060 filter
(as derived from the photometry)
against (F125W$-$F160W) colour for the basic sample
and the extended sample (marked XS in the legend).
}

\end{figure}

We note that ELG30 has the highest equivalent width (EW) emission
line of our sample, and that would suggest that it is an \fion{O}{iii}
emitter since they in general have large EW (see e.g., \citet{Penin14}).
Further insight into the redshift of ELG30 comes from Fig.~\ref{EWs},
showing the observed-frame EW of the line in the NB1060 filter
(as derived in Sect.~\ref{SFR_LINES}) against the (F125W$-$F160W) colour from
the G13 catalogue.
For $z=0.62$ (H$\alpha$ in NB1060), no strong emission lines will be
in neither F125W nor F160W\@.
For $z=1.12$ ([OIII]5007 in NB1060), H$\alpha$ will be in F125W
while no strong lines will be in F160W\@.
For $z=1.18$ (H$\beta$ in NB1060), no strong lines will be in F125W
while H$\alpha$ will be in F160W\@.
For $z=1.85$ ([OII] in NB1060), H$\beta$ will be in F125W and
[OIII]5007 will be in F160W\@.
These considerations indicate that a high-EW line emitter
with a blue (F125W$-$F160W) colour such as ELG30 is more likely to be
$z=1.12$ [OIII]5007 than $z=1.85$ [OII].

All things considered we are not able to assign a primary redshift
to ELG30.

\subsubsection{Cross-referencing with the MUSYC survey \label{Cross-Ref}}

\begin{figure}
\begin{center}
\epsfig{file=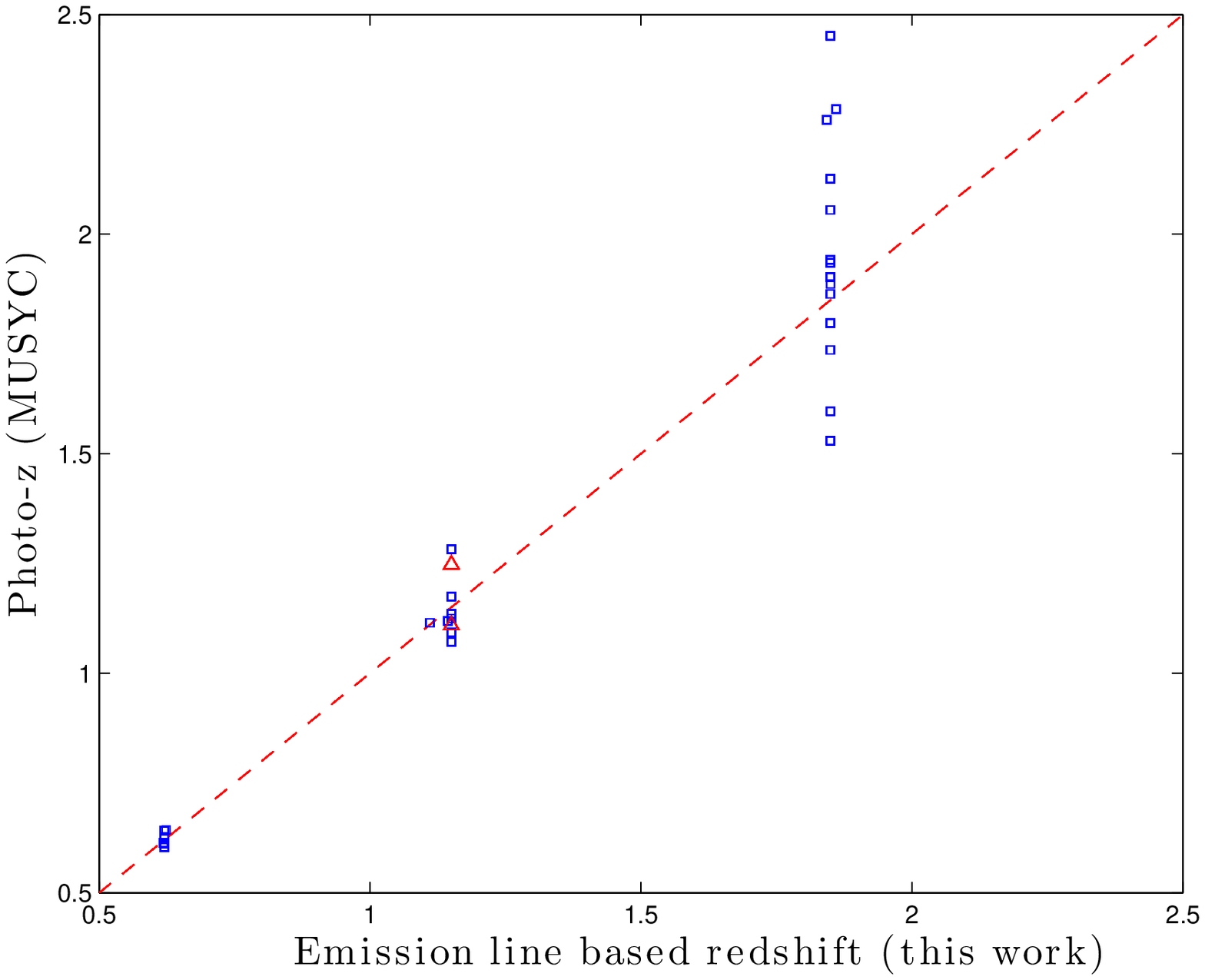,width=9.0cm}
\caption{
\label{MUSYC_Phot}
Redshifts from the MUSYC survey versus redshifts from this work as
listed in Table~\ref{Table1}. Secure redshift assignments are marked
by blue squares, two ``primary redshifts'' are marked by red triangles.
The agreement with MUSYC redshifts is seen to be good in the mean, but
the scatter of the MUSYC redshifts increase at higher redshifts. }
\end{center}
\end{figure}

In Fig.~\ref{MUSYC_Phot} we cross-check our final redshift assignments
against those of the MUSYC survey \citep{MUSYC_Cardamone'10}. The MUSYC
survey consists of imaging of the GOODS-South field in a wide range of
broad and medium-wide filters. The MUSYC catalogue contains photometry
for more than 84000 galaxies including the GOODS field. The catalogue
lists magnitudes, photometric and spectroscopic (when available)
redshifts and a large range of other characteristics. Photometric
redshifts have been obtained using the EAZY (Easy and Accurate Zphot
from Yale) photometric redshift code \citep{Brammer08}.  
In Fig.~\ref{MUSYC_Phot} we plot the MUSYC redshifts against our
redshifts, excluding six objects for which we could find no MUSYC
counterpart. Two primary redshift assignments (ELG51 and 58) are
marked with red triangles and the agreement is seen to be good. We
therefore conclude that our redshift assignments for those two objects are
secure. The last three non-secure redshifts have no counterparts in
MUSYC.

It is seen from Fig.~\ref{MUSYC_Phot} that there is a very good
agreement in the general trend, and the listed errors in the
MUSYC catalogue give mostly a reasonably distribution of $\chi^2$,
notably for the lower redshift slices. However, four of the 18 certain
\fion{O}{II} emitters are $\approx 2\sigma$ off, one is at $4.6\sigma$,
and one at $10.4\sigma$ (the latter being ELG12, which has a spectroscopic
redshift and which is detected in X-rays, and therefore possibly an AGN\@).
We therefore conclude that while the general trend
is in excellent agreement, and the errors for the $z=0.62$ slice are very
small, for the two higher redshift slices the errors become increasingly
larger, and for the $z=1.85$ slice the errors are in about 30\% of the
cases underestimated. Therefore galaxy scaling
relations derived from large statistical samples based on only photo-z
redshifts are probably reliable out to at least $z=0.6$, but at higher
redshifts there are significant, and in some cases significantly
underestimated, errors on the redshifts which will
propagate into errors on the derived physical parameters such as
stellar masses ($M_{\star}$) and star-formation-rates (SFR).
At higher redshifts one might therefore obtain more accurate results
from smaller samples but with more accurate redshifts.

\subsection{Broad band flux depth}

Our survey function is defined based on narrow band flux limit and
emission line equivalent width. This means that we do not have
any actual lower limit on broad band fluxes in our sample, and
therefore our survey differs significantly from spectroscopic
surveys where strict broad band flux limits are used for target
selection to ensure a good probability that a redshift can be
determined from the spectrum. We expect that our sample
is deeper than spectroscopic surveys in the same field, and
in order to assess how much we have extracted a complete
spectroscopic sample from the catalog of \cite{V08} (V08 hereafter).
The V08 survey targeted galaxies in the GOODS-S down to a limiting
magnitude of $z_{850}$(AB) = 26, making it one of the deepest existing
spectroscopic surveys (cf.\ Table~5 in \cite{LeFevre15}).

\begin{figure}
\begin{flushleft}
\includegraphics[trim = 9mm 1mm 0mm 0mm, clip, width=9.5cm]{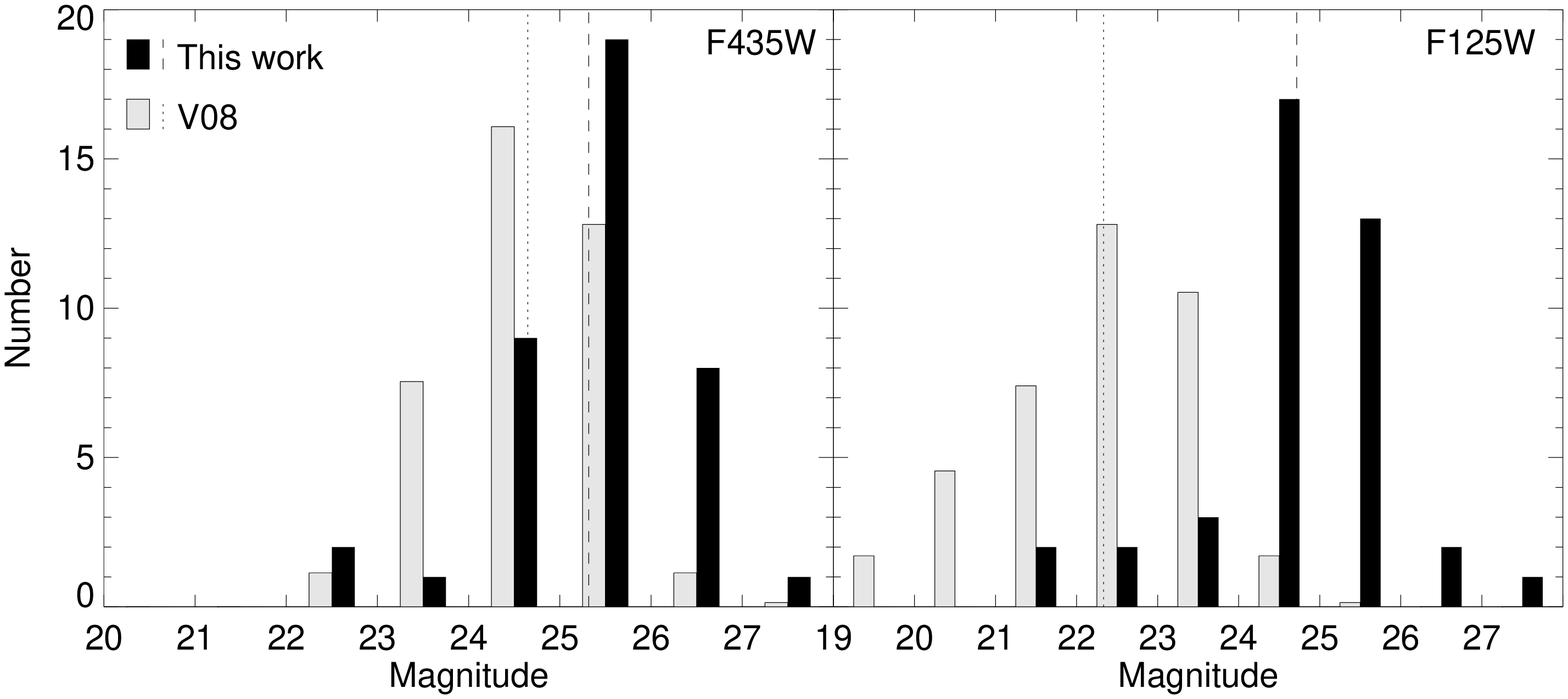}
\caption{
\label{V08_HIST}
The ''de facto'' broad band depth of our basic sample compared
to the sample from \citep{V08} (V08), which is one of
the deepest existing spectroscopic surveys.  The comparison is
done in two HST bands corresponding to B (left panel) and J (right
panel).  The medians of the samples are shown as dotted and
dashed vertical lines. Comparing the medians our sample is 0.8
and 2.3 magnitudes deeper than V08 in B and J respectively.
The height of the V08 histograms was divided by 7 for easy
comparison.}
\end{flushleft}
\end{figure}

From V08 we extracted all objects with redshift in one of our three
redshift slices.In order to obtain a comparison
 sample of a good size we used slices of width 0.4,
 centered at the same redshifts, i.e.\ $\pm0.2$
 around $z$ = 0.62, 1.15 and 1.85. In Fig~\ref{V08_HIST} we show the distribution 
 of two broad band magnitudes (F435W($\approx$B) and F125W($\approx$J))
for both our basic sample (black histogram) and the V08 sample
(grey histogram). In order to make the studies consistent, we obtained the photometry from 
the G13 catalog for all objects. It is seen that our sample is significantly deeper
in both bands. The median of the comparison sample 
is 24.71 and 22.56 ($B$ and $J$ respectively) while our sample has medians 25.49 
and 24.92, i.e. our sample goes around 0.8 and 2.3 magnitudes deeper.

For example in the overlapping region between our survey
and the recent catalog of HST grism spectroscopy \citep{Morris15} 
our sample has 33 objects at redshifts probed
by the HST spectroscopy ($z > 0.67$), but the HST catalog
contains only the seven brighter of those. The redshifts
are all in agreement.

\section{Results}
\label{results}
\subsection{The main sequence of star formation in three narrow redshift slices}
 \label{Sect4.1.1}
\begin{figure}
\epsfig{file=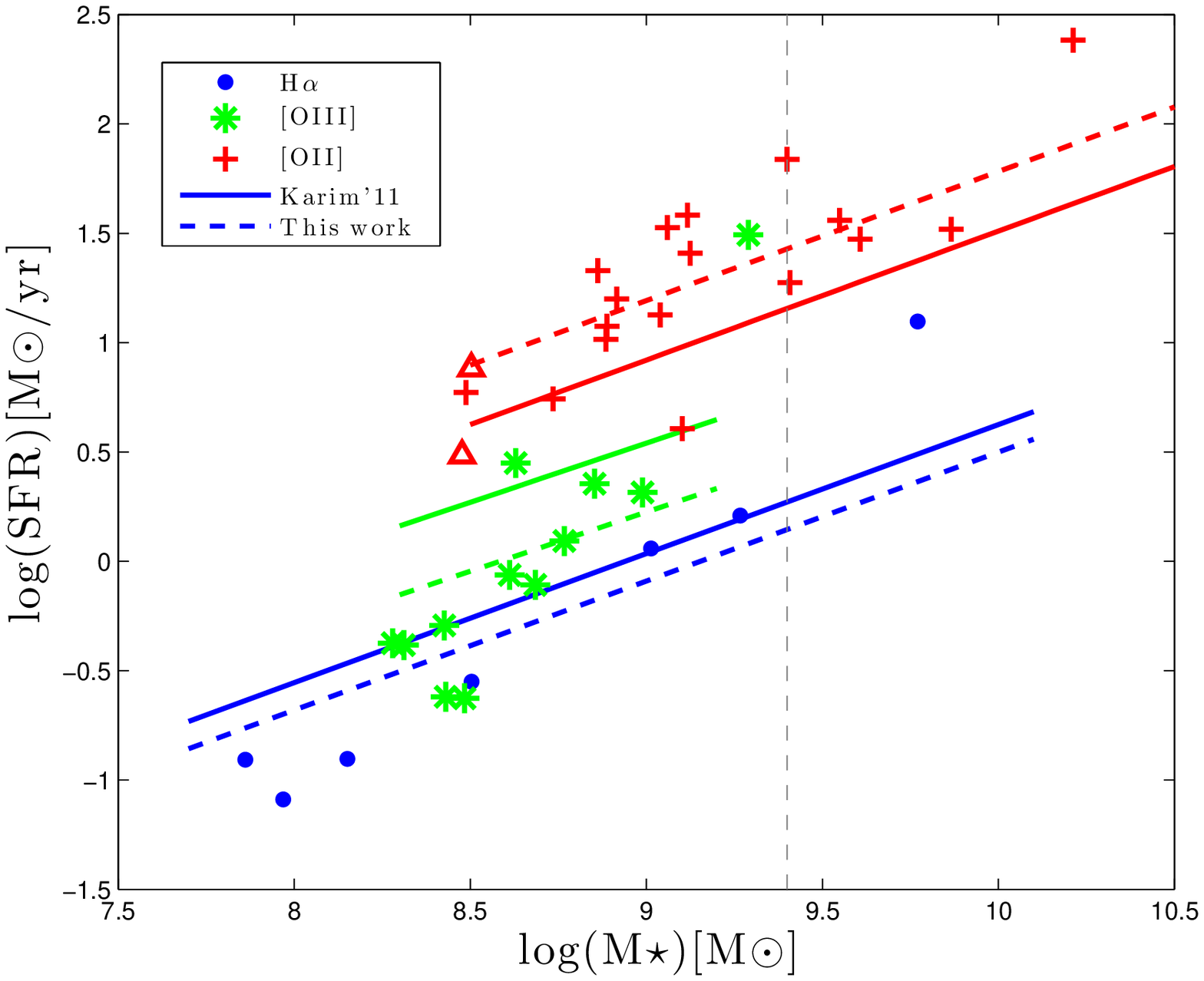, height=7cm,width=8.5cm}
\caption{\label{MSFR}
SFR vs stellar mass of emission line selected galaxies, color-coded according to
their redshift. The two red triangles mark objects with two redshift
solutions (only primary solution shown).  Solid lines
show the relations reported by \cite{Karim11}. Dashed lines are the
best fit of relations with the same slopes to our data. The vertical grey
dashed line marks the lower mass limit of the \cite{Karim11} sample.
}
\end{figure}

The SED fits described in Sect.~\ref{LePhare} also provide values for
$M_{\star}$ and SFR of each galaxy. We
list those values in Table~\ref{SED_param}, and in Fig.~\ref{MSFR} we
plot SFR vs $M_{\star}$.  Both in the local universe, and out to a
redshift of 3.5, it has been shown that SFR forms a tight correlation
with $M_{\star}$ \citep{Brinchmann04, Noeske07, Maiolino08}, the so called
main sequence of star formation (MS). The MS has been shown to evolve
with redshift and in Fig.~\ref{MSFR} we have overplotted the relations
from stacked radio data of star-forming galaxies reported in Table~\ref{estimates} 
of \cite{Karim11}
(full lines) at each of the redshifts of our three redshift slices.
From \cite{Karim11} we take the mean of their $z$ = 0.4--0.6 and 0.6--0.8
bins to represent $z=0.62$, their $z$ = 1.0--1.2 bin to represent $z=1.15$,
and their $z$ = 1.6--2.0 bin to represent $z=1.85$.
Both
the data and the relations are colour coded according to redshift slice
as in Fig.~\ref{ZJ-VI}. We also plot a vertical dashed
line at log($M_{\star}$)=9.4 which is the lower limit of the samples
considered by \cite{Karim11}. One object (ELG14) turned out to provide 
unstable physical parameters in the sense that leaving out a single
photometric point would severely change the output parameters. Upon
checking the HST image we noted a close neighbour galaxy of different
colour which presumably could have affected the photometry and caused
this. The redshift is good so we
keep it in the sample, but we exclude it from the analysis of the MS
relation. We also exclude ELG30 from this analysis since we do not have
a redshift for it. We use the primary redshift solutions for ELG66 and
75, but repeat the analysis using the secondary solutions. No
significant difference is found using the secondary solutions (see
Table~\ref{estimates}).

\begin{table}[h!]
\centering
\caption{Physical parameters resulting from SED fitting with fixed
redshift.
\label{SED_param}
}
\begin{tabular}{l c c c c c cl}
\hline
ID & log(mass) & log(SFR) & Redshift\\ [0.5ex] 
\hline
ELG\# & $logM_{\odot}$ & $log(M_{\odot}/yr)$ &fixed& \\ [0.5ex] 
\hline
3			& 9.01	  &  0.06		&	0.619	\\
4			& 8.63   & 0.45		&	1.144	\\
5$^c$  & 	9.12  &  1.41		&	1.86	\\
6			&  9.04  &  1.13	  	&	1.86	\\
9			&  8.15  & -0.90	    &	0.62	\\
10		&  9.27  &  0.21	 	& 	0.62	\\
11		& 	7.86  & -0.91		&	0.62	\\
12$^c$& 	10.21 &   2.38		&	1.843	\\
14$^c$ & 8.67  &  0.76		&	1.85	\\
15$^c$ & 9.12  &  1.58		&	1.85	\\
16$^c$ & 9.87  &  1.52		&	1.85	\\
20$^c$ & 8.92  &  1.20		&	1.85	\\
21$^c$ & 8.89  &  1.07		&	1.85	\\
22$^c$ & 9.40  &  1.84		&	1.85	\\
23	 	 & 8.74  &  0.74	 	& 	1.85	\\
25$^c$ & 8.86  &  1.33		&	1.85	\\
26	 	 & 8.85  &  0.36		& 	1.15	\\
28	 	 &	8.28  & -0.37		&	1.15	\\
34	 	 & 8.48  & -0.63		&	1.15	\\
35	 	 & 9.41  &  1.27		&	1.85	\\
36	 	 & 8.31  & -0.38		&	1.15	\\
37		 & 8.49  &  0.77		&	1.85	\\
41	 	 & 8.89  &  1.02		&	1.85	\\
43	 	 & 8.43  & -0.29		&	1.15	\\
45	 	 & 8.50  &  -0.55		&	0.62	\\
51               & 8.77  &  0.09    &	1.15	\\
52	 	 & 8.43  &  -0.62		&	1.15	\\
53		 & 9.61  &  1.47		&	1.85	\\
54	 	 & 9.06  &  1.53		&	1.85	\\
55	 	 & 9.29  &  1.49		&	1.15	\\
58               & 8.68  & -0.11	&	1.15\\
62	 	 & 7.97  & -1.09		&	0.62	\\
65 	 	 &	8.61  & -0.06		&	1.15	\\
66$^1$ & 8.48 &  0.48    &	1.85 \\
68	 	 & 9.10  &  0.61		&	1.85	\\
70		 & 8.99  &  0.32		&	1.15	\\
75$^1$& 8.50  &  0.88  	&	1.85\\
76	 	 & 9.55  &  1.56		&	1.85	\\  
78	 	 & 9.77  &  1.10		&	0.624	\\
\hline
&&\textbf{Ambiguous cases}& \\ [0.5ex]
\hline
66$^2$     & 7.91  	 &  0.38	&	1.15	\\
75$^2$     & 8.02   	 & 0.49		&	1.15	\\
30$^3$  	 & 6.72	 	 &	-1.15	&	0.62	\\
30$^3$  	 & 7.31  	 & -0.82	&	1.15	\\
30$^3$	 & 7.83  	 & -0.39 	&	1.85	\\
\hline
\end{tabular}
\begin{flushleft}
$^c$ - Cluster member galaxy. \\
$^1$ - Primary fixed redshift solution used for ELG66 and 75.\\
$^2$ - Secondary fixed redshift solution used for ELG66 and 75.\\
$^3$ - No preferred redshift for ELG30, although $z=0.62$ H$\alpha$
       is disfavoured.
\end{flushleft}
\end{table}

\begin{figure}
\flushleft
\epsfig{file=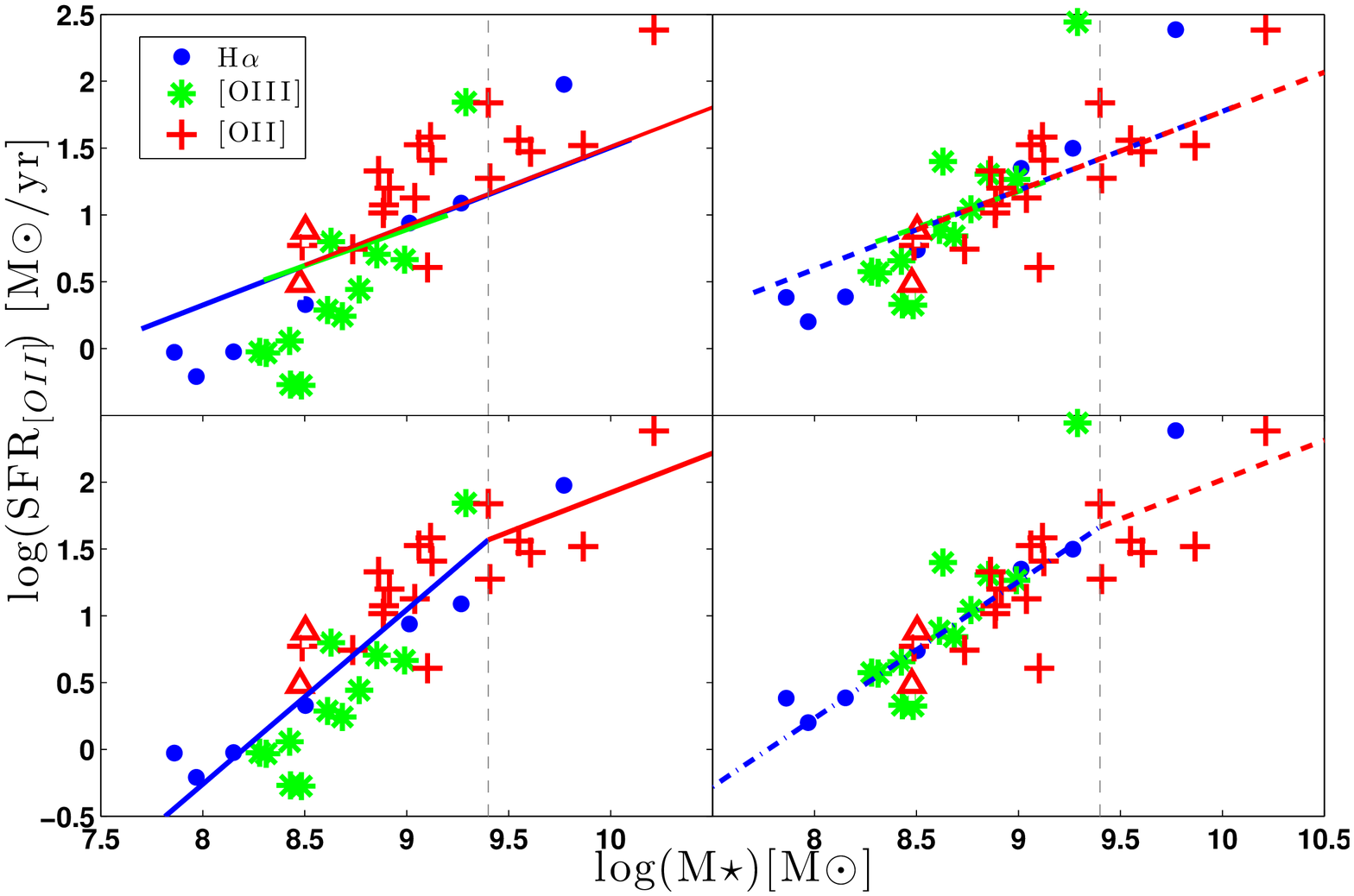,width=9.5cm}\\
\caption{\label{new_MSFRrelation} In the upper two panels we plot the same
data and relations as in Fig.~\ref{MSFR}, but we have here shifted each 
redshift slice to remove the effect of redshift evolution. In the left
column we have applied shifts to bring the blue and green full lines
on top of the red (i.e. applied redshift corrections as reported in
\cite{Karim11}), in the right we have done the same but used the
dashed lines. In the lower panels we provide the best fit of broken
MS relations. It is seen that under both assumptions the relation
steepens towards lower stellar masses.}
\end{figure}

From Fig.~\ref{MSFR} we see that our data roughly are in agreement
with the relation from \cite{Karim11}, i.e. that there is a MS and
that it evolves with redshift in the sense that galaxies of a given
stellar mass have lower SFR at lower redshifts. Our data points are
somewhat offset from the expected relations, but this could possibly
be due to the fact that our objects sample a much lower stellar mass
range than the relations we compare to. If the MS e.g. is steepening
at the low mass end, it would cause our low mass galaxies to drop below
the relations. In order to test this we first
assume that the slopes reported by \cite{Karim11} at each of our redshift
slices are correct for all masses and then we determine the offsets
to our data. The best fit offsets are shown as dashed coloured lines in
Fig.~\ref{MSFR} and provided in Table~\ref{estimates}. We then remove the 
effect of redshift evolution in two different ways. First we assume that
the evolution from \cite{Karim11} is correct and we apply a shift which
brings all galaxies (and the relations) to what they would have been
in the \fion{O}{ii} redshift slice (upper left panel of
Fig.~\ref{new_MSFRrelation}). We then fit
a broken linear relation to the data points with the following two
conditions: ($i$) at log($M_{\star}$) larger than 9.4 it must have the
slope of 0.59 (from \cite{Karim11}); and ($ii$) it must be continuous in
log($M_{\star}$)=9.4. The resulting best fit is shown in
Fig.~\ref{new_MSFRrelation}, lower left panel, and the best fit slope
is found to be $1.31$ with an rms of 0.31. In the two right panels of
Fig.~\ref{new_MSFRrelation} we show the same
as in the left, only here we have applied redshift correction shifts
such that the dashed lines in Fig.~\ref{MSFR} are lined up rather than
the full lines. In this case the best fit gives a slope of $1.02$ with
an rms of 0.29.

\begin{table}[h!]
\caption{\label{estimates} Offsets of SFR($M_{\star}$) relative to
\cite{Karim11}. The first 5 lines report the offset of individual
redshift sub-samples assuming for each the slope found by \cite{Karim11}.
The last two are best fit offset of the entire sample assuming now
a slope of 1.17 for the galaxies with mass below the mass completeness
limit ($10^{9.4} M_{\odot}$) of the  \cite{Karim11} sample.
In both cases we repeat the fit using secondary redshifts for ELG 66
and 75 but no significant change is seen.
}

\begin{tabular}{c c c c}
\hline \hline
 $z$ & $N_{\rm obj}$ &  SFR  offset   &  rms \\
\hline
0.62 &  7     & $-0.13\pm0.16$  & 0.35 \\
1.15 & 12$^1$ & $-0.33\pm0.13$  & 0.43 \\
1.15 & 14$^2$ & $-0.22\pm0.16$  & 0.52 \\
1.85 & 19$^1$ & $ 0.26\pm0.07$  & 0.27 \\
1.85 & 17$^2$ & $ 0.29\pm0.07$  & 0.27 \\
\hline
All &  38$^1$ & $0.33\pm0.05$  & 0.32 \\
All &  38$^2$ & $0.35\pm0.06$  & 0.34 \\
\hline
\end{tabular}
\begin{flushleft}
$^1$ - Primary redshift solution used for ELG 66 and 75.\\
$^2$ - Secondary redshift solution used for ELG 66 and 75.
\end{flushleft}
\end{table}

Our sample reaches stellar masses 1.5 decades lower than the sample
of \cite{Karim11} and we see that in the range below their lower mass limit
our sample follows a significantly steeper MS no matter how we correct
for the redshift evolution. Previous analyses of the derived stellar masses 
from SED fits with exponential declining and increasing star-formation rates 
in a population of star-forming galaxies at z=1-2 have shown that the stellar masses vary within $\sim$0.1 dex \citep{christensen12}. As we noted above, the offsets we reported
in Table~\ref{estimates} may in this case be dominated by this steepening of the
slope, and we shall therefore repeat the fit using a more realistic
assumption. Rather than assuming a constant slope we now use a slope
with a break at log($M_{\star}$)=9.4. For the high mass end we use the
slope of 0.59 from  \cite{Karim11}, for the low mass end we use the
mean of the slopes we found above, which is 1.17.

\begin{figure}
\flushleft
\epsfig{file=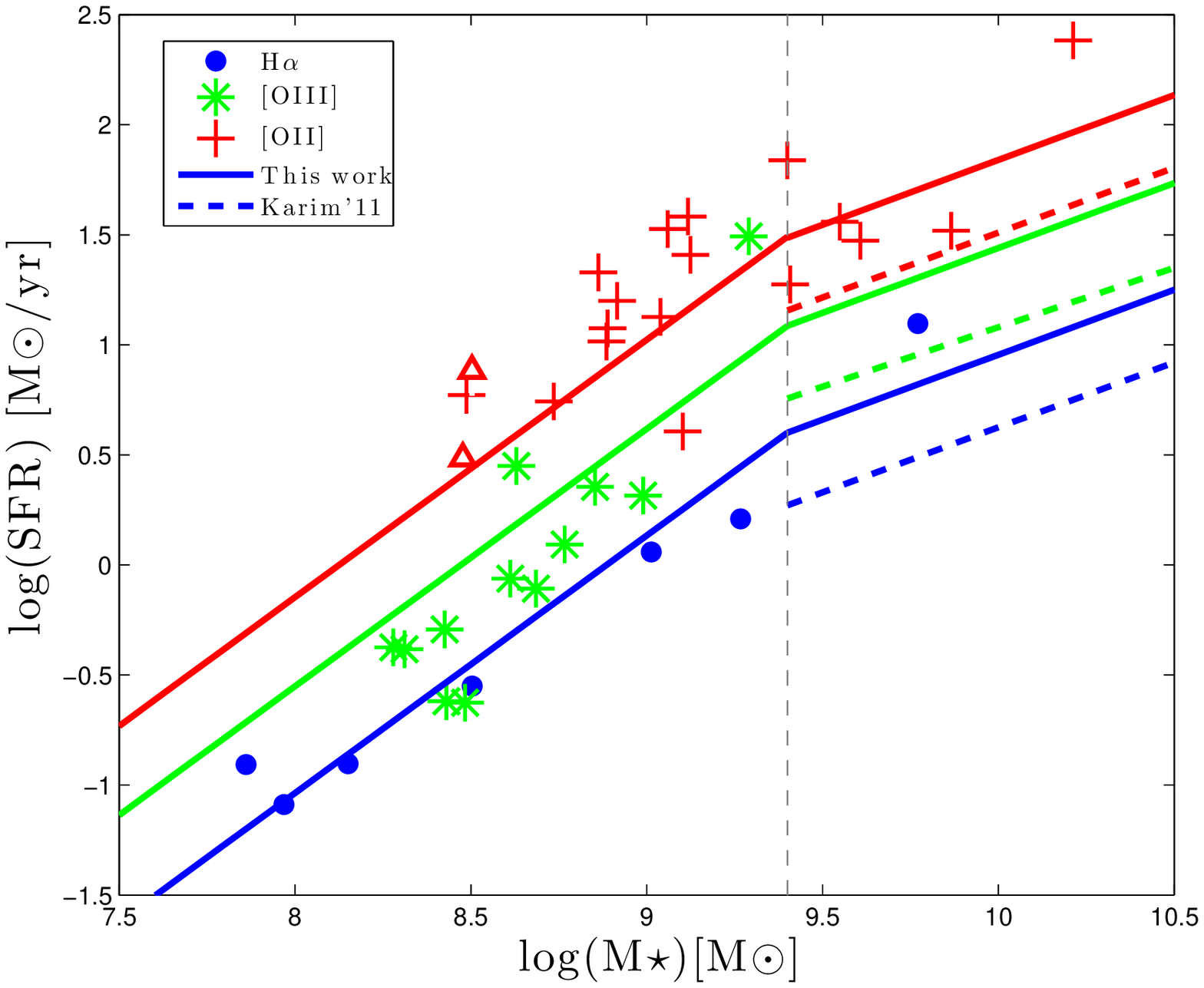,width=9cm}\\
\caption{\label{MSFR_fig12} Similar to Fig.~\ref{MSFR} but here we
show only the \cite{Karim11} fits (dashed lines) in the range above
their lower mass bound. The full lines now show the best fit to our
data of a ``broken'' MS with a steeper low mass slope.}
\end{figure}

\begin{figure}
\flushleft
\epsfig{file=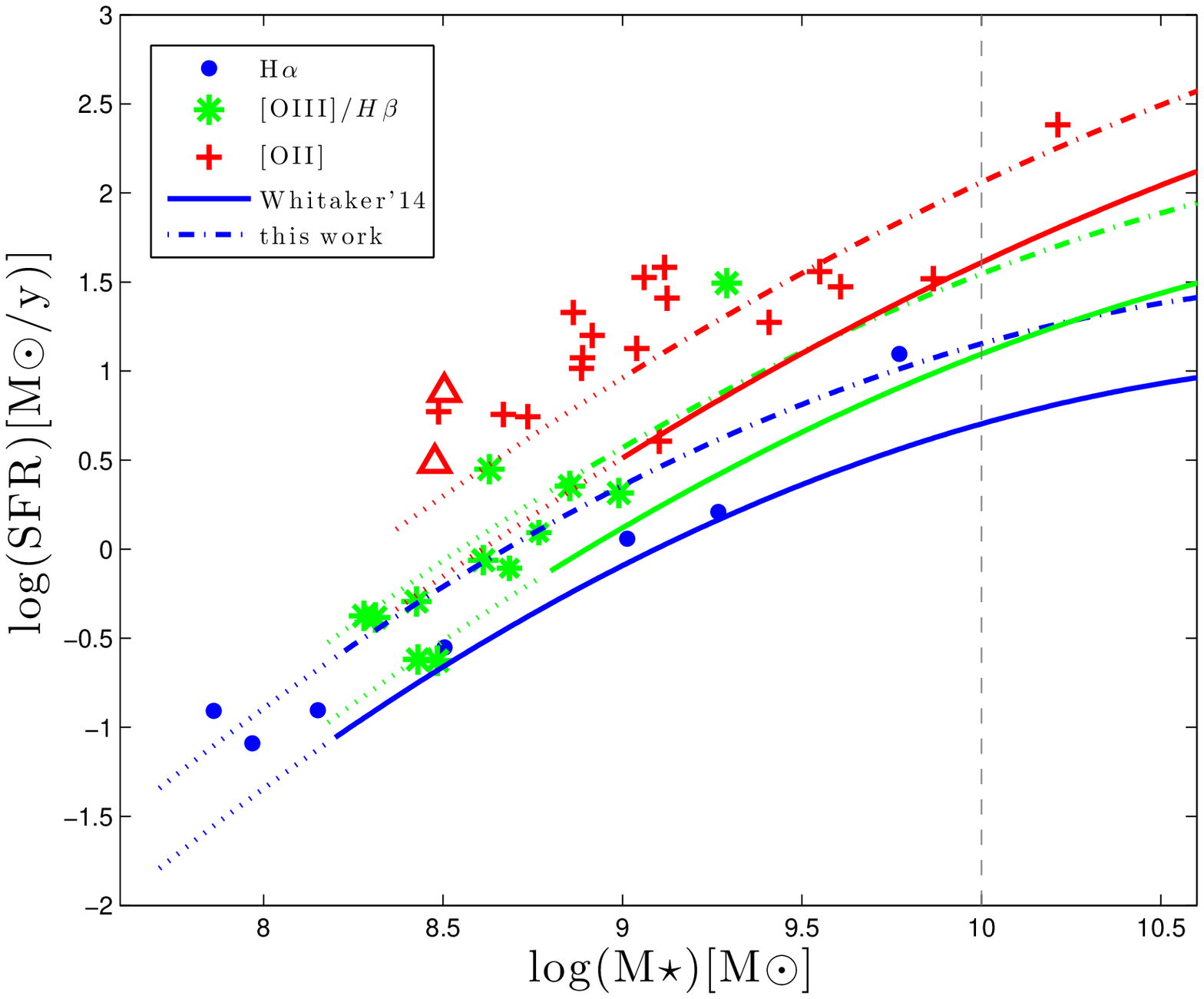,width=9cm}\\
\caption{\label{Whitaker} Similar to Fig. \ref{MSFR_fig12} but here 
we compare to the study by \cite{Whitaker14} (full lines) who also
reported a steepening towards low stellar masses.
Their SFRs are seen to be lower, but adding 0.45 to their fits we
obtain a better fit to our data (dash-dot curves).
We do not see any evidence for shallower
redshift evolution at low masses as they report.
The dashed vertical line marks the division between
their individual object (above $10^{10}$) and stacked
object (below $10^{10}$) fits. Dotted curves are
extrapolations of their fits where they had no data.
}
\end{figure}

The resulting best fit is shown in Fig.~\ref{MSFR_fig12} and again
reported in Table~\ref{estimates}. We see that allowing for the change
of slope we now get a
consistent positive offset towards higher SFR in all three redshift
slices. This is no great surprise as one would expect samples selected
by narrow band techniques to select the objects with the strongest
emission lines in any stellar mass bin, and consequently to contain
the highest SFR galaxies of any mass at any redshift. In that sense
our sample defines the upper envelope of the MS for low to
intermediate mass galaxies.

In conclusion of this section we first tested if our sample was
offset (up or down) in SFR compared to \cite{Karim11} and using their
reported slope. We found an inconsistent scatter with both positive and
negative offsets, but this could be because the median $M_{\star}$
is different in the three redshift slices. We then removed the effect
of redshift to make them more easy to compare, and noted evidence that
the slope is steeper at low masses. Assuming a steeper slope in the
low mass end we find that our data are consistent with a constant offset
from the \cite{Karim11} data (at 6.6$\sigma$) with an internal scatter
of 0.32. Performing the same fit to the data, but instead using the
constant slope of \cite{Karim11} at all masses gives a zero offset with
an internal scatter of 0.43 which is a significantly poorer fit even
allowing for the one degree of freedom less.

\subsubsection{Comparison with other studies}

\cite{Whitaker14} present MS fits from a study of galaxies in the
CANDELS fields. At stellar masses larger than $\sim 10^{10 }$ $M_{\odot}$
they use a UV+IR SFR indicator on photometry of individual photo-z
galaxies, at lower stellar masses they do the same on stacked
photometry and reach stellar masses of $10^{8.4}$ (at z=0.5) to $10^{9.2}$
(at z=2.5). Similar to our results of the previous section they report
a steepening of the slope at lower masses but they fit it with a
polynomial rather than a broken powerlaw. They also report a
shallower redshift evolution of the MS at lower masses than at high
masses. \cite{LeeN15} also report a steepening of the MS below 
$M_{*}=10^{10}M_{\odot}$, in agreement with our results.

We interpolate the polynomial fits of  \cite{Whitaker14} (their equation 2)
to our three redshift slices and plot them with our data in Fig.~\ref{Whitaker}
(full curves in their range of stacked data - dotted curves are
extrapolations of their polynomials). It is seen that the steepening
is in good agreement with what we have reported, but the normalization
is again lower than our data. Also in Fig.~\ref{Whitaker} we show the \cite{Whitaker14} 
models where we have added 0.45 to the log(SFR) (dash-dot
curves) which provide a better fit to our data, but it is seen that
they find much less redshift evolution than seen in our sample. In
particular we do not see any evidence for less evolutuion of the MS at
low stellar masses and our sample appears in stark disagreement with that
result. We note however that our data are from SED fits to
individual galaxies while \cite{Whitaker14} were fitting to
stacked data in the regime of comparison. \cite{Nilsson11}  performed a test fitting
40 emission line selected galaxies both individually and as a
stack, and concluded "Stacking of objects does not reveal the
average of the properties of the individual objects". The
difference could therefore be related to the stacking.

\subsection{SFRs from SED fitting and from emission lines}
\label{SFR_LINES}
From the NB magnitude we can calculate the emission line
fluxes since the flux density in the narrow-band is equal to the sum
of the emission line flux density and the continuum flux density:
 $f_{\nu,NB} = f_{\nu,line} +f_{\nu,cont}$. 
For each galaxy, we derive the underlying continuum
flux density from the best fit SED model by interpolating the flux density 
in adjacent 50 $\AA$ intervals blue and redwards of the NB filter. 
The continuum flux density is subtracted from the NB flux density taking 
into account the NB transmission curve. The derived emission line fluxes 
and equivalent widths (EWs) in the observed frame are listed in Table \ref{Table1}.
The results are consistent if we chose to derive the continuum flux density by 
interpolating between the observed magnitudes in the ACS/F850LP 
and WFC3/F125W bands and assuming a power law spectral slope
between the bands.

\begin{figure}
\centering
\includegraphics[trim = 7mm 3mm 0mm 0mm, clip, width=9.5cm]{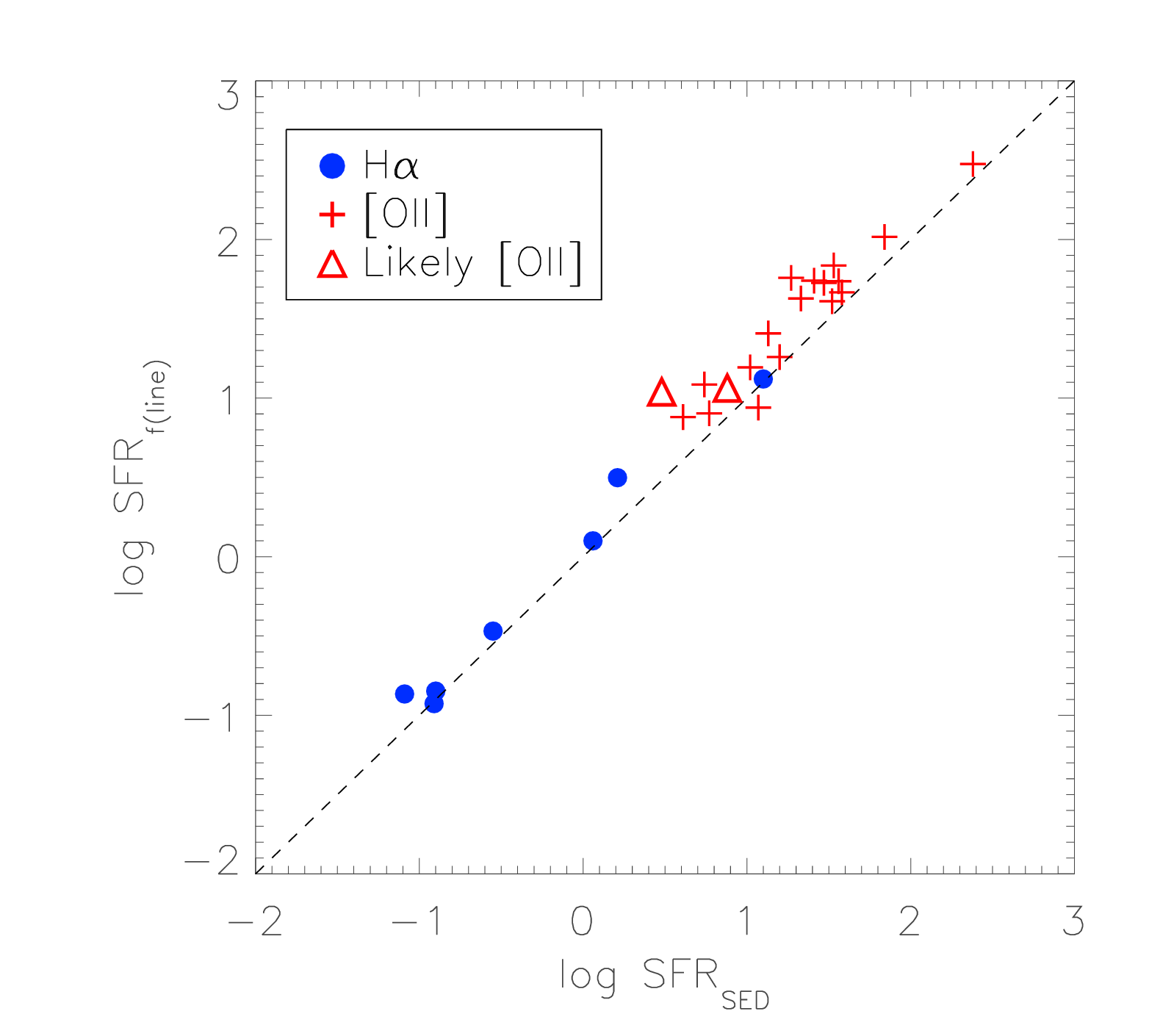}
\caption{ \label{SFR_lines} SFRs derived from the emission line flux plotted
against the SFR values obtained from the SED fitting method. Symbol shapes and 
colours are similar to those in Fig.~\ref{MSFR}. The two methods show the offset of 
$0.19\pm0.05$ dex which means that the values 
derived with two different methods are in excellent agreement for the entire sample.}
\end{figure}

\begin{figure}
\centering
\epsfig{file=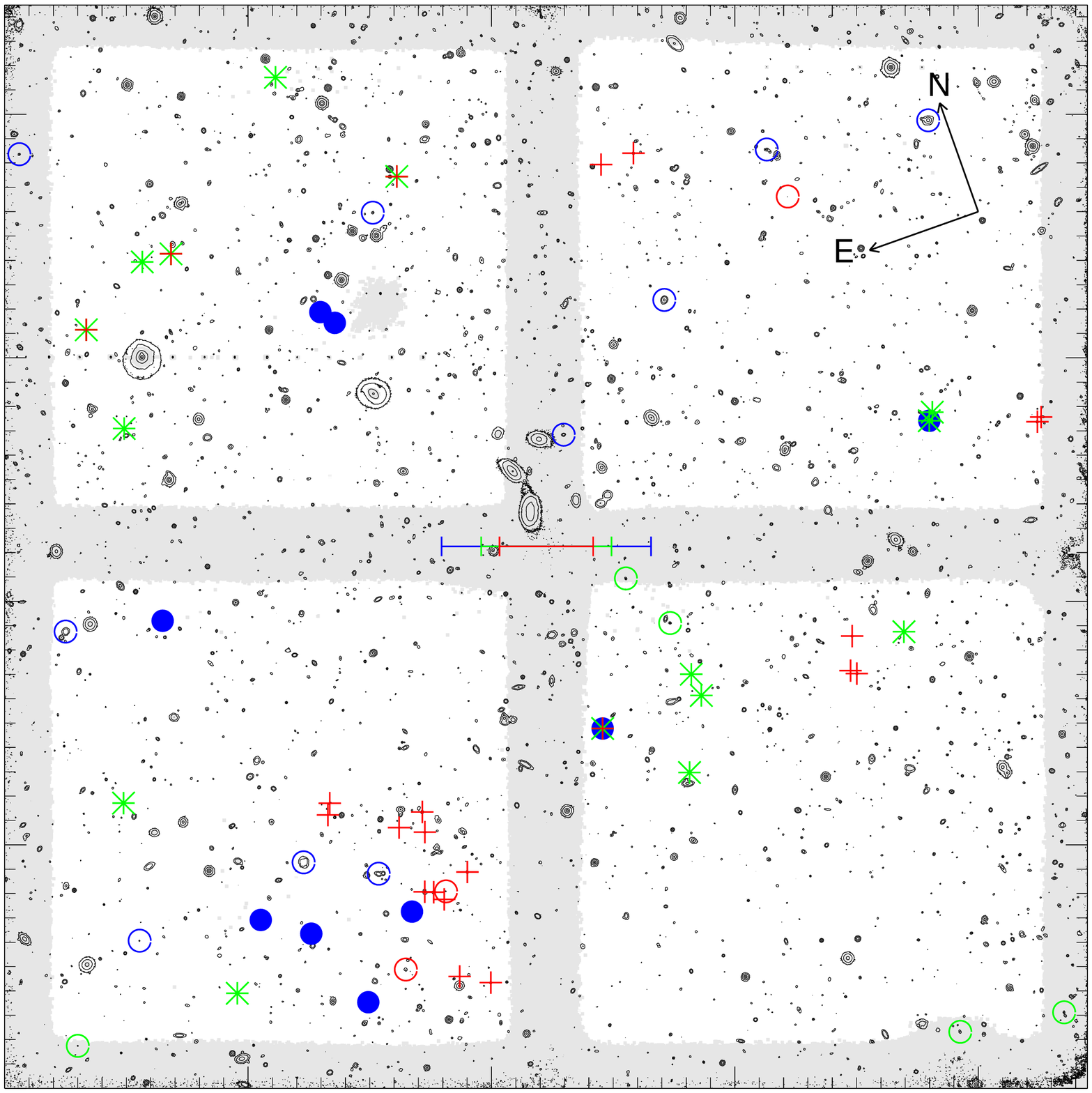,width=9cm}
\caption{\label{SKYMAP} The objects identified in the three redshift
slices overplotted on the narrow band image (black contours). Blue dots are 
$H\alpha$ emitters, green asterisks are \fion{O}{iii}/$H\beta$ and red 
crosses represent the \fion{O}{ii} emitters. Lower S/N areas of the image which 
were excluded from the basic sample are shaded grey. Multiple symbols 
over-plotted on each-other represent galaxies with multiple redshift solutions.
Open circles represent galaxies from the ''extended sample''.
Bars of length 1 comoving Mpc at the given redshift is over-plotted in in the
centre of the image with same color-coding as for the objects.}
\end{figure}

For ELG 30, where we do not have a preferred redshift, the line flux for $z=0.62, 
1.15, 1.85$ is $3.17\pm0.16, 2.96\pm0.16, 2.97\pm0.16\times10^{-17} erg/s/cm^2$, 
and the EWs are $2011.4\pm106.4, 1851.6\pm97.9, 1969.8\pm104.2 \AA$. 
In Table~\ref{Table1} we list the value for $z=1.15$.

Emission lines provide us with an alternative for measuring the SFR. We correct the emission line fluxes 
for intrinsic reddening using the best fit E(B-V) from the LePhare fits and a Calzetti extinction curve. 
We then calculate $H\alpha$ and \fion{O}{ii} luminosities which are converted to a SFR using the calibrations in
\citet{Kennicutt98}, and we include a downward correction of a factor of 1.8 to correct from a Salpeter 
to a Chabrier IMF. The result is shown in Fig. \ref{SFR_lines}, which demonstrates that there is an excellent agreement 
between SFRs derived from emission lines and from the SED fits. In fact the average offset in the SFR is just 
$0.19\pm0.05$ dex between the two different methods. Assuming a typical \fion{N}{ii}/H$\alpha$ ratio 
of 0.1 appropriate for low-mass galaxies, the emission line fluxes and the blue points in Fig.~\ref{SFR_lines} will have a downward correction of 0.05 dex.
Including this correction the offset between the emission line derived SFRs and the SED SFRs is $0.17\pm0.05$.

\subsection{Clustering and large scale structure in three narrow
redshift slices}
\label{Clustering}

\begin{figure}
\flushleft
\epsfig{file=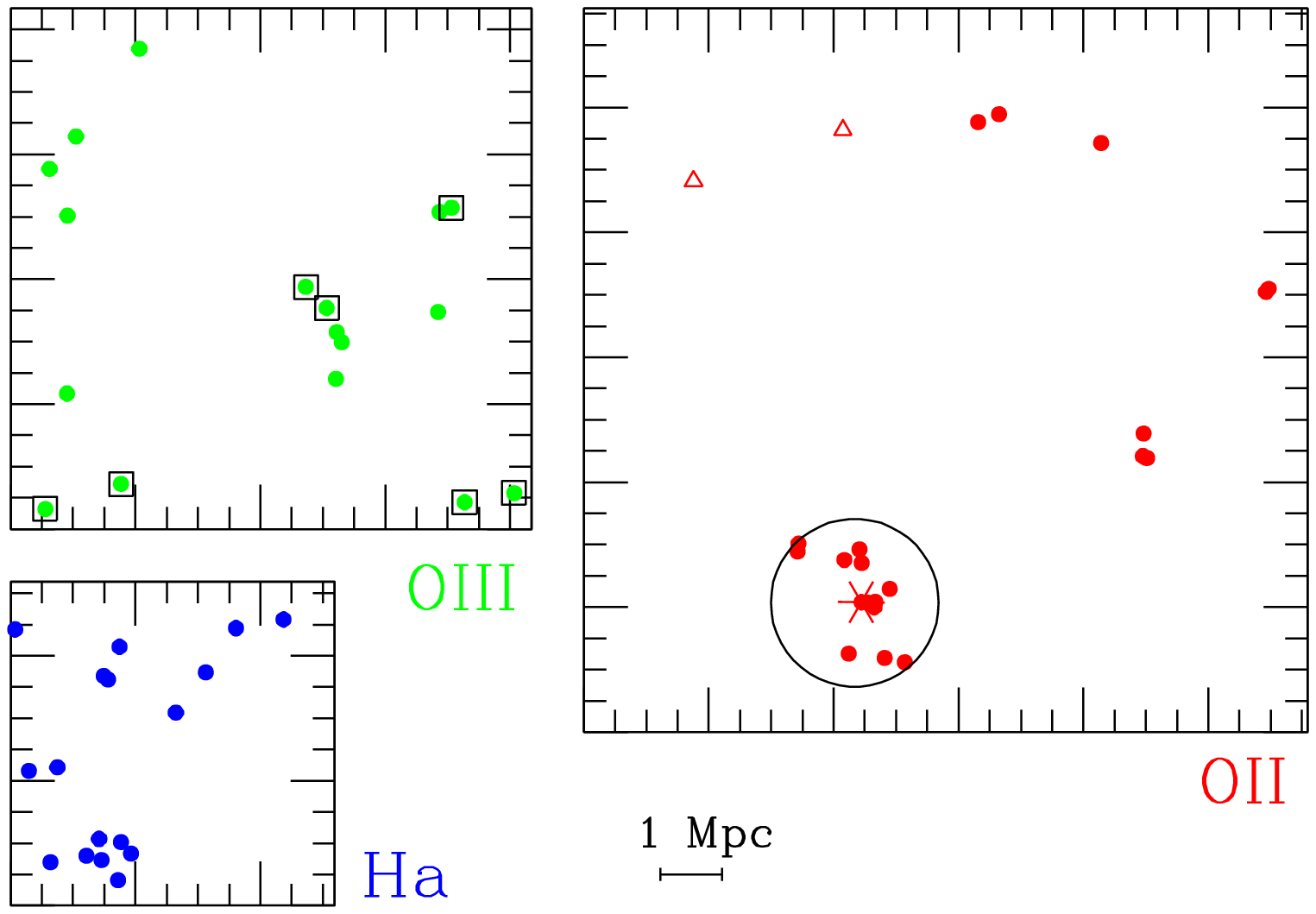,width=8.5cm}
\caption{\label{GC_fig.12} Objects detected in the three redshift slices
shown separately, and
scaled to the same co-moving scale. The colour coding is as in
previous figures, dots mark certain redshifts from the extended
sample, open triangles mark primary redshifts for the two uncertain cases.
Open black squares mark objects with known spectroscopic redshifts in
the \fion{O}{iii} slice. The red star in the \fion{O}{ii} slice marks the
galaxy with the highest M$_{\star}$, while the large black circle marks
the cluster centered around it at z=1.85.}
\end{figure}

In this section we consider the extended sample of 58 objects in three
redshift slices. In Fig.~\ref{SKYMAP} we plot the objects in the three
redshift slices overlaid on
our narrow band image (in black contours). In this figure we also show
the masked lower signal-to-noise regions (shaded grey). The same field
covers different physical scales, and different comoving scales, in
the three redshift slices. In Fig.~\ref{GC_fig.12} we again plot the
three slices separately, but here we have scaled them all to the same
comoving scale. We have subsequently found that the z=1.84 cluster has been 
discovered independently in a study based on CANDELS and 3D-HST spectroscopic 
redshifts in the field \citep{Mei2015}. We refer the reader to this work for further 
discussion of this interesting structure.

One feature which is immediately visible is the concentration of
\fion{O}{ii} emitters in the lower left quadrant. In section~\ref{Sect4.1.1}
we found that the \fion{O}{ii} emitter sample on average has higher mass
than galaxies in the other slices, so because high mass galaxies are known
to cluster more strongly than low mass galaxies, this is indeed the slice
in which we would be most likely to find a galaxy cluster. In
Fig.~\ref{GC_fig.12} we have
marked a circle with a diameter of 2.55 comoving Mpc, which encloses
13 of the 23 \fion{O}{ii} emitters in our extended sample. We have also marked
the position of the highest mass galaxy in our sample, and it is seen
to fall very close to the centre of the circle. From Fig.~\ref{SKYMAP} we
see that there is indeed evidence for a higher density of both optical
and X-ray sources \citep{Xue12} around the position of the clump of \fion{O}{ii} emitters.
Computing the surface density of galaxies inside the circle we find
2.5 per comoving $Mpc^2$ while outside of that it is 0.08 per
comoving $Mpc^2$. 
On the basis of the observations reported above, we here conclude
that we have identified a galaxy cluster at $z=1.85$
in our \fion{O}{ii} redshift slice.

Simulations of early galaxy and structure formation all share a common
prediction that the first structures to form are filaments who's ends
are connected in nodes. Young low mass galaxies form in the filaments,
and while they assemble further and grow, they also drift along the
filaments into the nodes where they form galaxy groups and
eventually clusters \citep{Monaco05}.  Samples of high mass galaxies
are therefore strongly clustered and well suited to identify the nodes
as we showed in the previous paragraph, but in order to identify
filaments one needs samples of lower mass galaxies covering volumes large
enough to cover the expected sizes of filaments (20-25$h^{-1}$ Mpc),
\citep{DemDor99}. The end product of the evolution of this cosmic web
has been well studied at low redshift, and recently a large catalogue
of filaments in the redshift range $z=0.009-0.155$ was published
\cite{Tempel14}, but at higher redshifts than
0.155 this becomes much very difficult.
\cite{Warren96} argued that Ly$\alpha$ emission line selected galaxies have
lower masses than continuum flux selected samples, and suggested that
they could be used to identify filaments. \cite{moller98}
showed on a statistical basis that Ly$\alpha$ emitters do tend to line
up in strings. Nevertheless, the actual mapping of filaments is hampered
by two issues: mostly the observed volumes are too small, and mostly
there is no follow-up spectroscopy which is required to provide the
3-D mapping of the volume.

In one case a fully resolved filament mapped in Ly$\alpha$ was
identified at z=3.04 \citep{Moller01} where a total of eight objects were
found to be enclosed in a cylinder with proper radius 400 kpc
which in the cosmology we use here corresponds to also 400 kpc.
In Fig.~\ref{GC_fig.12} we see that 10 of 17
galaxies at $z=1.15$ lie close to a line going almost diagonally from the
lower left corner of the field towards the upper right. This could be
a chance alignment of galaxies at mixed redshifts, but it could also
be a filament seen under some inclination angle. As in the work by
\cite{Moller01} our field is too small to identify a filament
which lies in the plane of the sky, we would see too few objects in
such a small filament section. To test if we do indeed have enough 3-D
information we have also in Fig.~\ref{GC_fig.12} marked (black squares)
those objects in the \fion{O}{iii} slice for which we have spectroscopic
redshifts, and we see that we have 5 spectroscopic redshifts
covering the entire length of the diagonal.

\begin{figure}
\flushleft
\epsfig{file=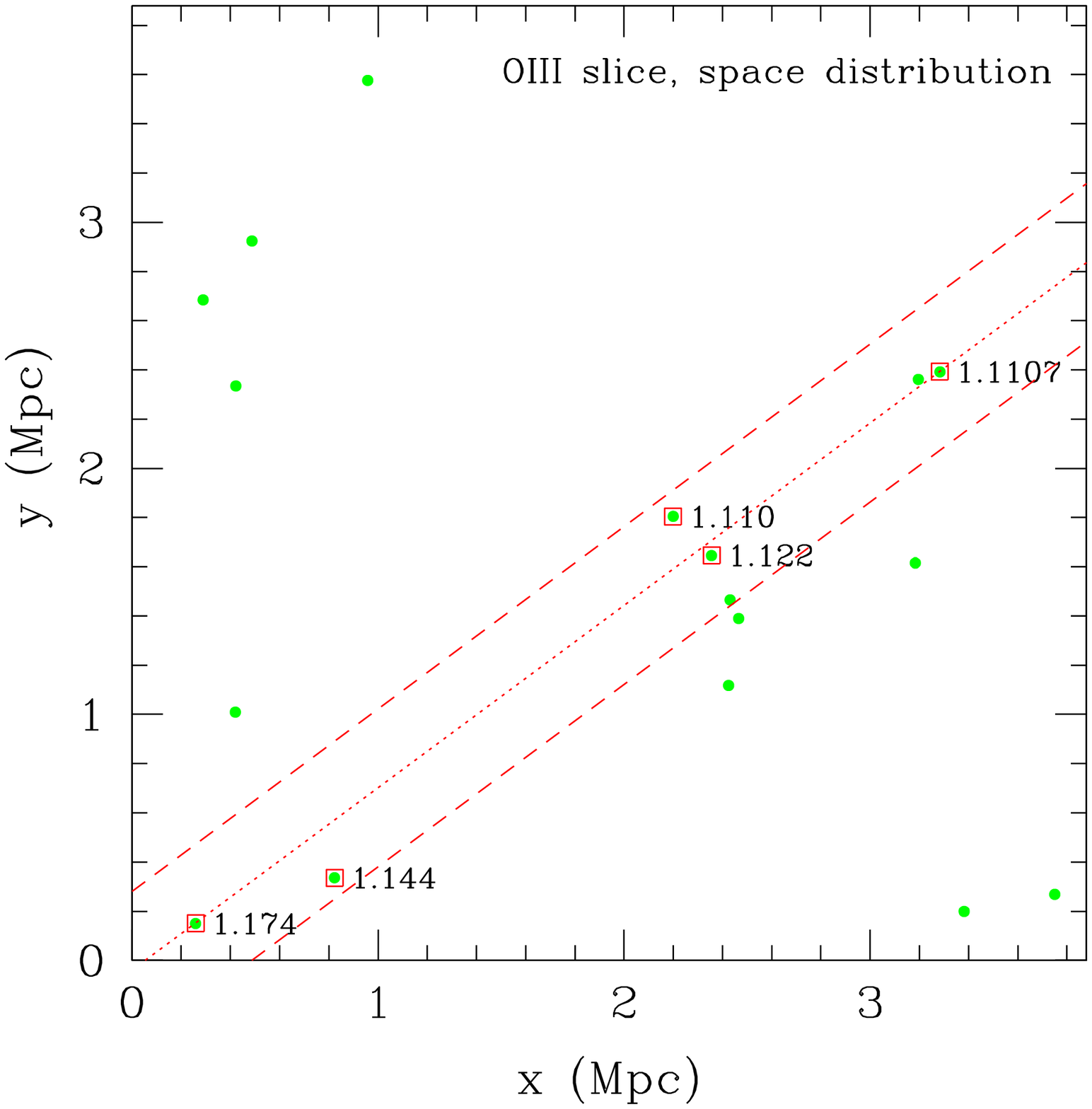,width=8.5cm}
\caption{\label{GC_fig.13} Here we plot again objects detected in the
\fion{O}{iii} slice but now on proper scale. The red dotted and
dashed lines provide the size-scale of the filament of emission
line galaxies at $z=3.04$ reported by \cite{Moller01}. Also we mark
the spectroscopic redshifts of five galaxies which may outline a
similar filament at $z=1.15$ in this field.}
\end{figure}

In Fig.~\ref{GC_fig.13} we again plot the objects in
the \fion{O}{iii} slice, but here in
proper length scale, and with the redshifts of the 5 galaxies on the
diagonal line marked. We see that the redshifts in general grow from the
upper right towards the lower left, so this does indeed appear to be a
filament pointing from the upper right towards the lower left away
from us. In order to compare to the previously reported Ly$\alpha$ filament
we have marked the width (400 kpc) of that filament on top of this one
by dashed red lines, and all 5 objects are seen to fit well within
this cylinder in this projection.
Availability of spectroscopic redshifts allow us to also compute the
arrangement of the objects along the line of sight.

\begin{figure}
\flushleft
\includegraphics[trim = 2.5mm 80mm 0mm 0mm, clip, width=9cm]{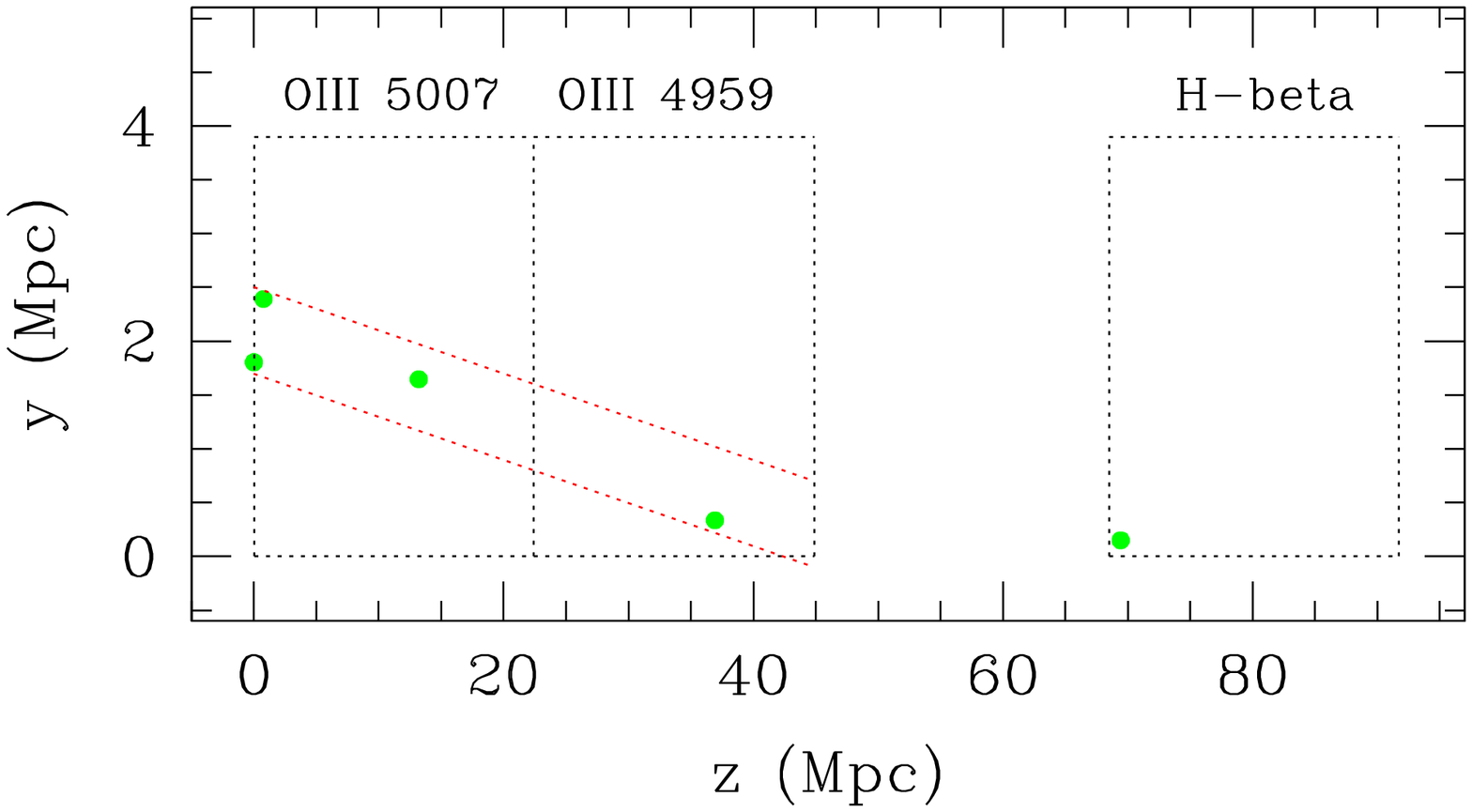}
\caption{\label{GC_fig.14} Similar to Fig.~\ref{GC_fig.13} but here we
have converted redshifts to proper distance, and show the projection
onto the (y vs distance) plane. For comparison we again mark a filament of
the same proper width as in Fig.~\ref{GC_fig.13}. The \fion{O}{iii}
slice covers a $z$-range about four times wider than the other slices
because of the three emission lines. Selection by each of the three lines
is marked by the dotted lines.}
\end{figure}

The \fion{O}{iii} redshift slice is thicker than the other two slices because
we here have three individual lines (\fion{O}{iii} 5007, \fion{O}{iii} 4959, and
H$\beta$) either of which could fall into the narrow pass band. We
visualize this in Fig.~\ref{GC_fig.14} where we have kept the field
y-axis of Fig.~\ref{GC_fig.13}, but have turned the volume 90 degrees and replaced the x-axis by
the z-axis (i.e. redshift converted to proper distance). The three
dotted boxes here represent the volumes sampled by each of the three
emission
lines, green dots are the galaxies, and the diagonal dashed lines again
mark out a filament of thickness as in Fig.~\ref{GC_fig.13}. We see that the first
four galaxies would indeed fit into a straight, cylindrical filament of
this thickness, but it would be somewhat longer than the Ly$\alpha$
filament at redshift 3.04.
The last galaxy seen in H$\beta$ may well belong to the same filament, but it would have 
to be bent or thicker in that case. The length of the filament is in excellent agreement with 
the detection of the Ly$\alpha$ filament at $z=3.01$ reported by \cite{Matsuda05} and the 
recent work at low redshifts \cite{Tempel14}.

In conclusion, we have shown that emission line
selected galaxies at those redshifts are well suited to perform
observational tests of simulations of large scale structure. The
H$\alpha$ field size is in this case too small, but surveys over
larger fields like UltraVISTA \citep{Henry,Milvang} will provide fields of
sufficient size. The \fion{O}{ii} slice has a larger volume, and in general the
\fion{O}{ii} selected galaxies have higher mass than the lower redshift slices,
making the \fion{O}{ii} slice ideal for rich group and cluster statistics.
The \fion{O}{iii} slice is extremely well suited for filament searches
because the depth allows to identify filaments at any inclination angle.
This promises that it may soon be possible to perform the alternative and
''purely geometrical'' cosmological test and determine $\Omega_{\Lambda}$
using filaments as described in detail by \cite{Weidinger02}.
Identifying filaments require spectroscopic redshifts, or some other
diagnostic for more accurate redshift determination. One such novel method
using only VISTA narrow band data has recently been described
(Zabl et al. in preparation).

\section{Discussion and conclusions}

\subsection{Galaxy scaling relations at low masses}

Understanding the scaling relations of galaxies of all masses is
fundamental to understand galaxy formation and evolution. Yet, galaxy
samples selected in all emission bands ranging
from X-rays over UV, optical, IR, sub-mm, and mm, to radio, all form
flux limited samples of galaxies being the most luminous, and
presumably the most massive, of their kind. Such samples are,
by definiton, the easiest to obtain, and by right large fractions of
our knowledge of high redshift galaxies originate from such samples.
However, to explore the low mass range of galaxies, notably at high
redshifts, other selection techniques are required. One such technique
is emission line selection via deep narrow and broad band imaging.

We are involved in several narrow/broad band imaging surveys, and in
this paper we have reported on a pilot project to study the feasibility
of using such surveys to trace low mass galaxy scaling relations and
their redshift evolution.
Simple narrow/broad emission line selection allows to select galaxies
with strong emission lines, thereby providing a deepening of the flux
limited samples, and in this present study we have specifically chosen a
broad-narrow-broad selection that results in a selection of the highest
emission line equivalent width galaxies.
Two simple predictions for a study of this kind would be

({\it i}) that our sample in the mean could have higher SFR for any
given galaxy stellar mass, and

({\it ii}) that our sample in the mean will select galaxies down to lower
stellar masses than continuum flux limited samples. \\
We carry out a detailed
comparison of our dataset to previous studies and find that both
of those predictions have been confirmed in this work. We
thus provide an ``upper boundary'' to the main sequence of star formation
(MS) at each of the three redshifts we study.

Our comparison to previous work also show that the MS has a
significantly steeper slope at the low mass end (below
$M_{\star} = 10^{9.4}$) than at higher masses.

\subsection{Narrow band selection as cosmological tool}

Any narrow/broad band survey carried out at a
wavelength in excess of the rest wavelength of H$\alpha$ provides a
roughly even coverage of three widely separated narrow redshift
slices corresponding to the redshifted wavelengths of H$\alpha$,
\fion{o}{iii}/H$\beta$, and \fion{o}{ii}. A few additional species at other wavelengths
will also on occasion appear, but only rarely, due to the much weaker
strength of their transitions. The exact ratio of detected objects
between the three main slices depends on their relative equivalent
widths (as a function of redshift), their relative number density (as
a function of redshift), and of the ratio of the surveyed volumes (as
a function of narrow band wavelength and assumed cosmology).

In this work we have surveyed comoving volumes of 1221 Mpc$^3$
(H$\alpha$), 3092 Mpc$^3$ ($\times$3 due to H$\beta$,
\fion{O}{iii}$\lambda$4959, and \fion{O}{iii}$\lambda$5007) and
5536 Mpc$^3$ (\fion{O}{ii}).
Down to our conservatively chosen narrow-band AB magnitude limit of
24.8 they distribute in the
following proportions: H$\alpha$ emitters 20\%, \fion{O}{iii}/H$\beta$
emitters 30\% and \fion{O}{ii}-emitters 50\% (see Fig.~\ref{histogram}).
We compare our redshifts to previous photo-z redshifts from the
literature and show that narrow band selection allows a much more
accurate redshift assignment, notably in the highest redshift
slices. The errors on redshift assignment from photo-z will
propagate into errors on the physical parameters ($M_{\star}$ and
SFR) so smaller, but more accurate, samples of narrow band selected
galaxies will provide checks to see if the propagated errors simply
add scatter, or if they add systematic effects.

We show that the galaxies can be classified fairly robustly based on
two broad
band colours (Fig.~\ref{ZJ-VI}) confirming the earlier study by
\cite{Bayliss'11}.  Therefore, we conclude that emission-line selected
galaxies do indeed split into the evolutionary groups according to
their color.
In Fig.~\ref{MSFR} we see that the galaxies in our lowest redshift
slice on average have the lowest masses, and that galaxies then become
progressively more massive at higher redshifts. This could possibly
be related to the selection via
different emission lines in the three slices, but is more likely a
result of using the same observed magnitude limit for all slices. One
very interesting thing to note is that we are able to select star
forming galaxies of stellar masses down to $10^{8.5} M_{\odot}$ at a redshift
of 1.85, and well below that in the other two slices. With emission
selected samples it is very difficult to study low mass galaxies
beyond the critical redshift of ''cosmic high noon'' at z=2.5, but
absorption selected galaxy samples and samples selected as gamma ray
burst host galaxies (GRBs) have been shown to reach much lower
masses (\citet{Moller13, Christensen14, Arabsalmani14}). Therefore,
in order to be able to connect absorption and GRB selected samples
(with median $M_{\star}$ of $10^{8.5} M_{\odot}$) with continuum emission selected
samples at high redshifts, it is important to create well studied
samples with a wide overlap in stellar masses. Absorption selected
galaxies are in general more easily identified via line emission than
via continuums emission \citep[e.g.][]{Weatherley05,Rauch08,Fynbo10,Fynbo11,Fynbo13},
and the ongoing UltraVISTA \citep{Henry} narrow-band survey
covering $\approx$0.8 deg$^2$ at slightly higher redshifts
(for H$\alpha$: z=0.815, for \fion{O}{iii}/H$\beta$: z=1.38/1.45 and
for \fion{O}{ii}: z=2.19) will create a large sample of low mass
emission line selected galaxies in those three slices \citep{Milvang}.
The UltraVISTA sample will be well suited to connect the current flux
limited galaxy samples out to the highest redshifts (z=6-8) currently
explored by DLA galaxies and GRBs.

One of the objectives of this paper is to derive more robust forecasts on and
what will be found in ongoing or upcoming deep surveys, in particular the
UltraVISTA survey \citep{Henry, Milvang}. The UltraVISTA survey uses a slightly
redder narrow-band filter centered at $1.19\mu \text{m}$ \citep{Milvang}, but
this difference is sufficiently small that evolutionary effects on the
population of $z < 2$ emitters (H$\alpha$, \fion{o}{iii}/H$\beta$ and \fion{o}{ii})
should be small. We can hence make forecasts
for which numbers of the most common types of such emitters
we expect to find in the
UltraVISTA survey based on the present work. Scaling with the area we expect to
detect $\gtrsim$1000 of each of H$\alpha$, \fion{o}{iii}, and \fion{o}{ii}
emitters.  Given the large area of the UltraVISTA we predict to find more rare
line-emitters that are not represented in the more than 70 times smaller area
sampled in the present work.

\subsection{Structure formation traced by emission line selected galaxies}

In Fig.~\ref{MSFR_fig12} we show that objects in the \fion{O}{ii}
slice on average have higher masses than those of the other two
slices. As argued by \cite{Moller01} and \cite{Monaco05}, the lowest
mass galaxies at any redshift are the best candidates for mapping out
the filamentary structure of the cosmic web, while the higher mass
galaxies will be more clustered around the nodes of the web, and
could hence mark the sites of early cluster formation.

In this paper
we have pursued their line of thought and identified a galaxy cluster
(or proto cluster) at $z=1.85$. The cluster has an elliptical shape as
predicted by N-body simulations and has no extended X-ray emission so
it is probably in its early stages of formation.
The galaxy with the highest mass of our entire sample lies in the
centre of the forming cluster, and has been identified as an X-ray
emitter. This makes this galaxy of special interest since it is a very
good candidate for the pre-stage of a central cluster cD galaxy.

Secure identification of filaments is more difficult since it requires
even better redshifts than the narrow band data alone can provide.
We identified a possible filament lying diagonally across the field of
the \fion{o}{iii} slice, and enough of the objects had known
spectroscopic redshifts for a 3D mapping. The candidate filament has
width and length in good agreement with simulations, and with the
previous detection of \cite{Moller01}. Obtaining a few more redshifts
would be good in order to securely confirm the identification, but the
detection of a forming cluster and a likely filament are examples of
the strong potential for tracing the formation of structure in the
early universe with deep narrow band data.

\begin{acknowledgements}
We thank Thomas Kr\"uhler and Jens Hjorth for useful suggestions. Special
thanks to Olivier Ilbert for immense help with LePhare code. We thank the referee 
for comments that helped improve the paper. The Dark
Cosmology Centre is funded by the DNRF. The research leading to these results
has received funding from the European Research Council under the European
Union's Seventh Framework Program (FP7/2007-2013)/ERC Grant agreement no.
EGGS-278202. BC acknowledges support from the ERC starting grant CALENDS. 
This research has made use of the SIMBAD database,
operated at CDS, Strasbourg, France.
\end{acknowledgements}

\bibliographystyle{aa}
\begin{bibliography}{99}

\end{bibliography}

\newpage

\appendix
\onecolumn
\section{Thumbnail images for the ''Basic Sample'' galaxies}
\begin{figure*}[!h]
\begin{center}
\epsfig{file=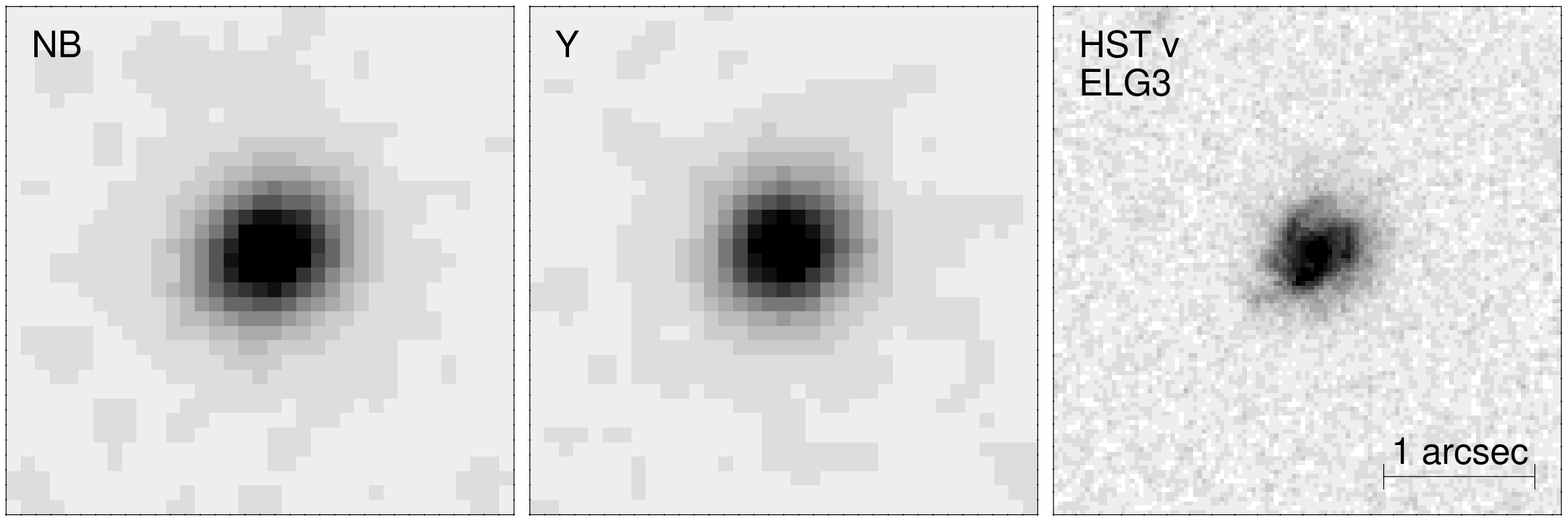, width=6.0cm}
\epsfig{file=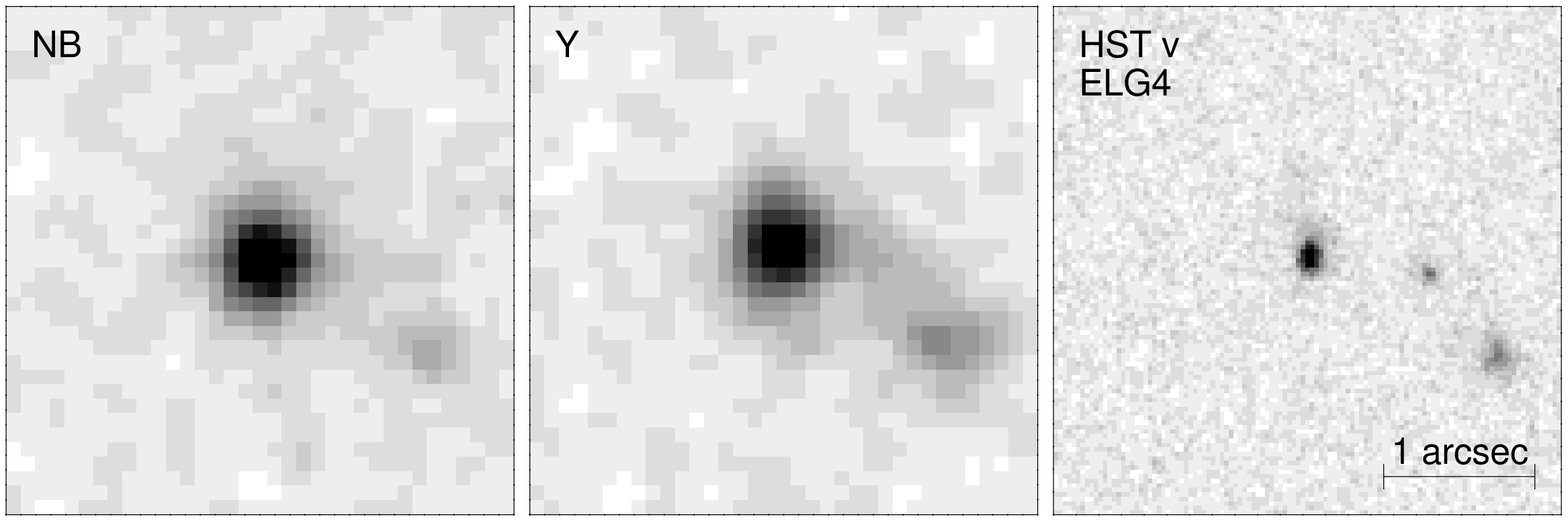, width=6.0cm}
\epsfig{file=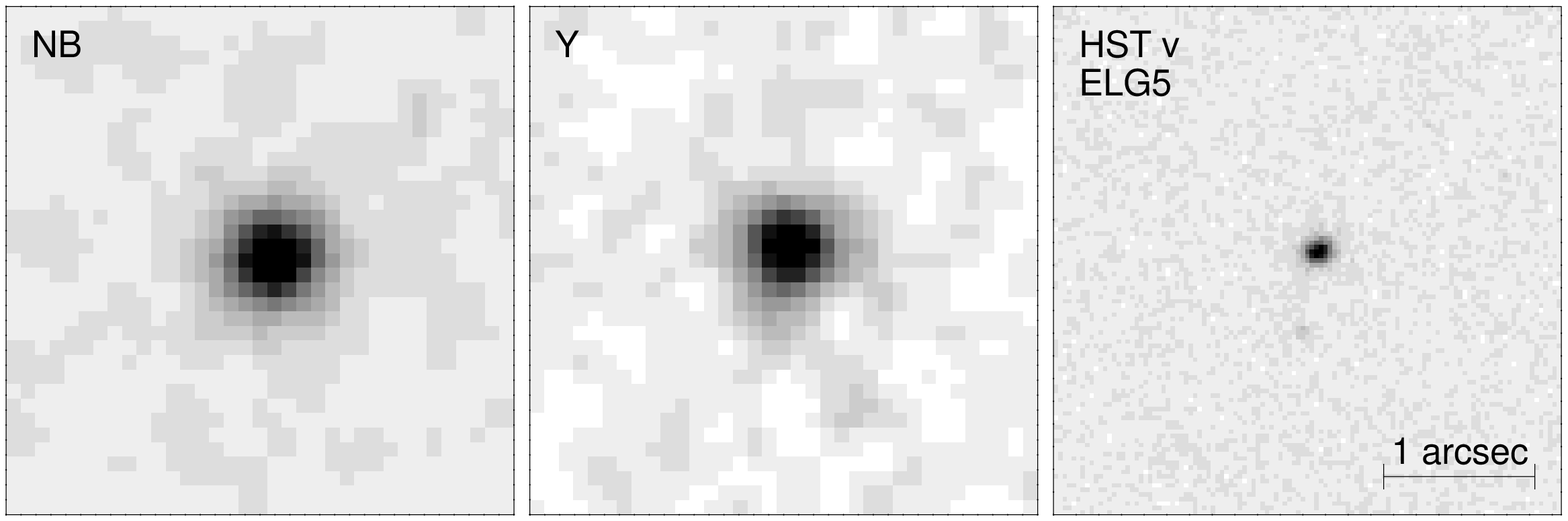, width=6.0cm}\\
\epsfig{file=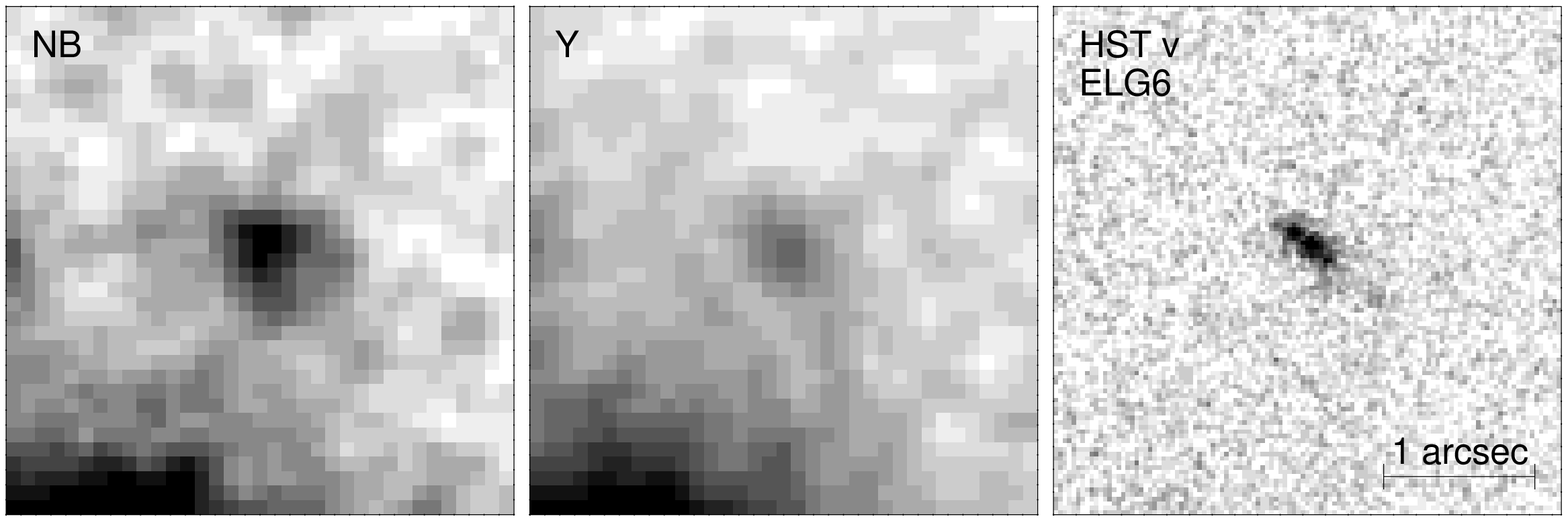, width=6.0cm}
\epsfig{file=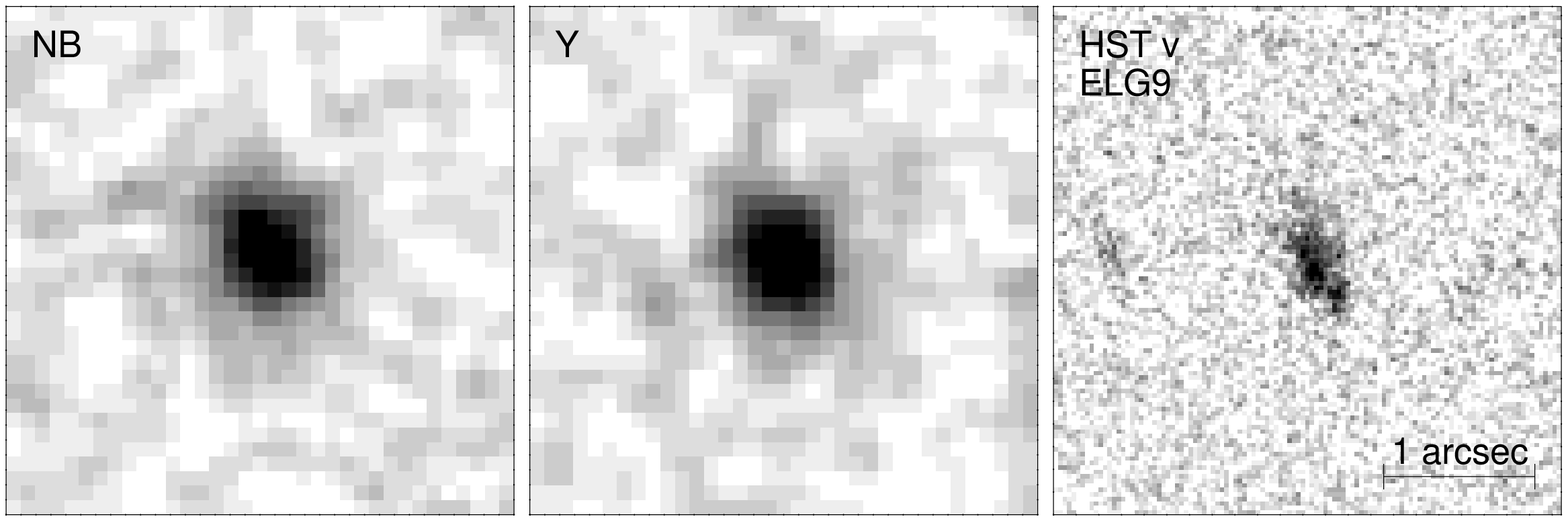, width=6.0cm}
\epsfig{file=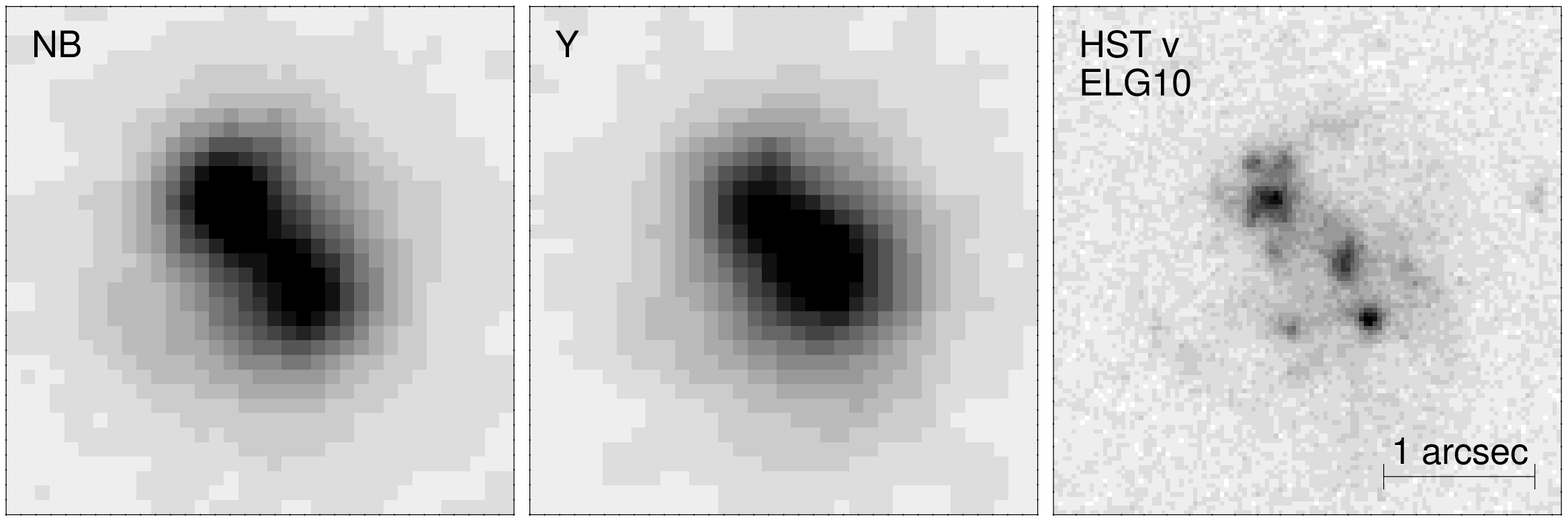, width=6.0cm}\\
\epsfig{file=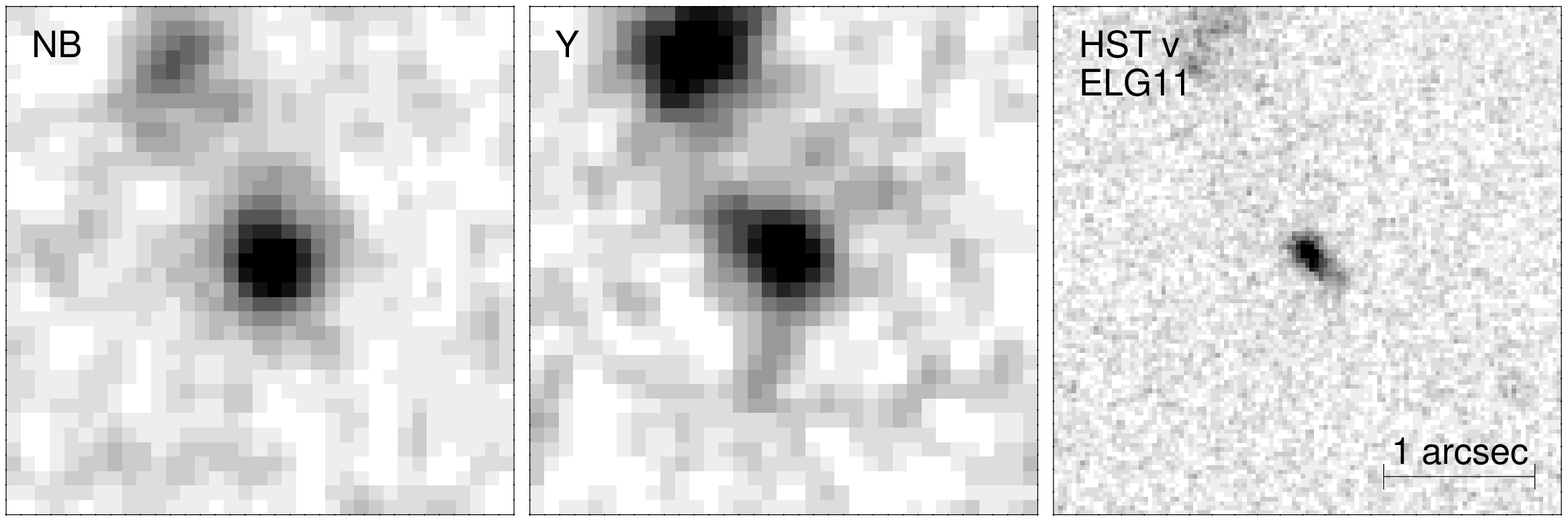, width=6.0cm}
\epsfig{file=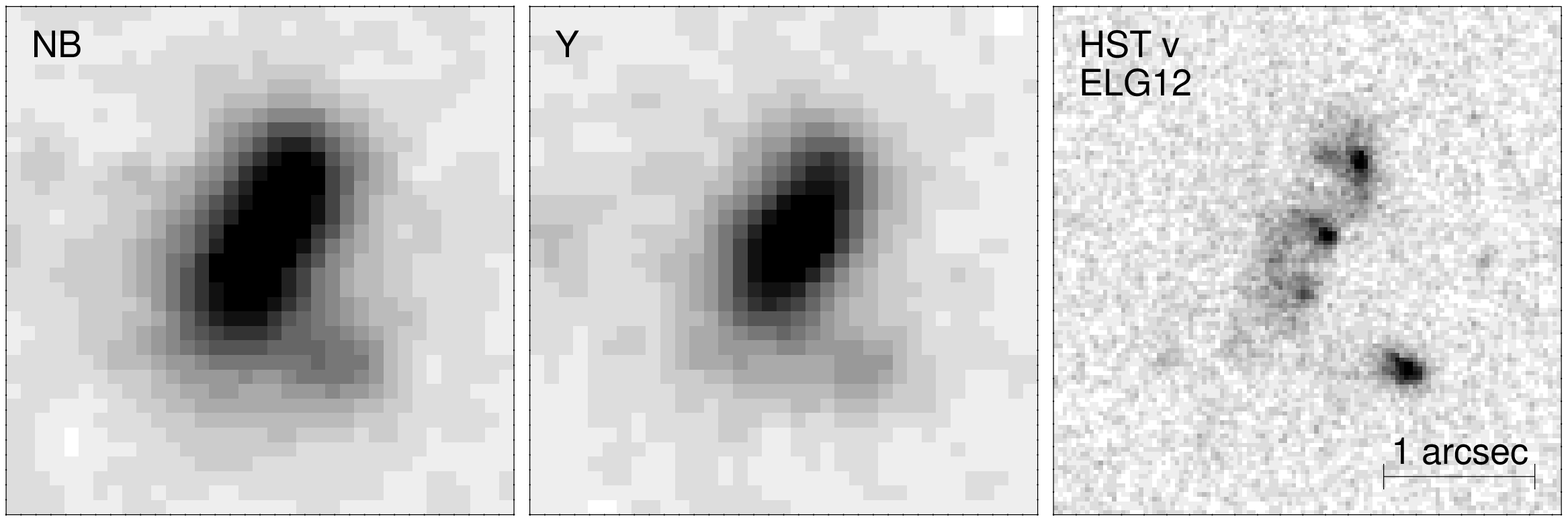, width=6.0cm}
\epsfig{file=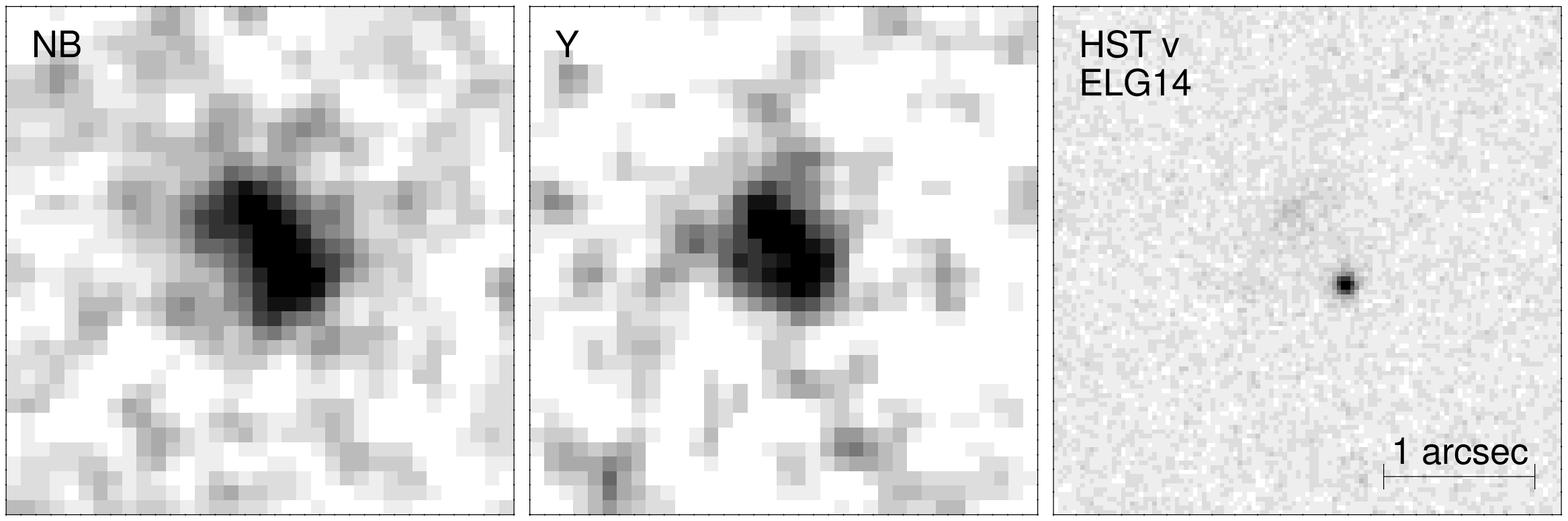, width=6.0cm}\\
\epsfig{file=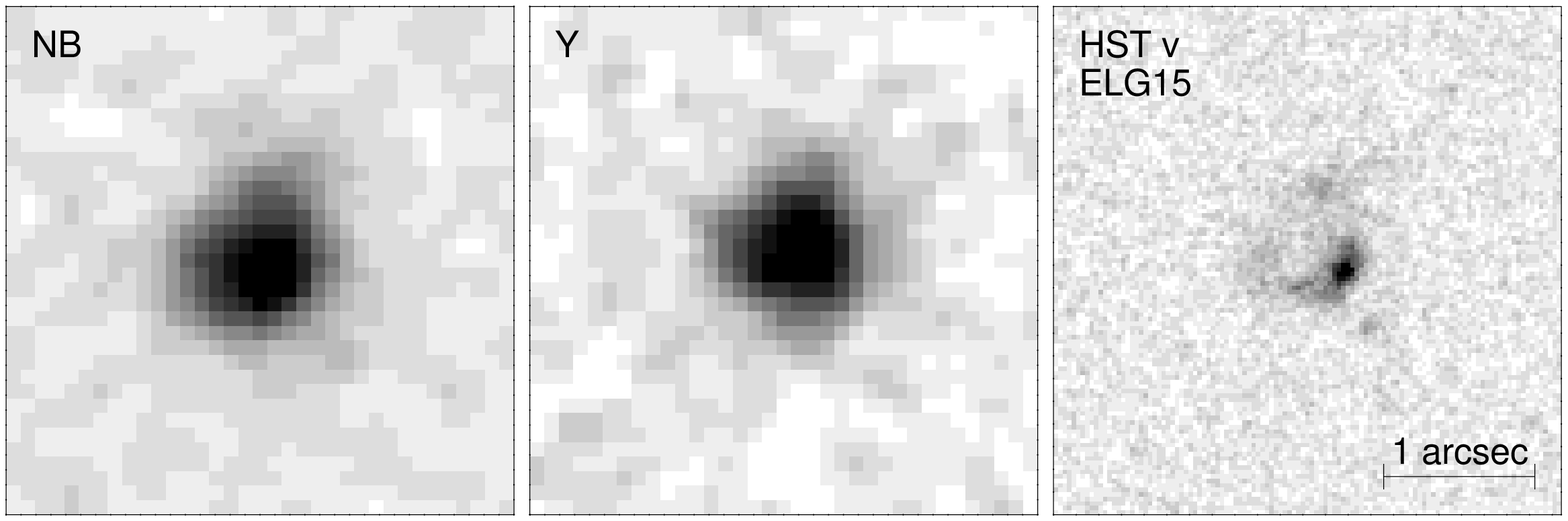, width=6.0cm}
\epsfig{file=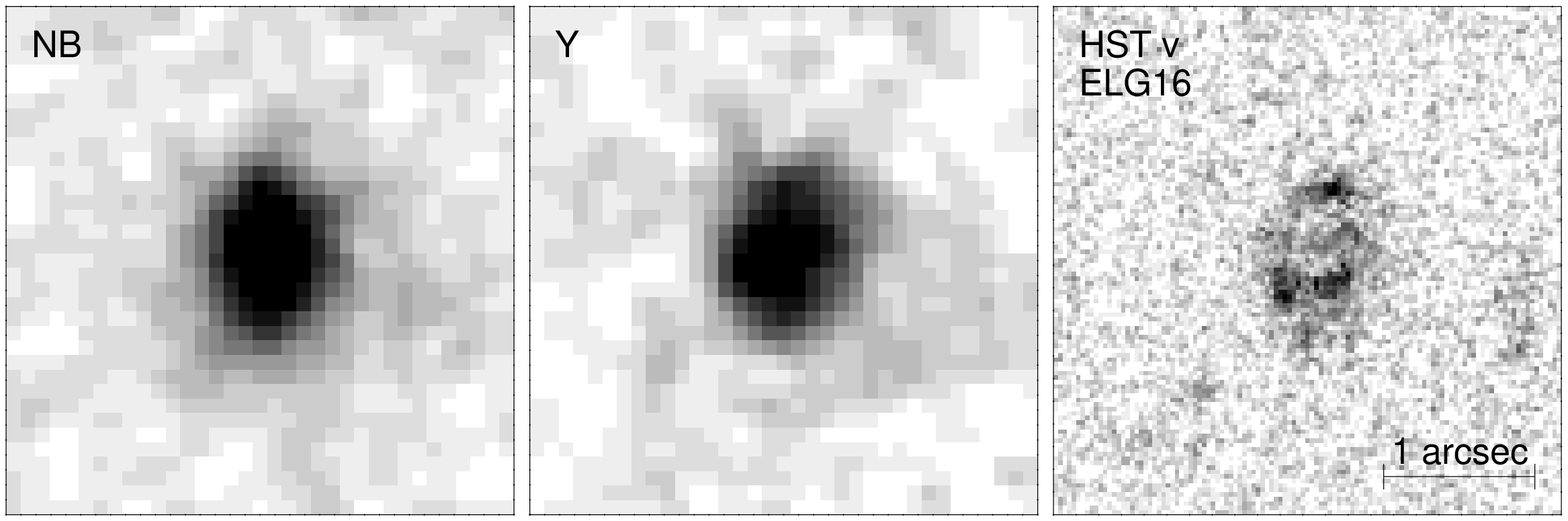, width=6.0cm}
\epsfig{file=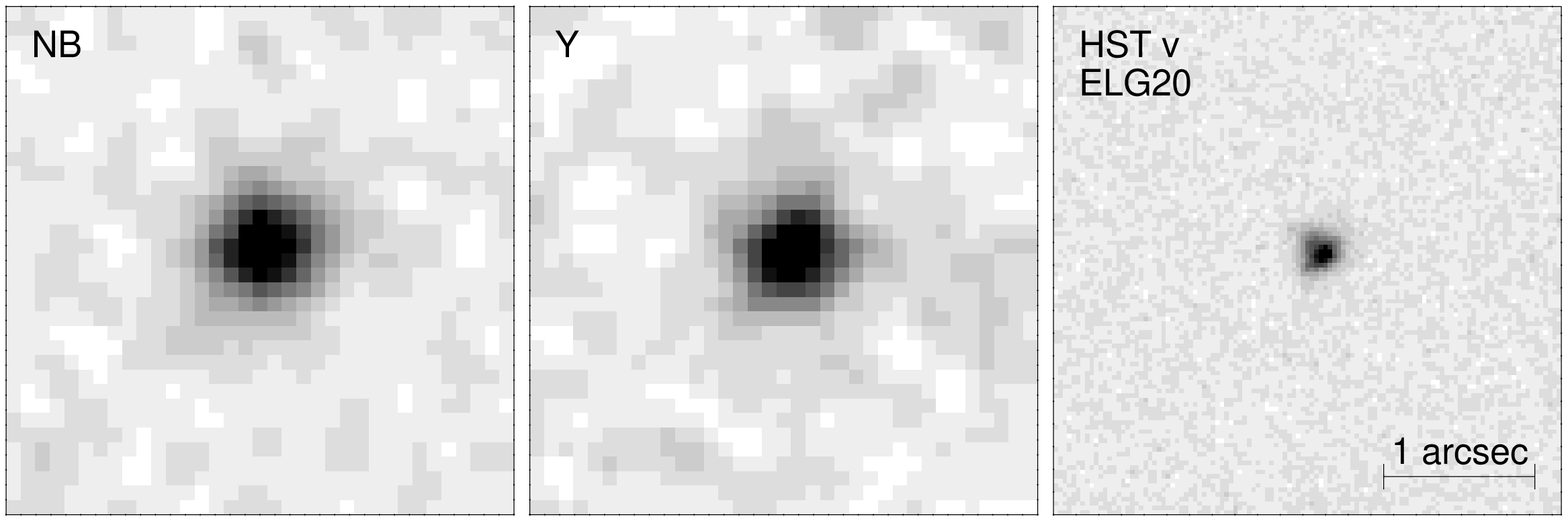, width=6.0cm}\\
\epsfig{file=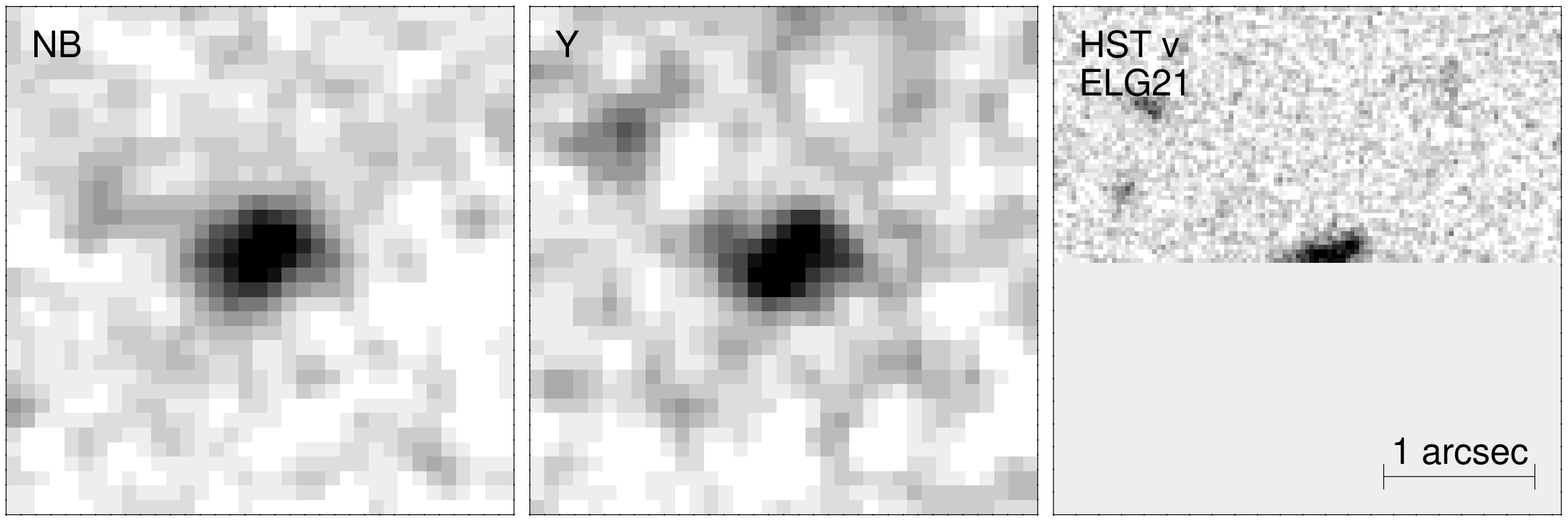, width=6.0cm}
\epsfig{file=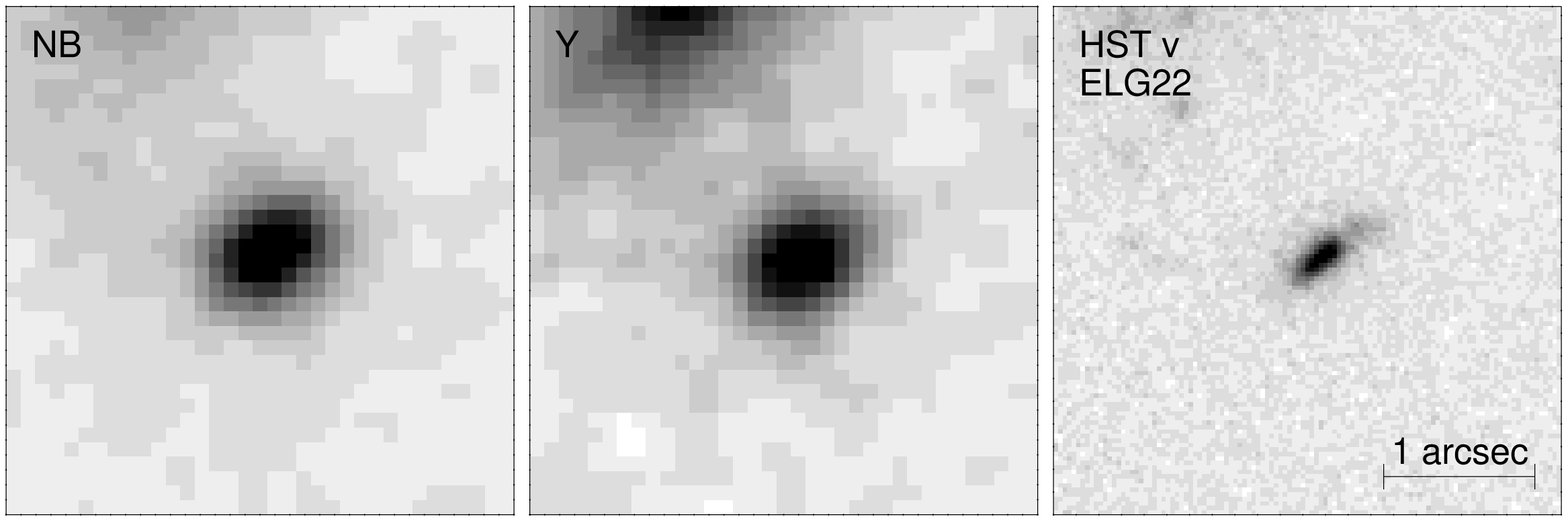, width=6.0cm}
\epsfig{file=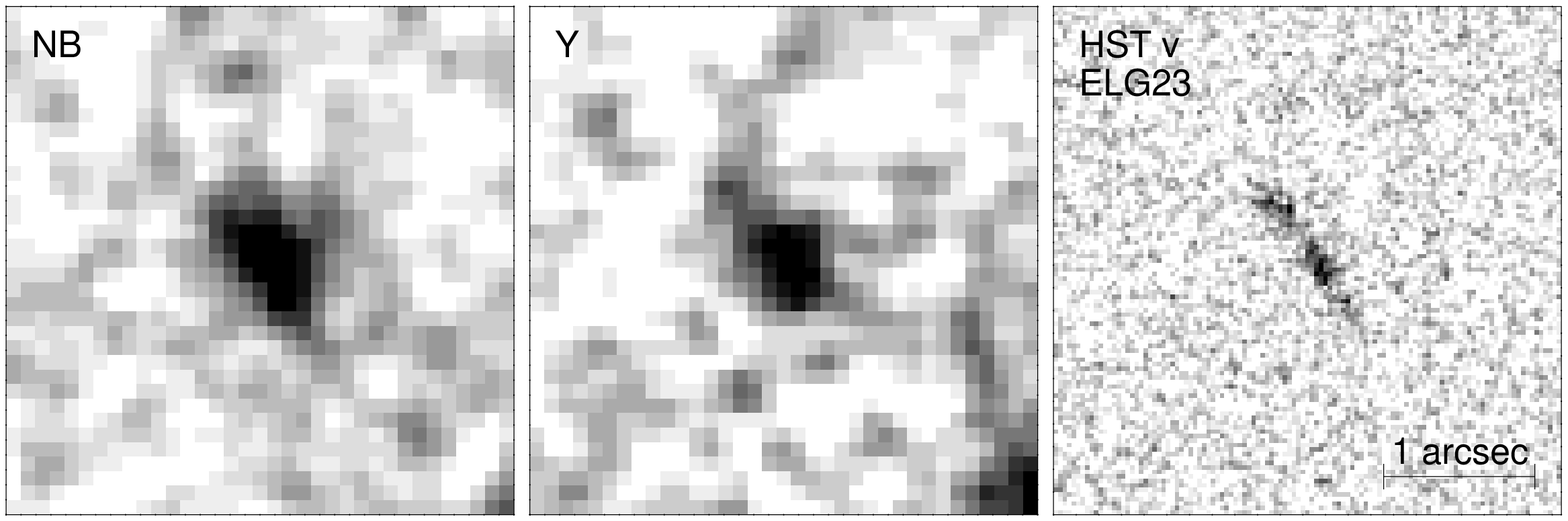, width=6.0cm}\\
\epsfig{file=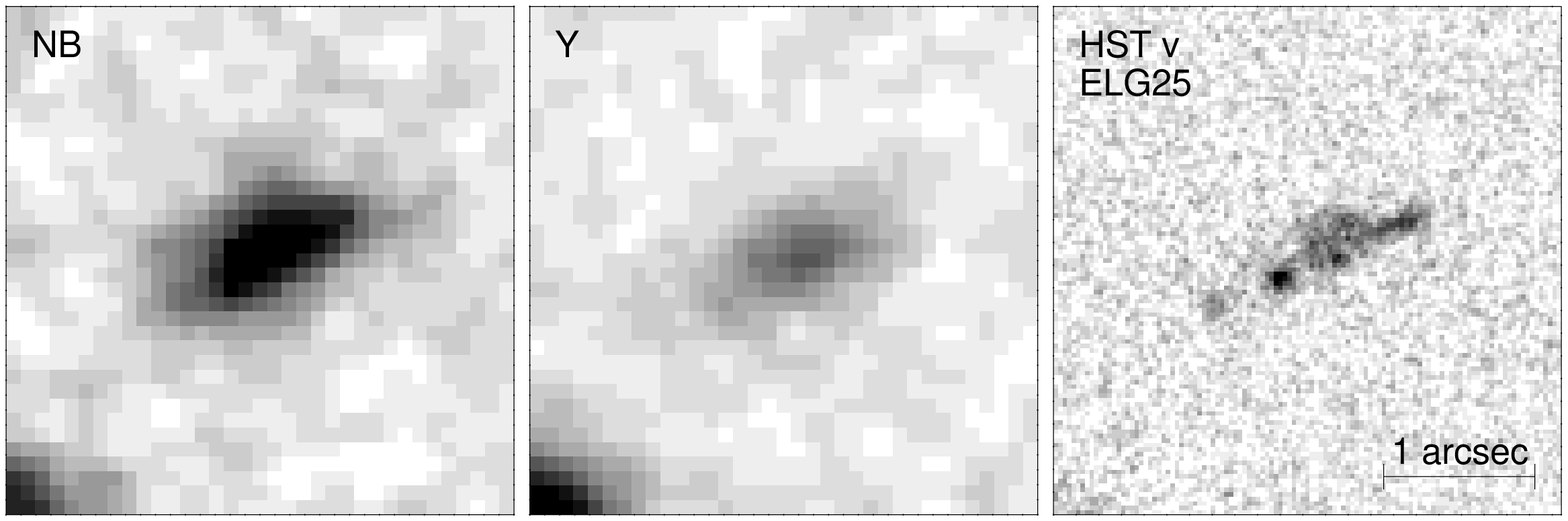, width=6.0cm}
\epsfig{file=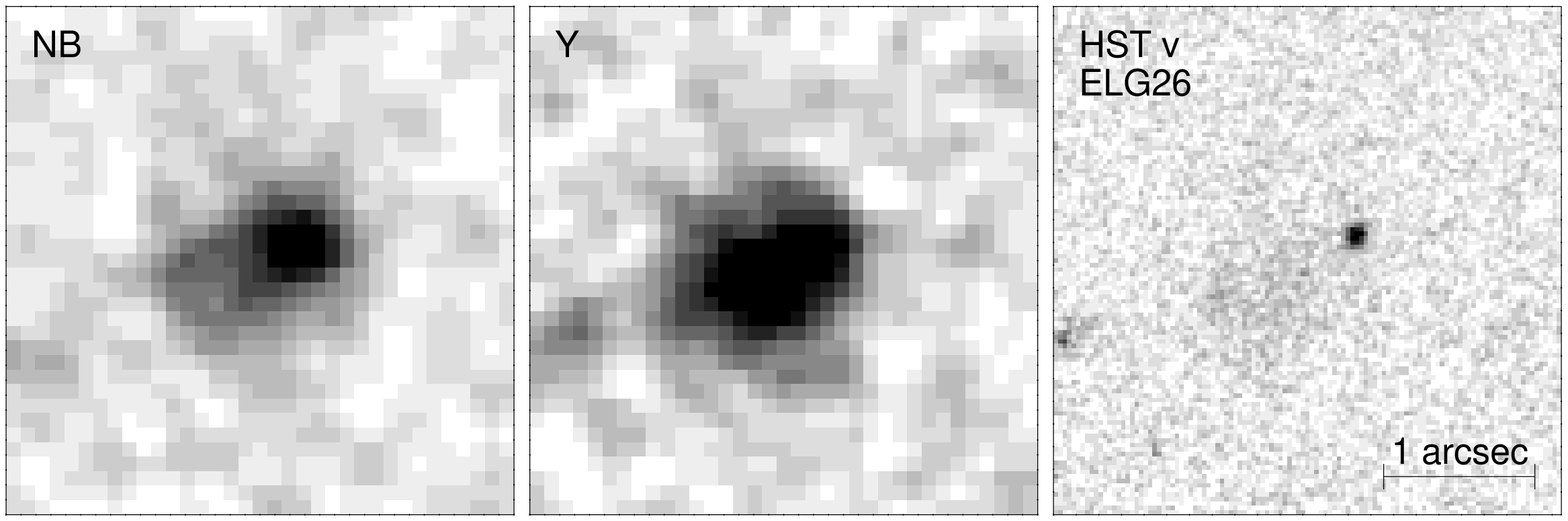, width=6.0cm}
\epsfig{file=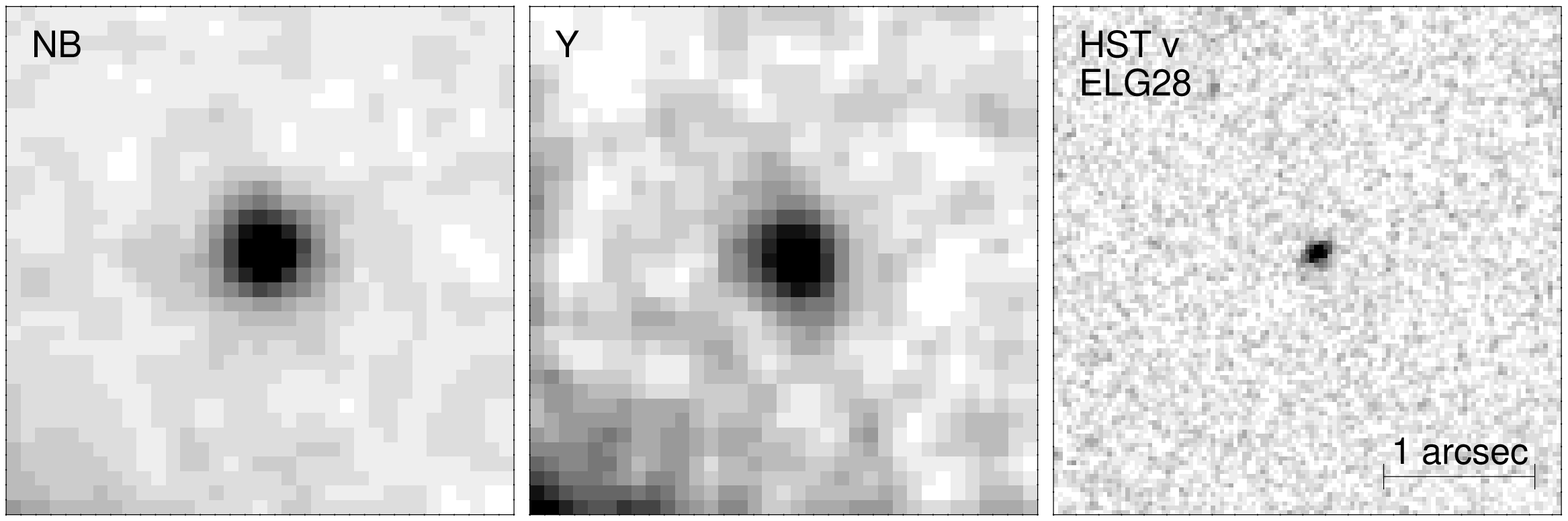, width=6.0cm}\\
\epsfig{file=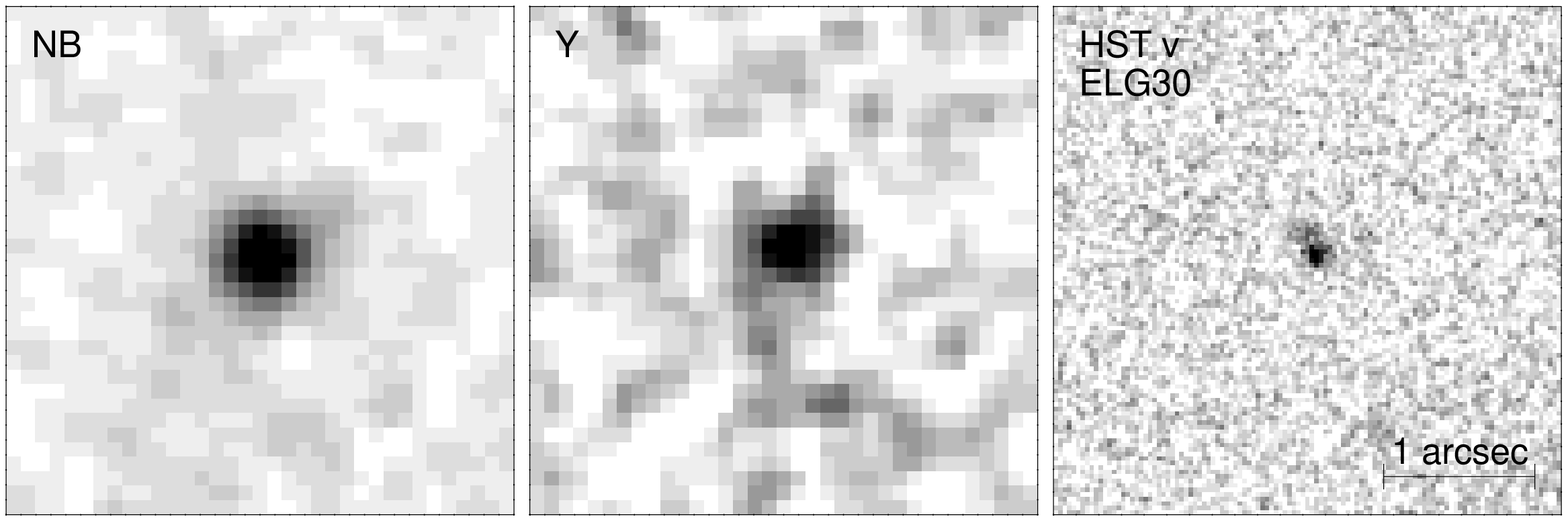, width=6.0cm}
\epsfig{file=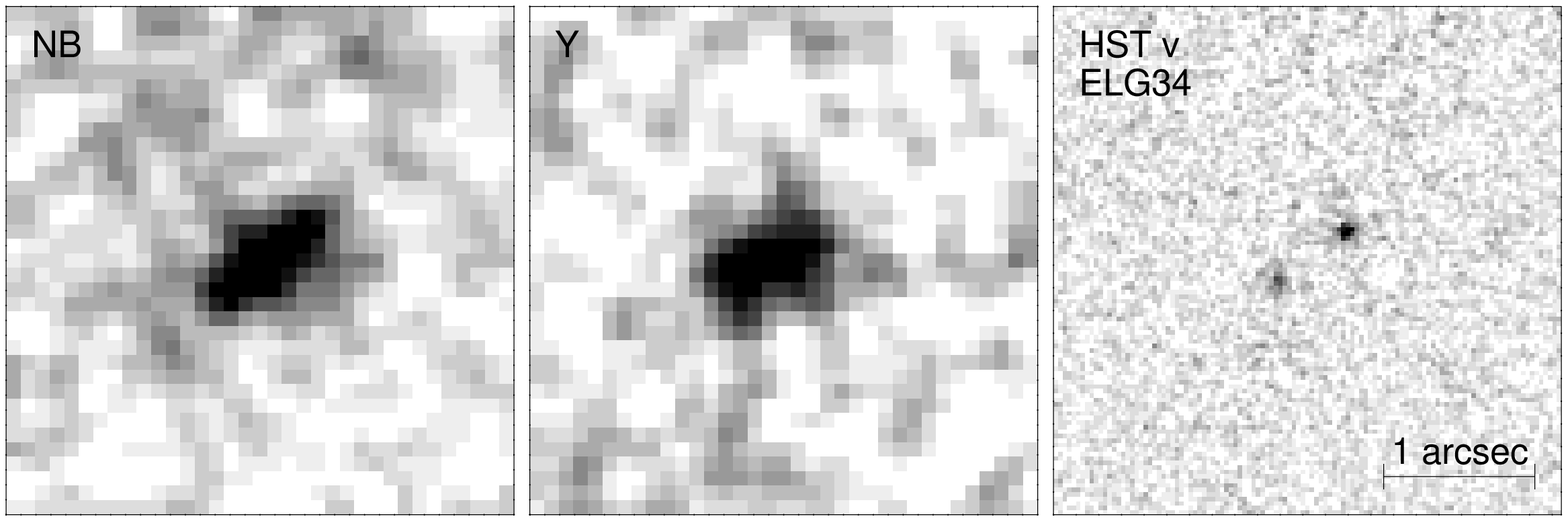, width=6.0cm}
\epsfig{file=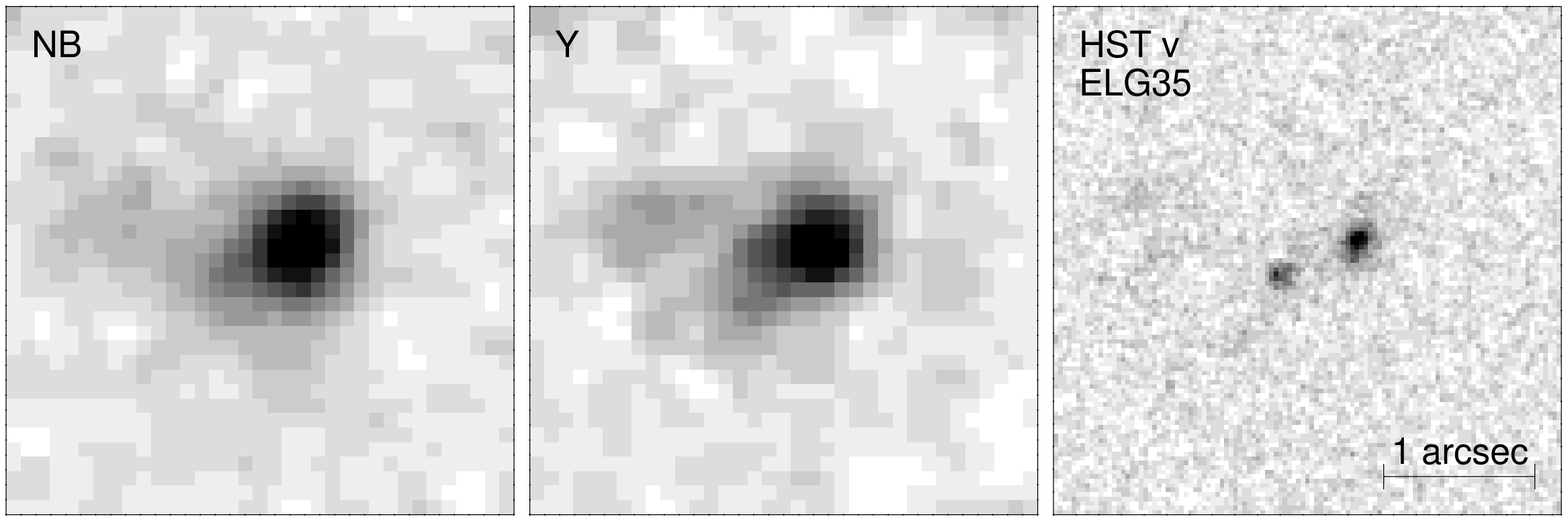, width=6.0cm}\\
\epsfig{file=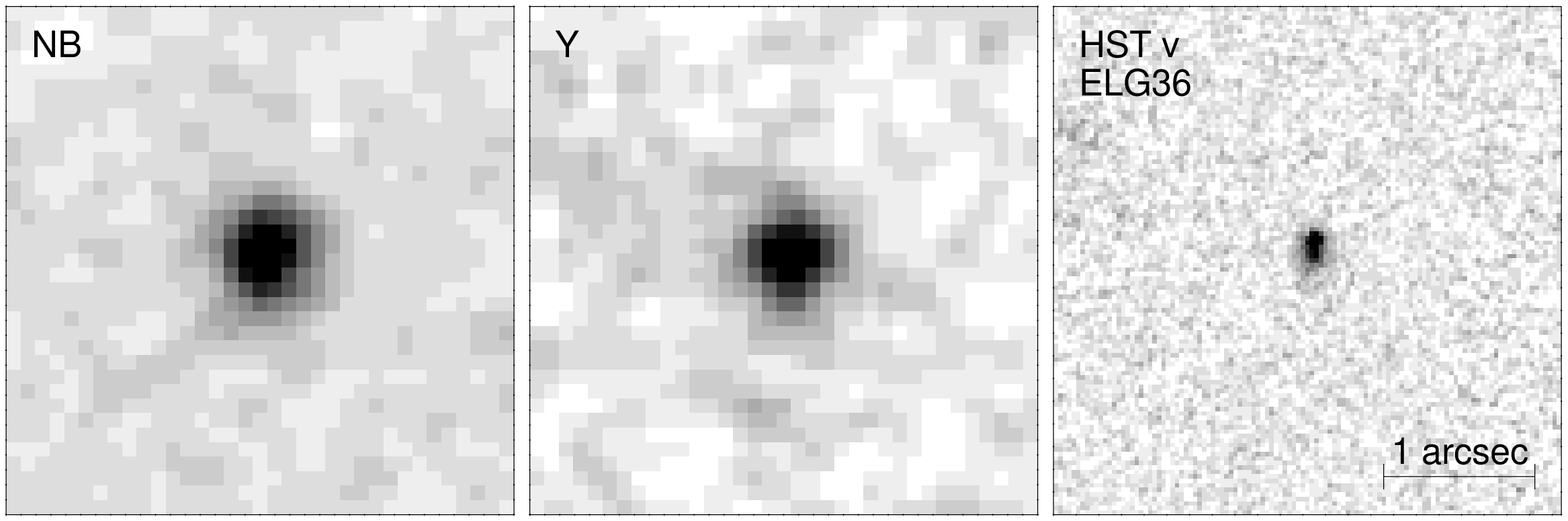, width=6.0cm}
\epsfig{file=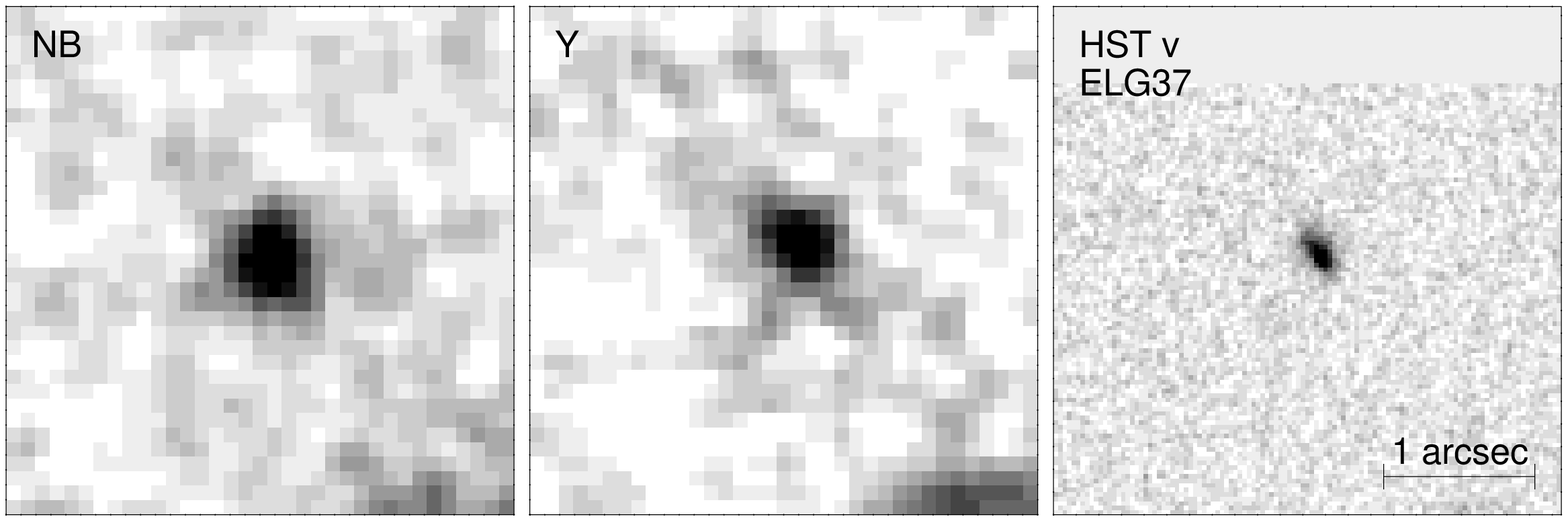, width=6.0cm}
\epsfig{file=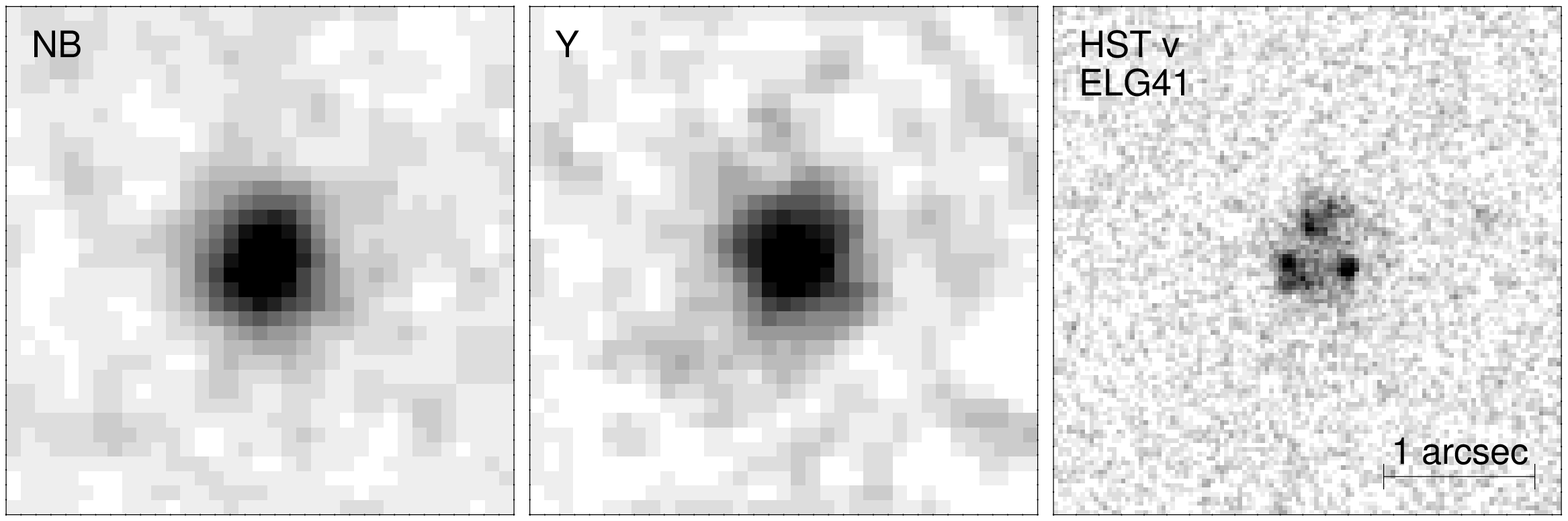, width=6.0cm}\\
\epsfig{file=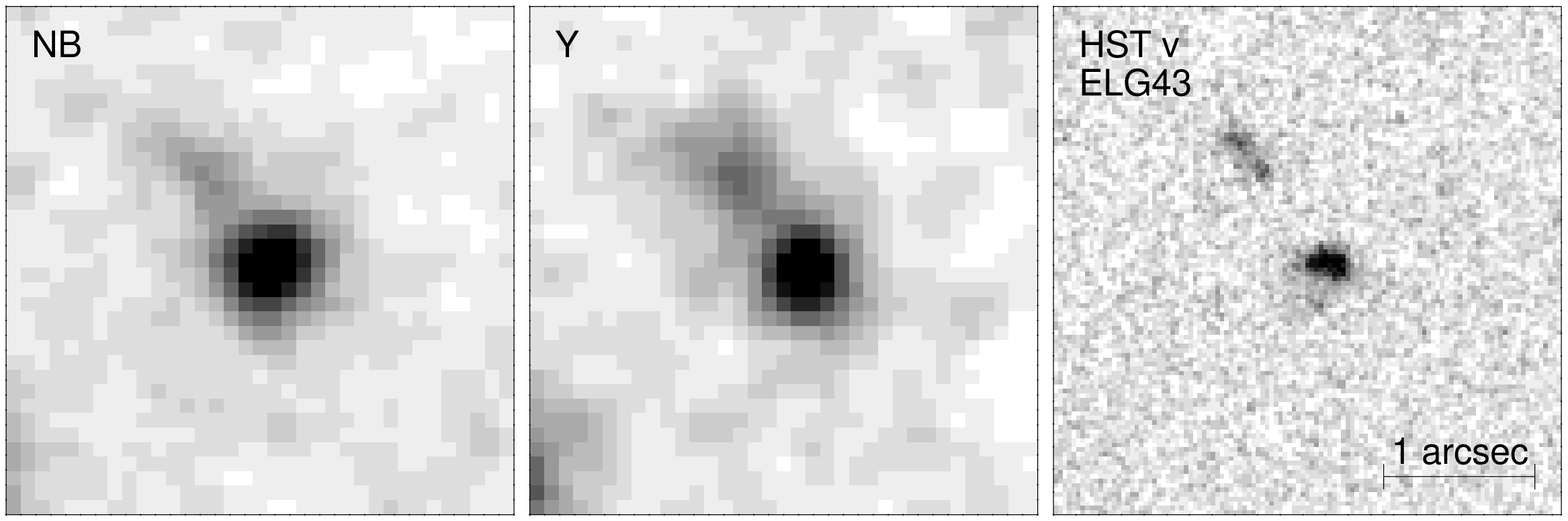, width=6.0cm}
\epsfig{file=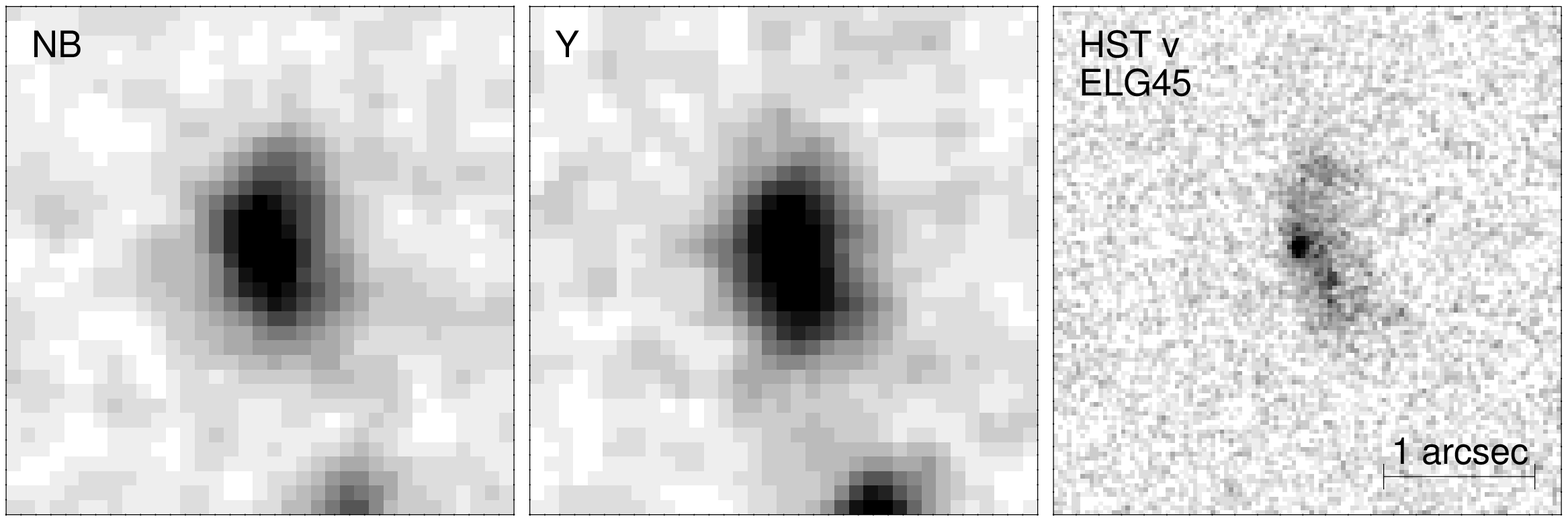, width=6.0cm}
\epsfig{file=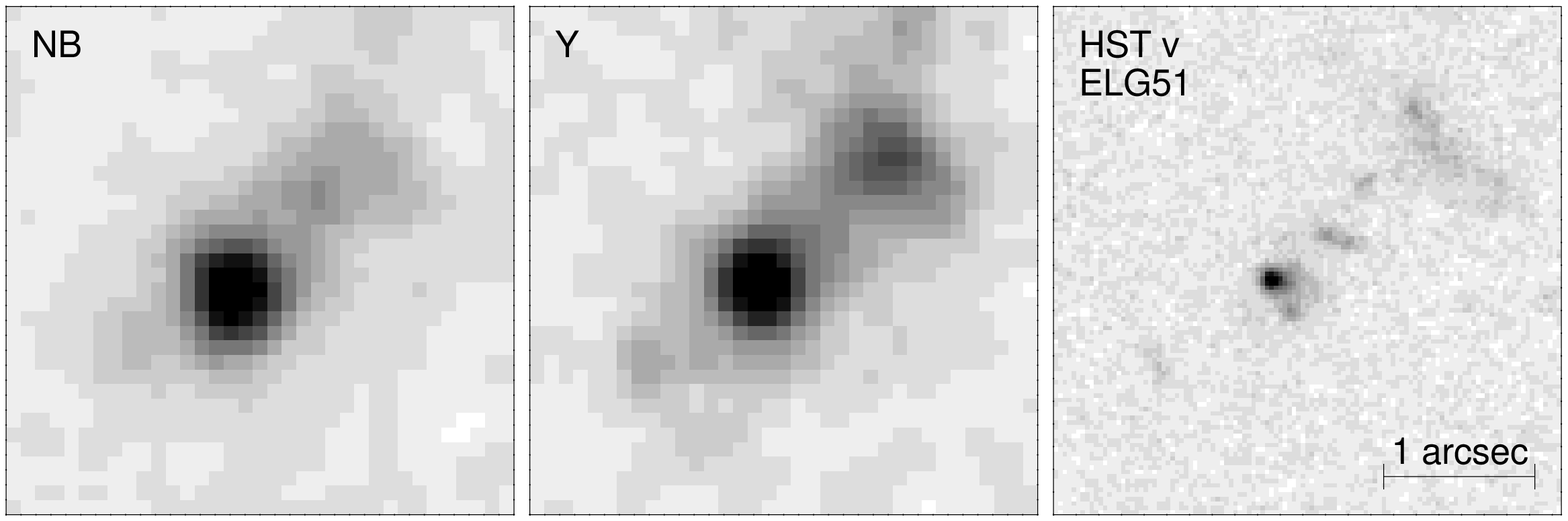, width=6.0cm}\\
\epsfig{file=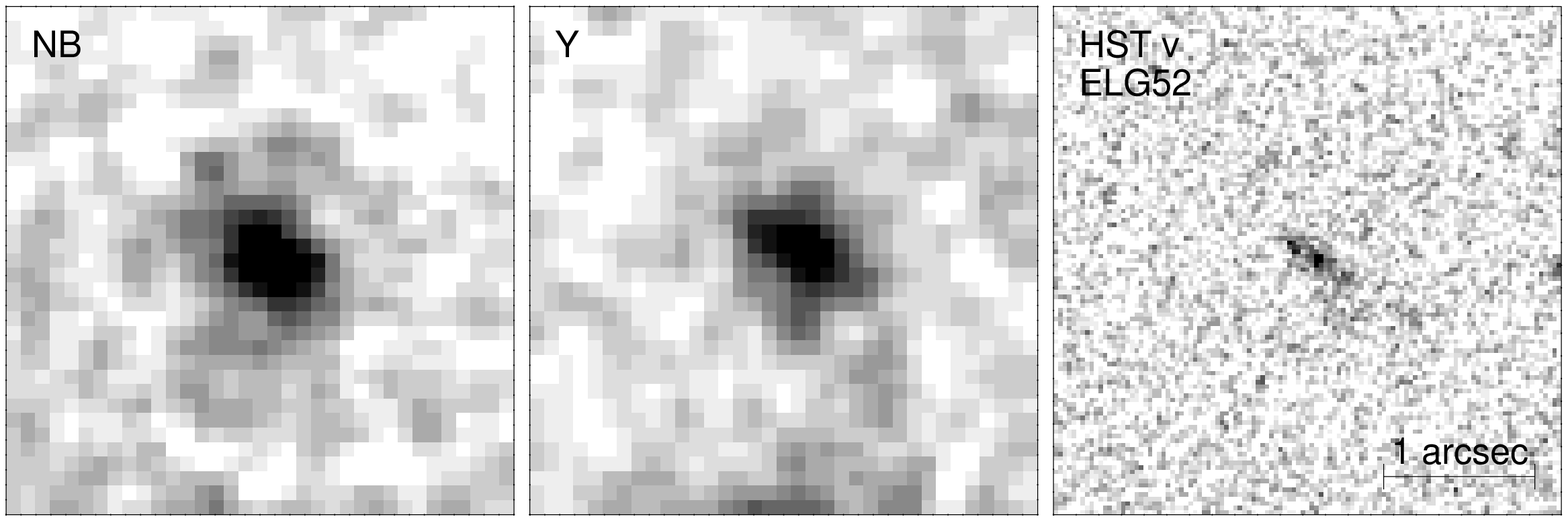, width=6.0cm}
\epsfig{file=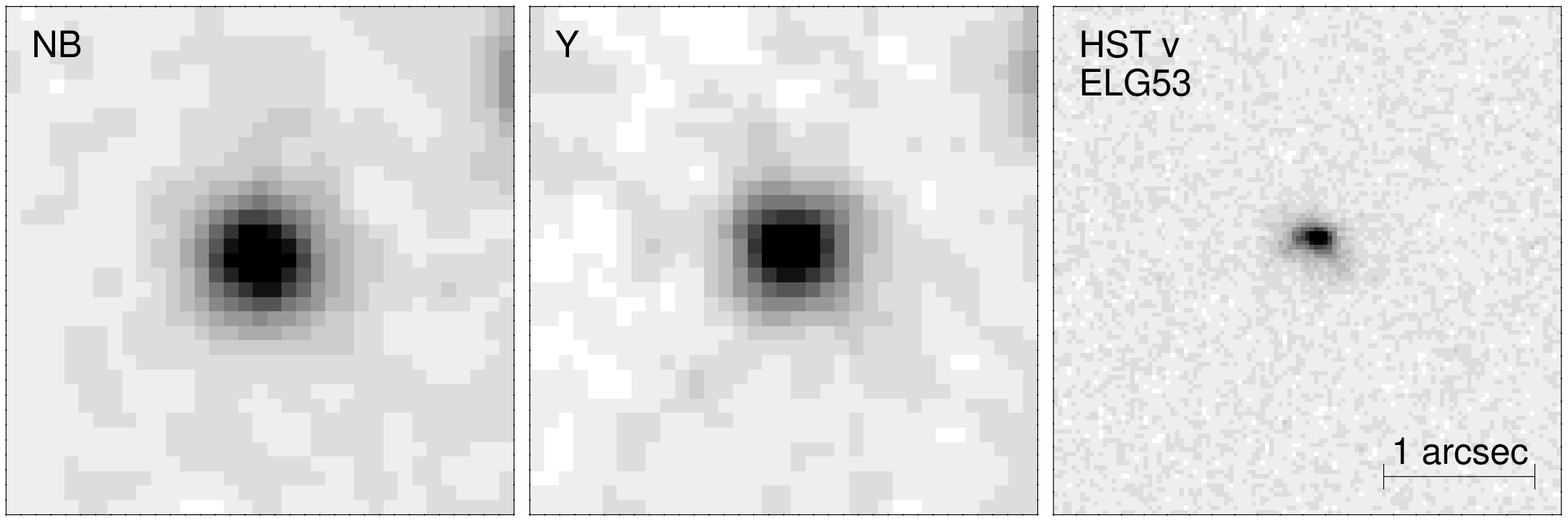, width=6.0cm}
\epsfig{file=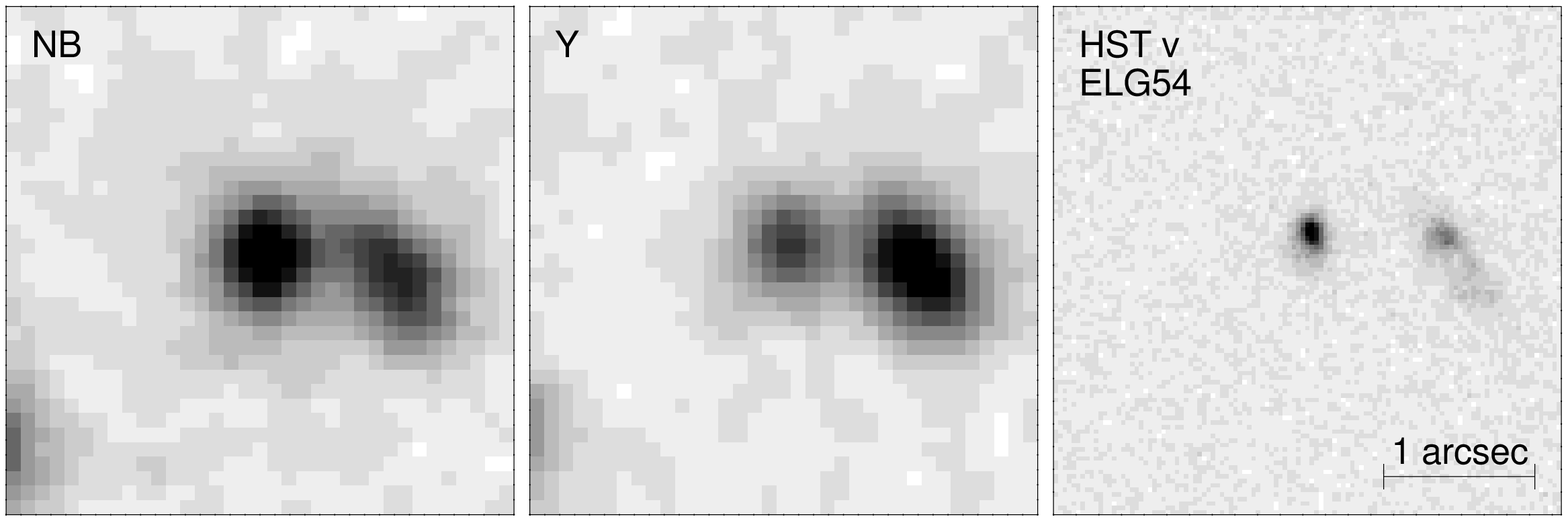, width=6.0cm}\\

\caption{
Thumbnail images of the $NB1060$, $Y$, and {\it HST} F606W (``$v$ band'')
filters for the candidates selected from
$NB1060-Y$ and $NB1060-J$ colours and the additional
source (ELG00) only detected including the $NB1060-J$ colour.
A 1$''$ bar is given on the panels.
}
\label{clipouts}
\end{center}
\end{figure*}

\begin{figure*}
\begin{center}
\addtocounter{figure}{-1}

\epsfig{file=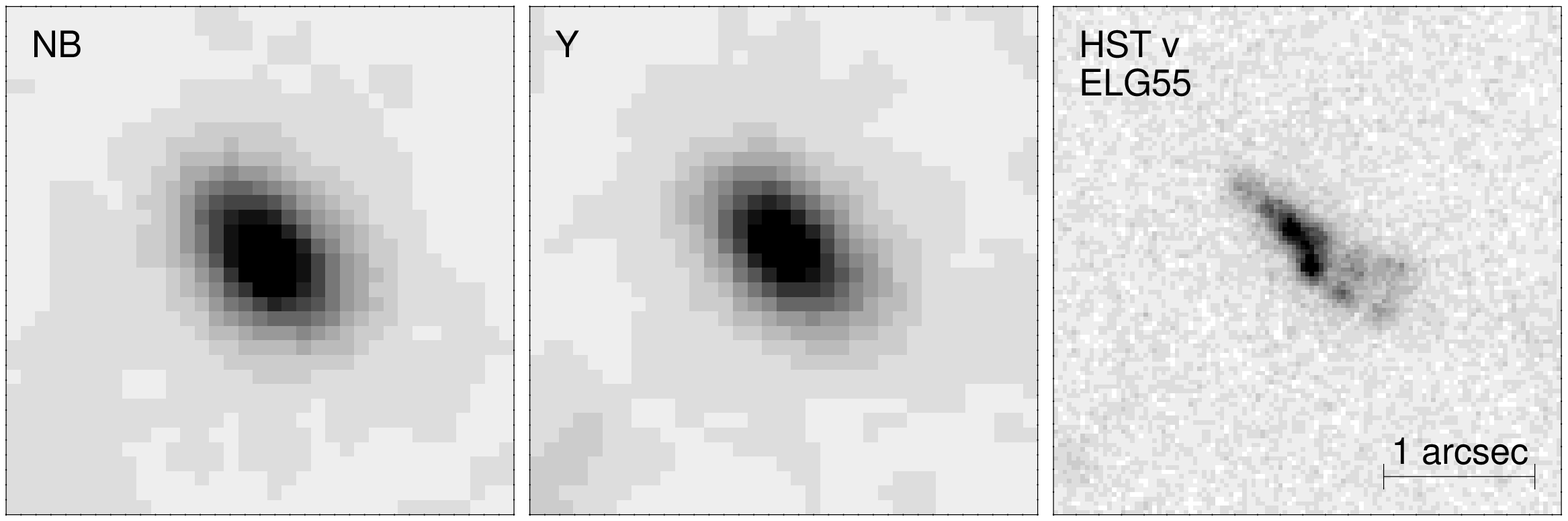, width=6.0cm}
\epsfig{file=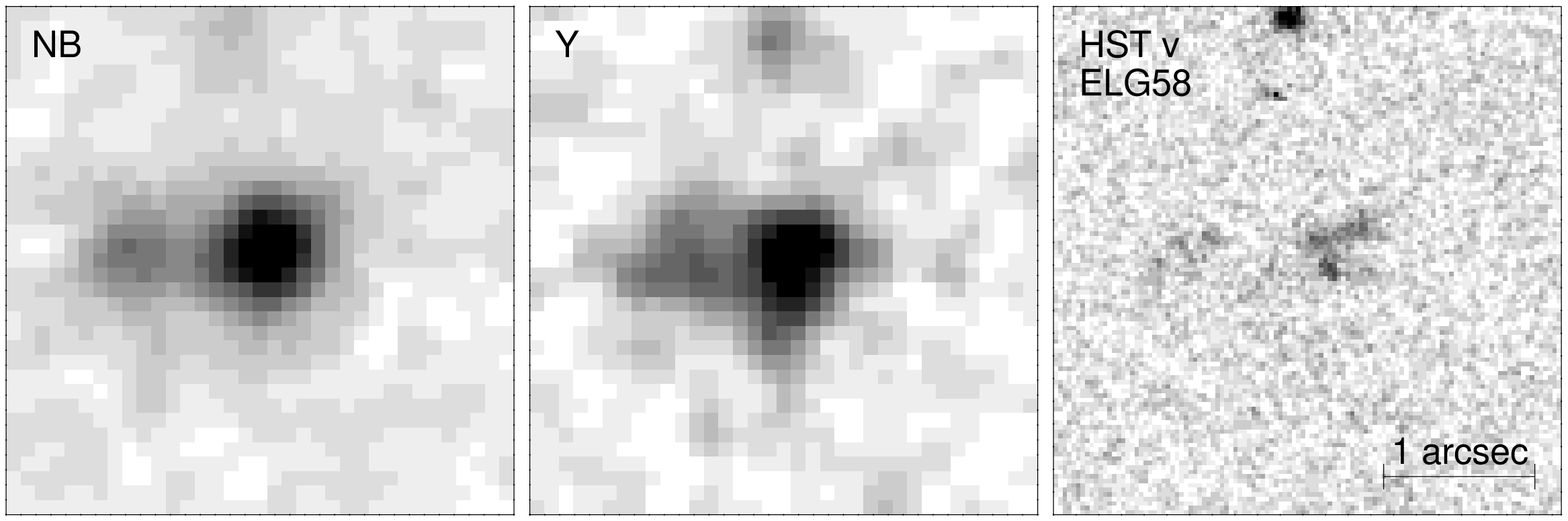, width=6.0cm}
\epsfig{file=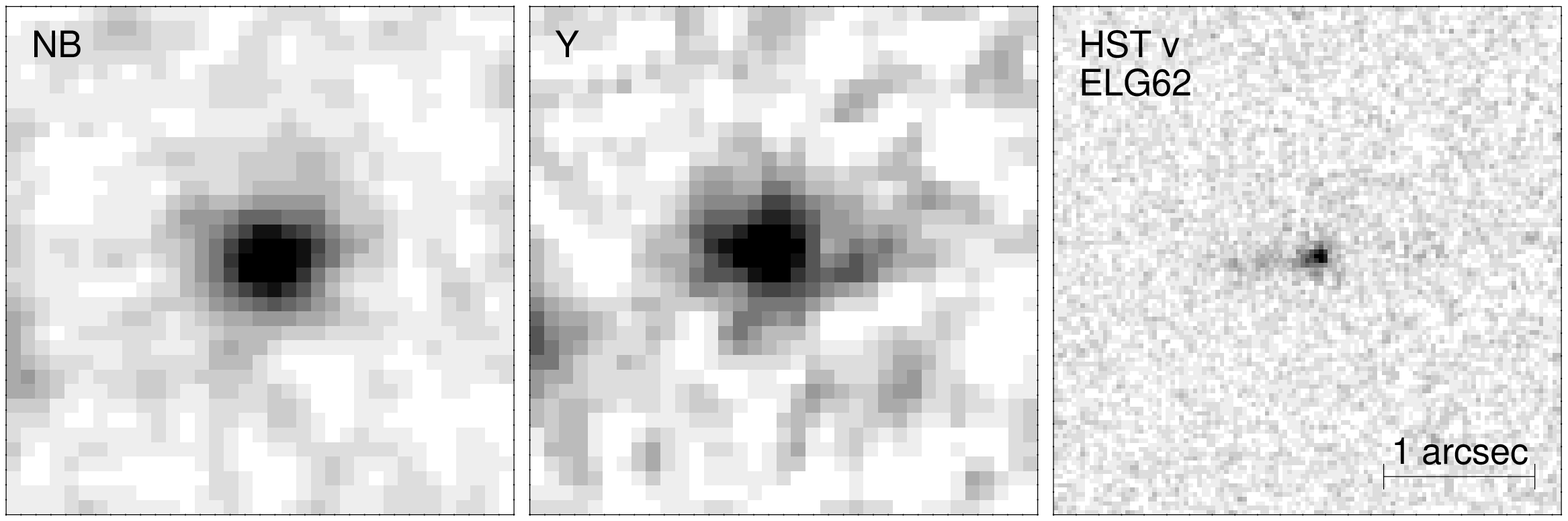, width=6.0cm}\\
\epsfig{file=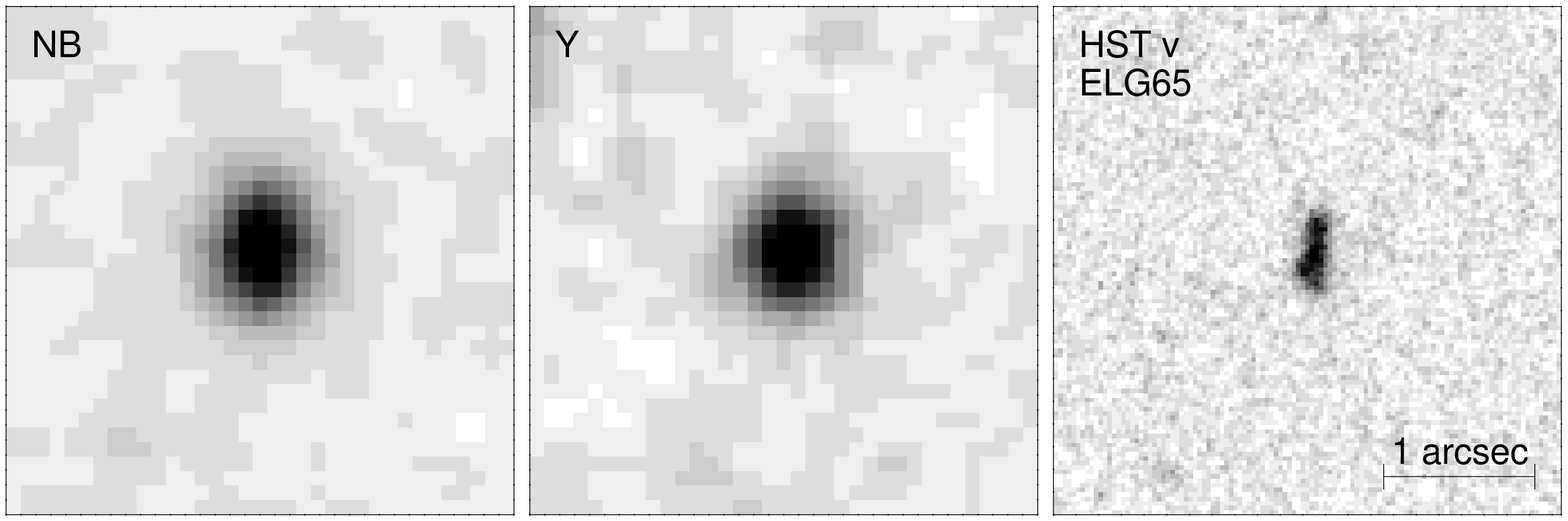, width=6.0cm}
\epsfig{file=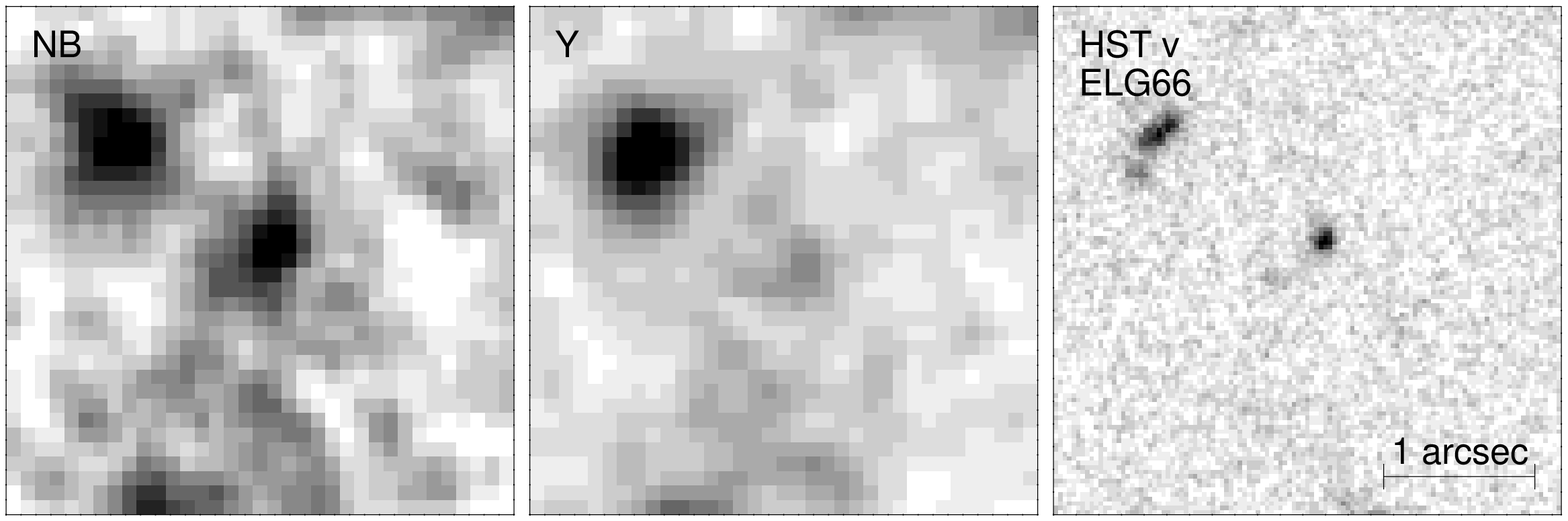, width=6.0cm}
\epsfig{file=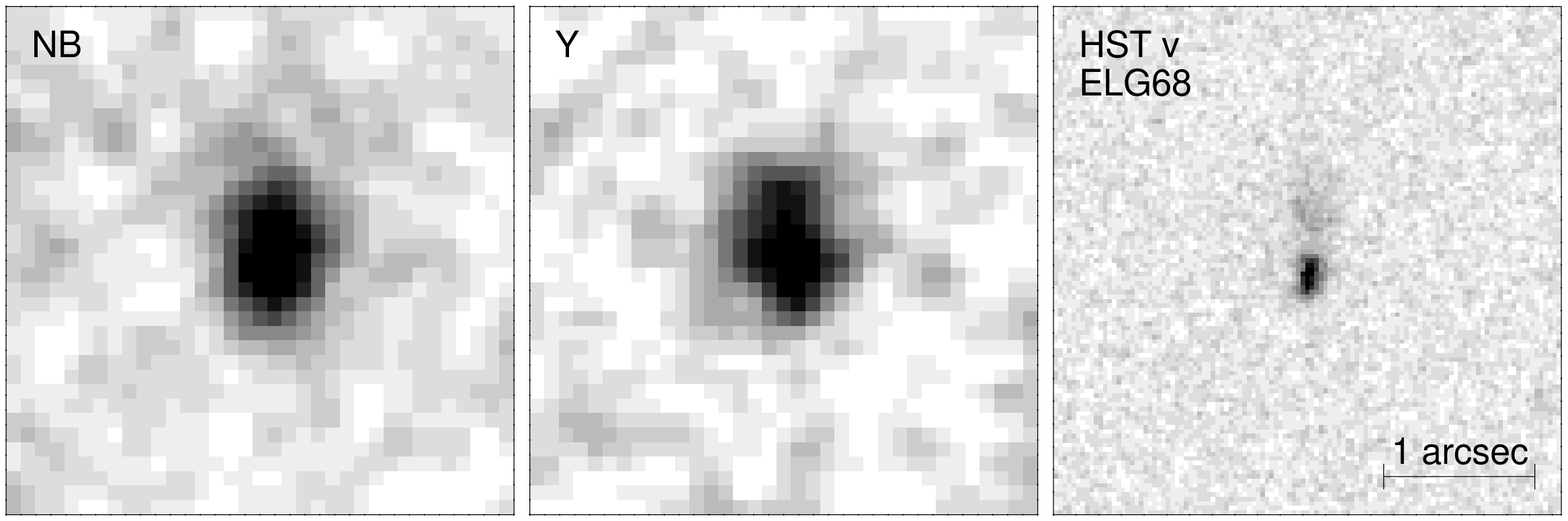, width=6.0cm}\\
\epsfig{file=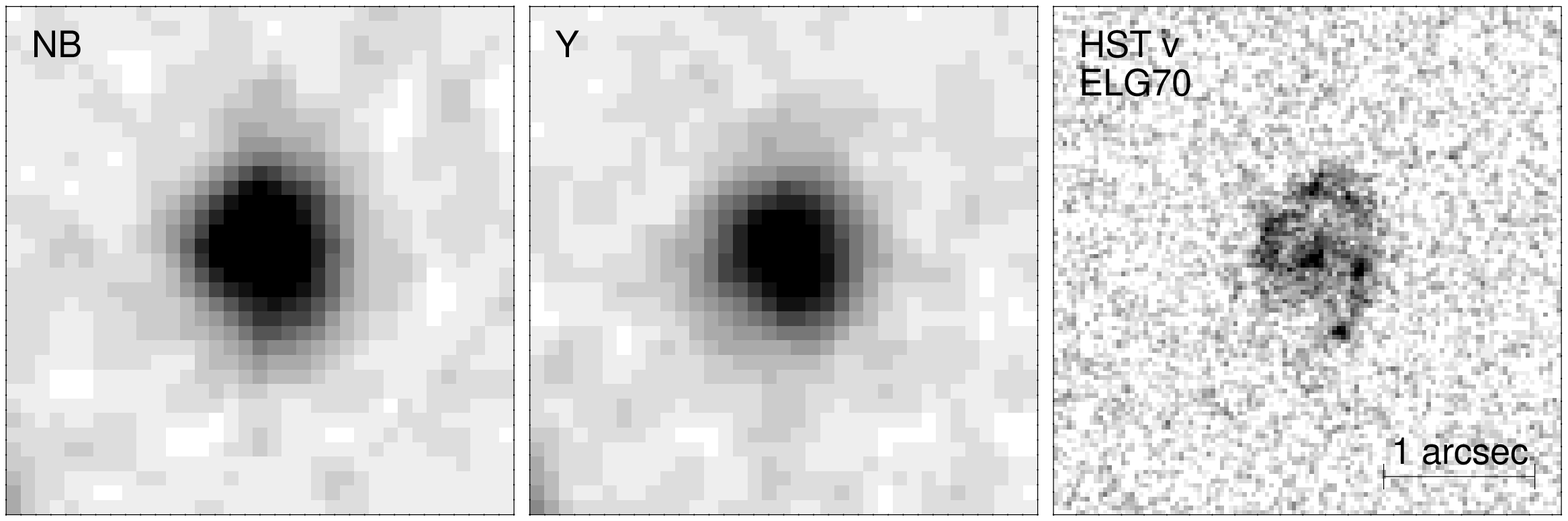, width=6.0cm}
\epsfig{file=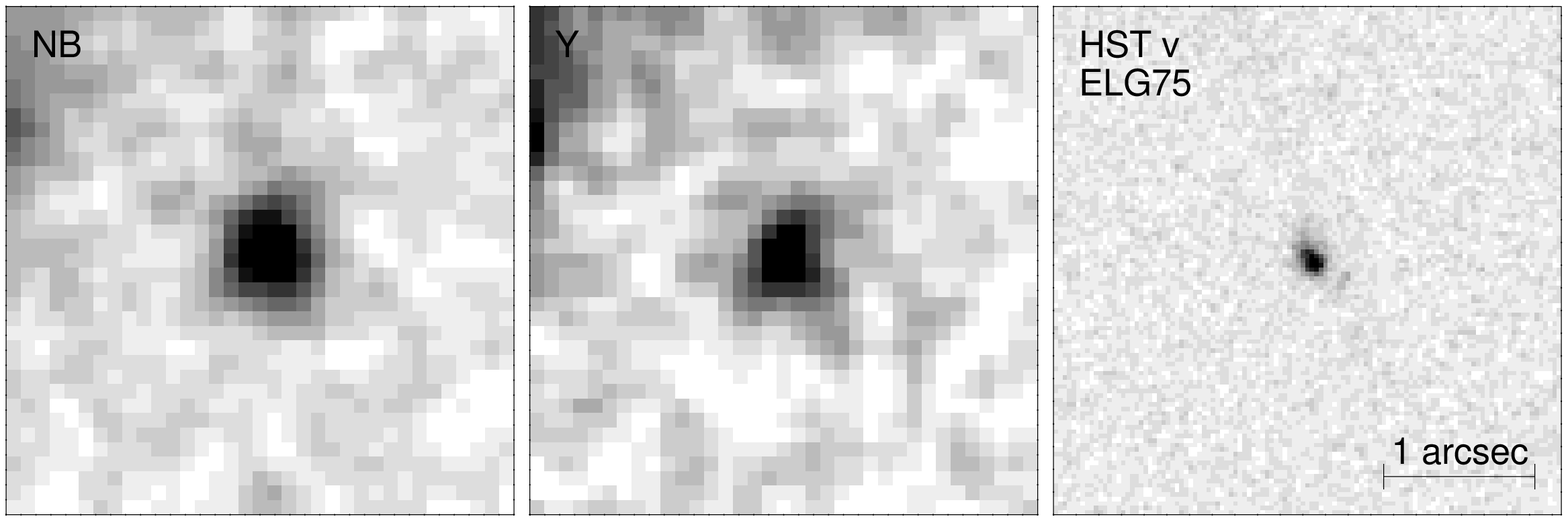, width=6.0cm}
\epsfig{file=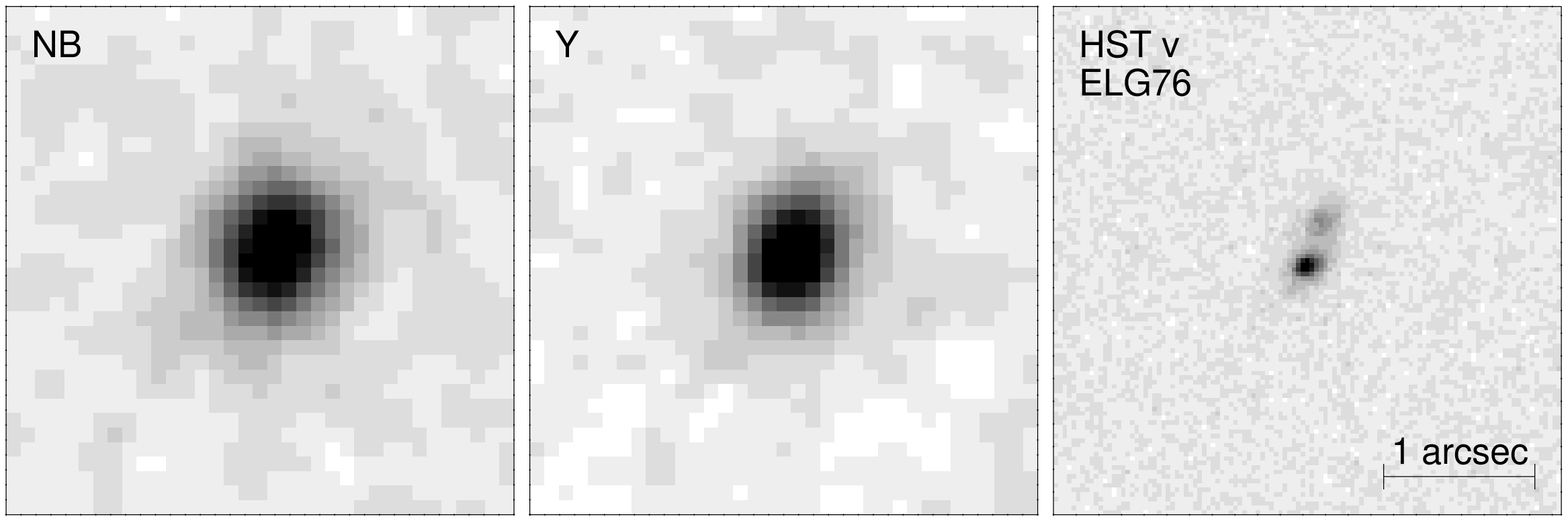, width=6.0cm}\\
\epsfig{file=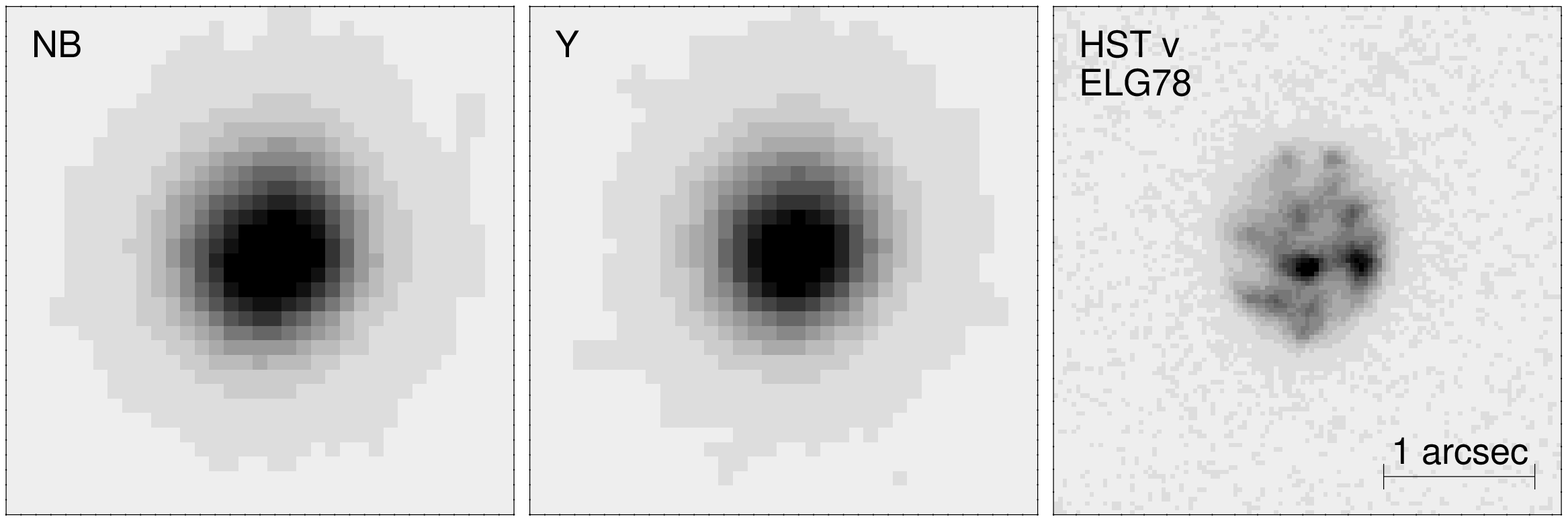, width=6.0cm}
\epsfig{file=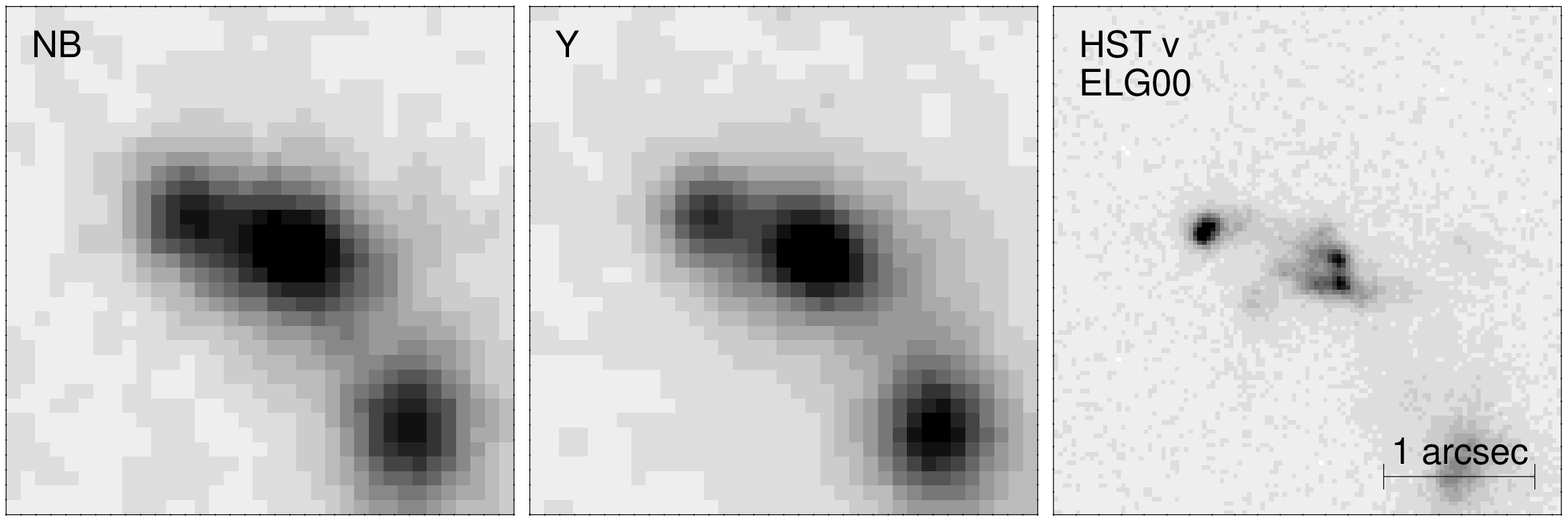, width=6.0cm}
\caption{
continued.
}
\label{clipouts2}
\end{center}
\end{figure*}

\end{document}